\documentclass[preprint,10pt,letterpaper]{aastex}

\usepackage{amsmath}
\usepackage{amssymb}
\usepackage{graphicx}

\def\inverf{\operatorname{erf}^{-1}}
\DeclareMathOperator{\covar}{cov}

\newcommand{\transp}[1]{#1^\mathrm{T}}

\newcommand{\plotthree}[5]{
\begin{figure}[#1]
#2
#3
#4
#5
\end{figure}
}

\newcommand{\plotfour}[6]{
\begin{figure}[#1]
#2
#3
 \\
#4
#5
#6
\end{figure}
}

\newcommand{\plotnine}[9]{
\begin{center}$
\begin{array}{ccc}
#1
#2
#3
\\
#4
#5
#6
\\
#7
#8
#9
\end{array}
$\end{center}
}

\slugcomment{Accepted for publication in \textit{The Astrophysical Journal}. \copyright\ 2014. The American Astronomical Society. All rights reserved.}

\begin{document}

\title{Simulated Performance of Timescale Metrics for Aperiodic Light Curves}

\author{Krzysztof Findeisen}
\affil{Cahill Center for Astronomy and Astrophysics, California Institute of Technology, MC 249-17, Pasadena, CA 91125}
\email{krzys@astro.caltech.edu}

\author{Ann Marie Cody}
\affil{Spitzer Science Center, California Institute of Technology,  MC 314-6, Pasadena, CA 91125}

\author{Lynne Hillenbrand}
\affil{Cahill Center for Astronomy and Astrophysics, California Institute of Technology, MC 249-17, Pasadena, CA 91125}

\begin{abstract}
Aperiodic variability is a characteristic feature of young stars, massive stars, and active galactic nuclei. With the recent proliferation of time domain surveys, it is increasingly essential to develop methods to quantify and analyze aperiodic variability. 
We develop three timescale metrics that have been little used in astronomy -- $\Delta m$-$\Delta t$ plots, peak-finding, and Gaussian process regression -- and present simulations comparing their effectiveness across a range of aperiodic light curve shapes, characteristic timescales, observing cadences, and signal to noise ratios. We find that Gaussian process regression is easily confused by noise and by irregular sampling, even when the model being fit reflects the process underlying the light curve, but that $\Delta m$-$\Delta t$ plots and peak-finding can coarsely characterize timescales across a broad region of parameter space. We make public the software we used for our simulations, both in the spirit of open research and to allow others to carry out analogous simulations for their own observing programs.
\end{abstract}


\def\dmdt{$\Delta m$-$\Delta t$}

\def\lcmcversion{2.3.0}
\def\timescversion{1.0.0}
\def\kpflibversion{1.0.0}

\section{Introduction}

Observational astronomy is currently undergoing a revolution following the advent in the 1990's of synoptic time-domain data sets. Led by the Massive Compact Halo Objects (MACHO) project and the continuing Optical Gravitational Lensing Experiment (OGLE) project, the time domain astronomy effort is now complemented by newer survey projects such as the All-Sky Automated Survey (ASAS), the Catalina Real-Time Transient Survey (CRTS), and the Palomar Transient Factory (PTF).  Additional work on variability is being performed at other wavelengths (e.g. the VISTA Variables in the Via Lactea or VVV survey) and with more precision and higher cadence from space (e.g. MOST and the K2 mission).  Future ground-based time domain projects in the optical include the Zwicky Transient Facility (ZTF) and the Large-Scale Synoptic Survey Telescope (LSST). The coming flood of time-domain data will enable new approaches to long-standing questions in all fields of astrophysics --- but only if we have the necessary tools to make full use of variability information.

Timescales are an important clue to the physics underlying variability, and therefore key to making the most of time-domain surveys.  Pulsating variables such as Miras, Cepheids, RR Lyrae, and ZZ Ceti stars have such a timescale, the pulsation period.  Periodicity or multiperiodicity is also the basis of asteroseismological studies and stellar rotation work. 
In the study of variable young stars, the topic that motivated this paper, the bulk of the published literature focuses on derivation of periods and scientific interpretation thereof.  Notable periodic variability work includes papers by, \citet[e.g.,][]{2000ASPC..219..216G, RSM2006, HEM2007, IHA2008, CB2007}, typically using periodograms \citep[e.g.,][]{LombPeriod, ScarglePeriod, Period04} to identify the dominant period. Most authors assume that the observed periodic signal reflects the rotation period of the star, regardless of whether the variability origin could be attributed to accretion-induced hot spots or to cool starspots \citep{HHG1994}.

However, not all variables are periodic, and aperiodic variables can probe different physics from periodic variables. In accreting young stars, complex physical processes also lead to aperiodic variability, which is cataloged but relatively uninterpreted in the young star literature. The suspected physical processes may occur at different locations (from the stellar photosphere, from the accretion flow region, or over a broad circumstellar disk) or driven by different physics (dynamical changes, re-radiation of variability at other wavelengths, magnetic field reorganization, or thermal evolution), and result in a range of different timescales, from hours to years. Similar complexities apply to other accreting systems such as cataclysmic variables, x-ray binaries, and active galactic nuclei.
\label{intro_whytime}

Unlike many periodic phenomena, the aperiodic variables' connections to specific physics are not well-established. It is, in general, not known whether a particular ``irregular'' signal probes a dynamical, radiative, or thermal timescale.
Roughly half to two-thirds of the variable stars in star-forming regions do not have well-defined periods \citep[e.g.,][]{SE2004, RME2009, SigOri}. As aperiodic variables constitute a large fraction of variable stars in star-forming regions, and are expected to probe variable accretion flow or geometry \citep[e.g.,][]{2003ApJ...595.1009R}, rapidly changing disk structures \citep[e.g.,][]{DustTurner}, and other key processes, characterizing and understanding aperiodic variability is essential to completing our understanding of young star and disk physics. Analogous arguments apply to aperiodic variables in other fields, such as active galactic nuclei where sizable data sets also exist \citep[e.g.,][]{Macleod2012}.

Our interest in aperiodic variability lies in the interpretation of several different time series data sets on young star samples.  First is the Palomar Transient Factory, which we have used for an unprecedented characterization of young star variability in the North America Nebula (NAN) complex, as described in \citet{KpfThesis}. The high cadence and long base line of the PTF-NAN data are unmatched among blind surveys of star-forming regions, and allow a variety of variability on timescales of days to years to be treated with a unified approach.  Second are two higher cadence but shorter duration data sets involving space-based platforms, the standard Young Stellar Object Variability (YSOVAR) data set discussed by \citet{RGS2011} encompassing a number of young clusters and the Coordinated Synoptic Investigation of NGC 2264 (CSI-2264) data set presented by \citet{AmcCorot}.  All are discussed in more detail in the next section. 

The key challenge in making full use of these data sets is that no standard metric analogous to periodograms exists for characterizing aperiodic signals. Therefore, we concentrate on finding a robust metric for characterizing the variability timescale of aperiodic variables and quantitatively distinguishing rapidly- from slowly-varying signals of this type.  Our aim is to establish how to quantify both aperiodic and periodic variability in a uniform manner.

In order to use any timescale metric to best effect, one must first understand the impact of measurement noise, limited cadence, and finite observing windows on the analysis. For example, the generation of periodograms is affected by aliasing, which is a strong function of the cadence and can be characterized by tools such as the window function \citep{WindowFunction}. To our knowledge, no analytical tool analogous to the window function has been developed for the analysis of aperiodic signals sampled at a particular cadence. In this work, we employ simulated data sets, which allow us to experiment with the effects of noise and cadence while giving us a ``true'' timescale to which to compare the output of each timescale metric.
Our work supplements recent publications from the extragalactic community \citep{2010MNRAS.404..931E, deDiego2010, deDiego2014} that characterize statistical tools popular in AGN research.

In this publication, we present simulations of both periodic and aperiodic signals across a grid of amplitude, timescale, noise, and observing cadence. We describe the signals in section~\ref{analytic_testlclist} and the simulations in section~\ref{siminfo}. Section~\ref{lcmc_dmdt} introduces three candidate timescale metrics -- \dmdt~plots, peak-finding, and Gaussian process regression -- that we then apply to the simulated data in section~\ref{timescale_examples}.
By observing how the computed timescale varies with input timescale, other light curve properties, and observing cadence, and how stable it is across multiple simulation runs, we evaluate the performance and reliability of each candidate metric in section~\ref{perform}. We finish with recommendations for characterizing aperiodic timescales for real data. Application to our own data on variability in the North America Nebula complex is deferred to a later paper.

\section{Test Signals}\label{analytic_testlclist}

Our simulations covered five types of signals: sinusoids, white noise, squared exponential Gaussian processes, damped random walks, and undamped random walks. These signals present a sufficiently diverse set that any timescale metric that works well on all of them is likely to also be reliable when applied to real data. The white noise case served as a control group, and is presented only briefly; the other light curves are illustrated in Figure~\ref{fig_analytic_lc_demos}, while the frequency distribution for each is presented in Figure~\ref{fig_analytic_powerspec}.

\plotfour{ptb}{
\includegraphics[width=0.50\textwidth]{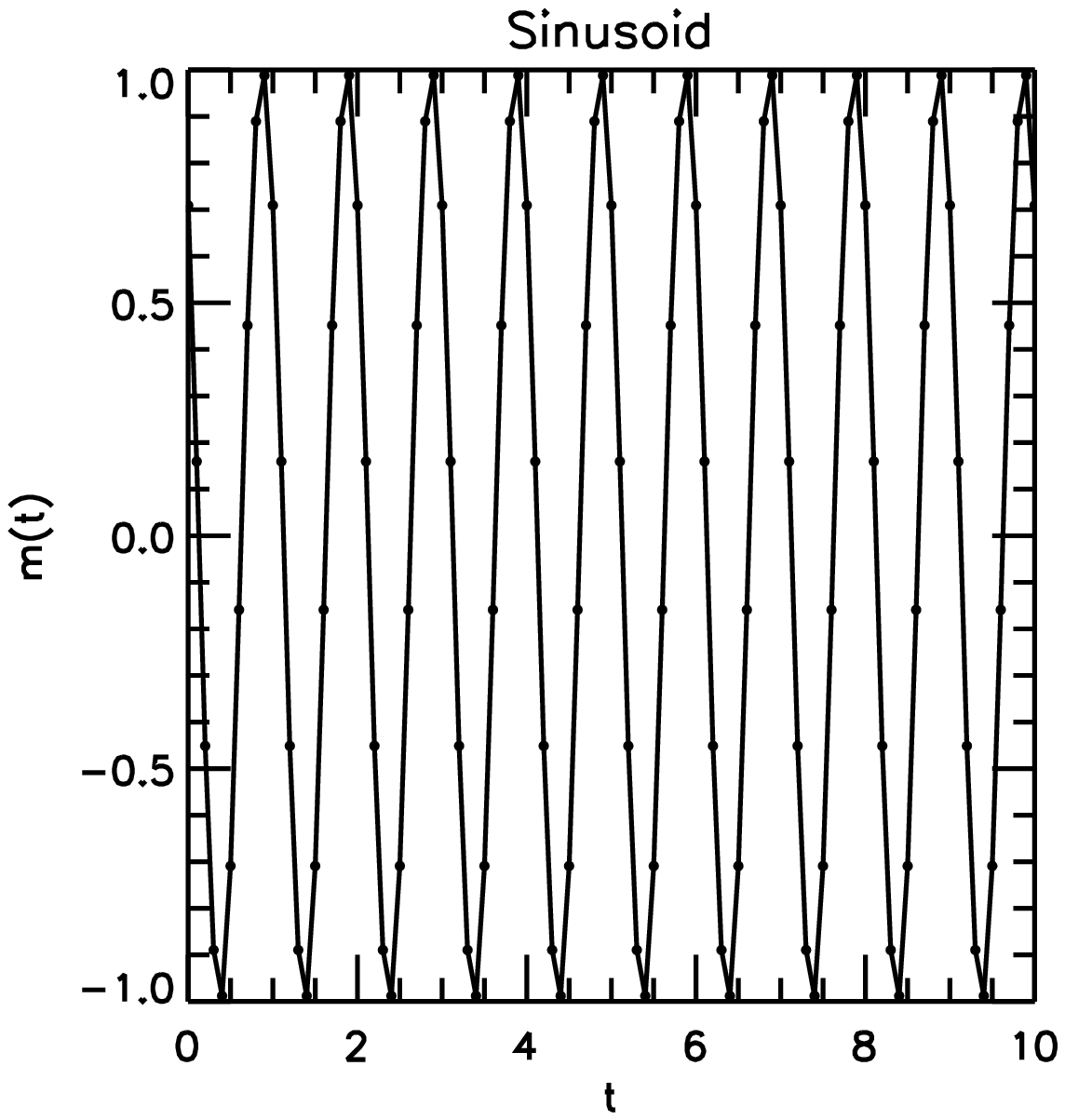}
} {
\includegraphics[width=0.50\textwidth]{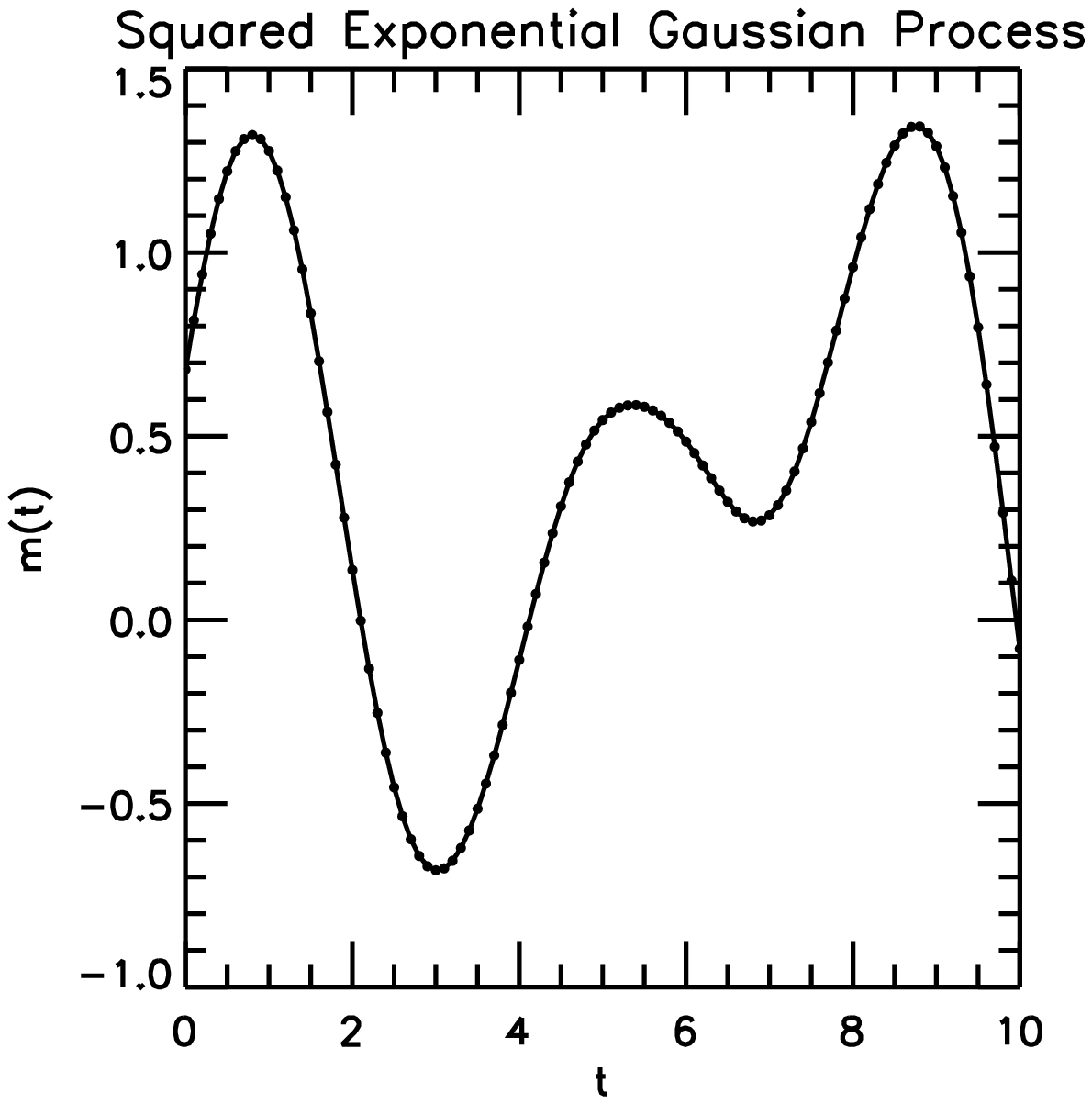}
} {
\includegraphics[width=0.50\textwidth]{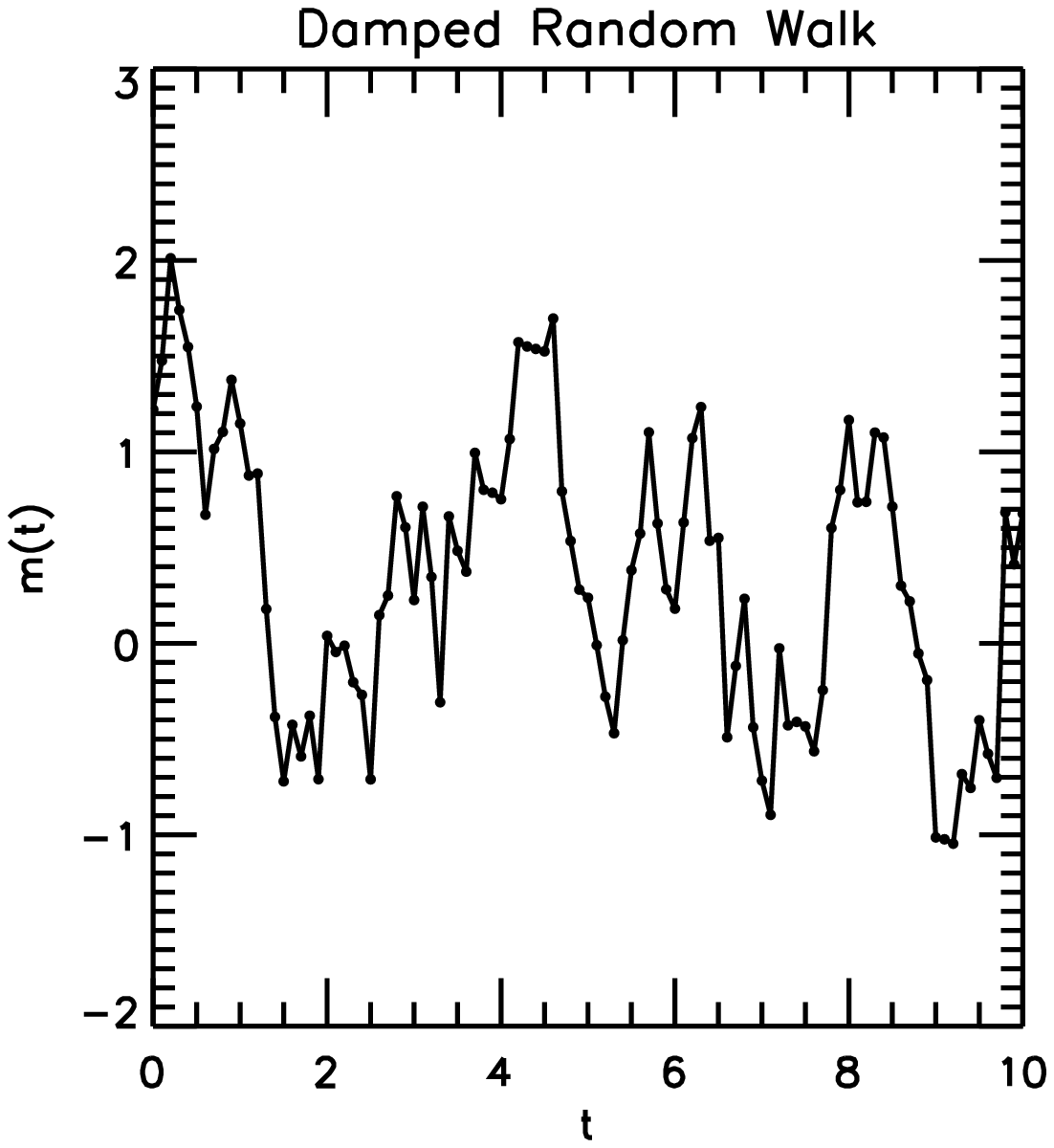}
} {
\includegraphics[width=0.50\textwidth]{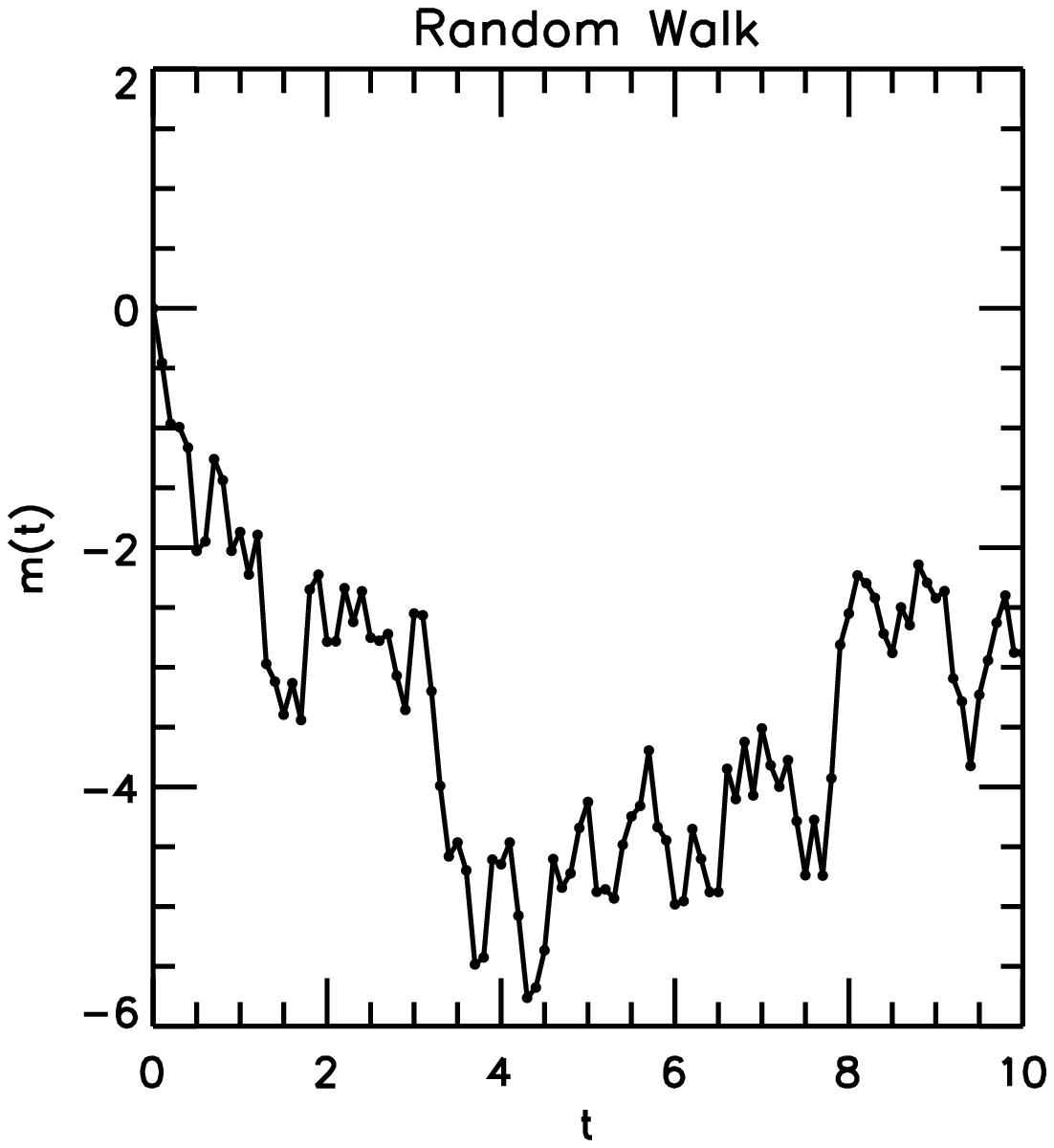}
} {
\caption{Examples of the light curve models discussed in this section, in arbitrary units. The random walk has no characteristic timescale, but has a diffusion constant equal to that of the damped random walk. All other light curves have a characteristic timescale of 1~unit.} \label{fig_analytic_lc_demos}
}

\plotfour{ptb}{
\includegraphics[width=0.48\textwidth]{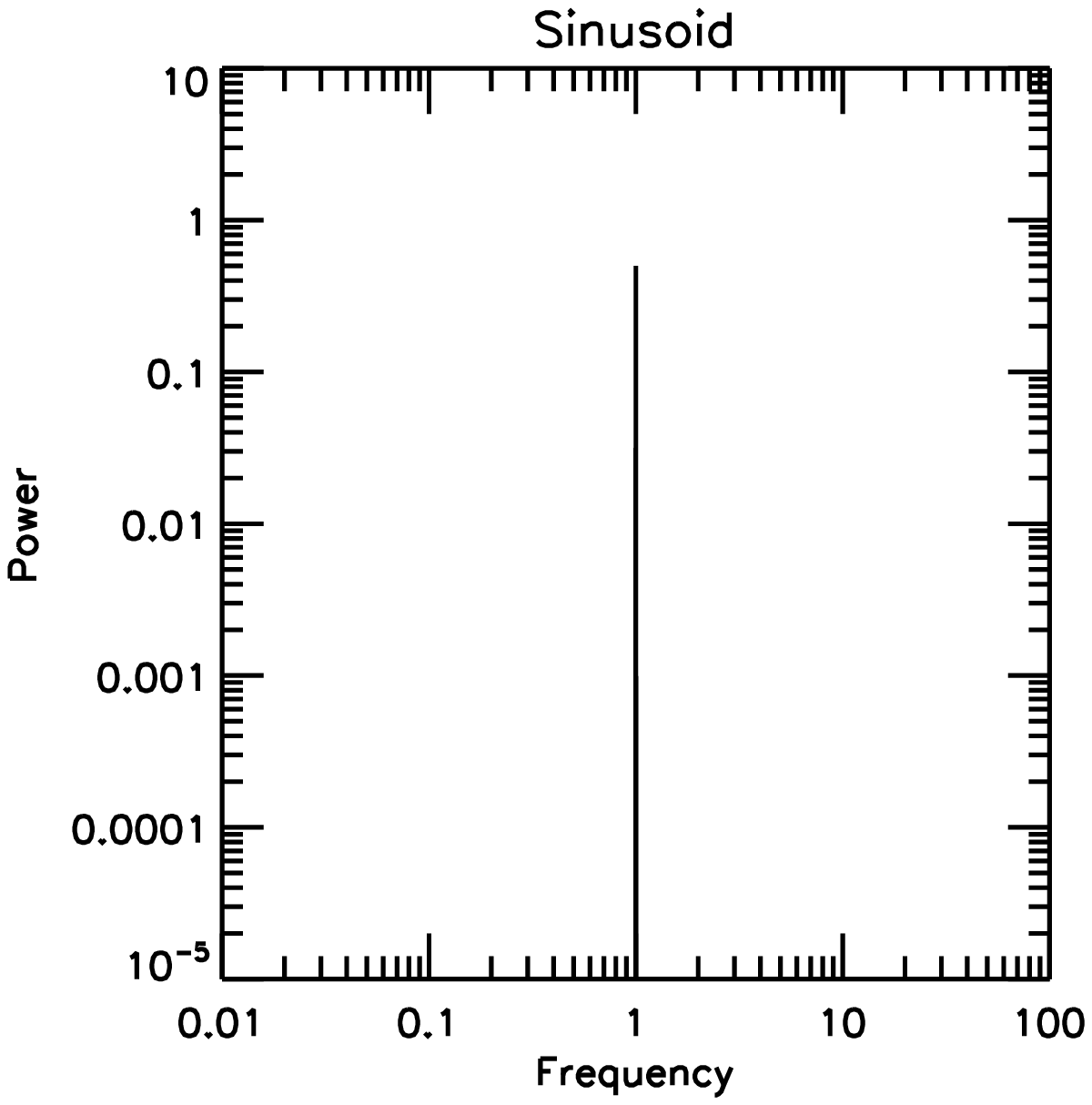}
} {
\includegraphics[width=0.48\textwidth]{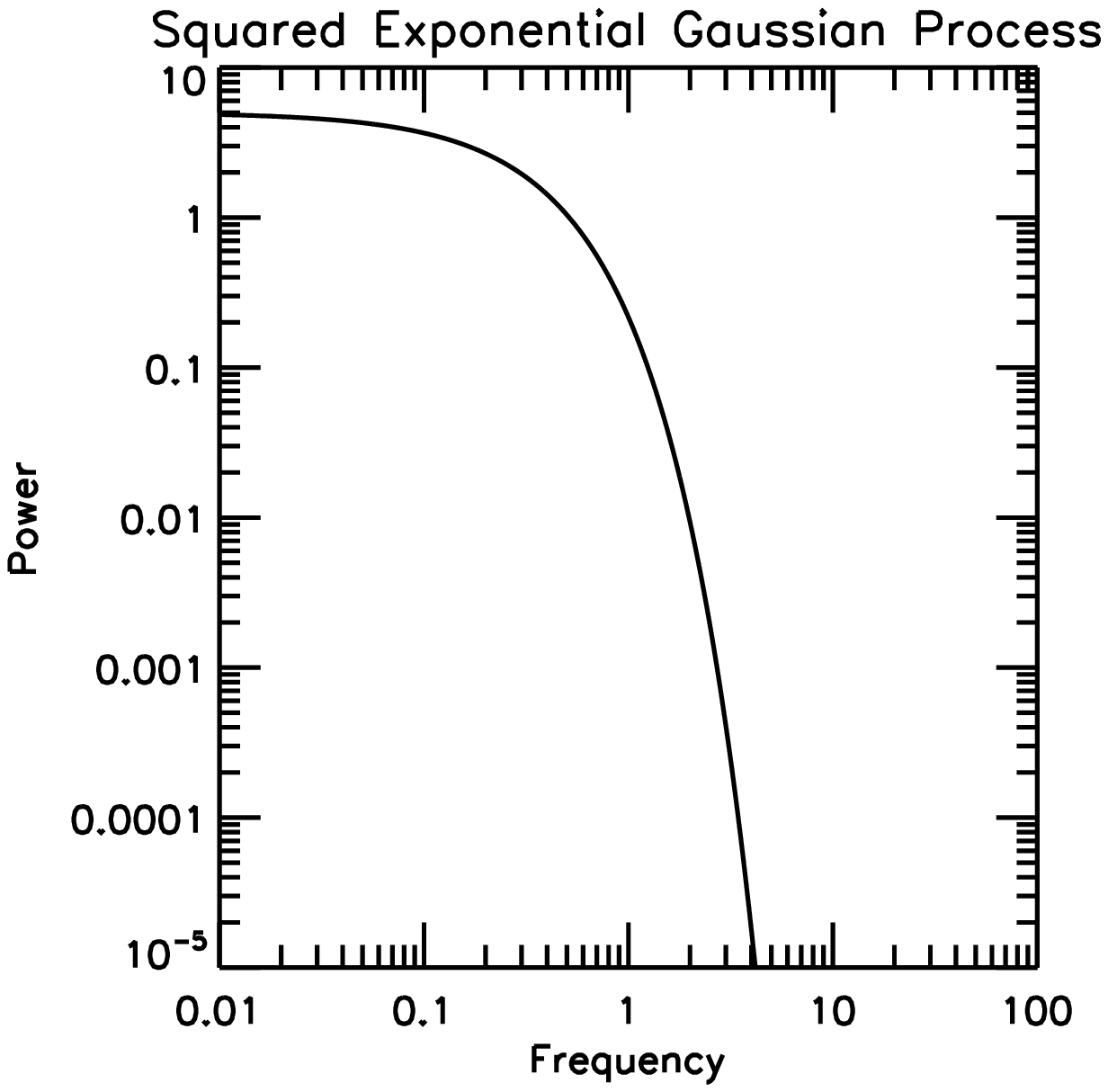}
} {
\includegraphics[width=0.48\textwidth]{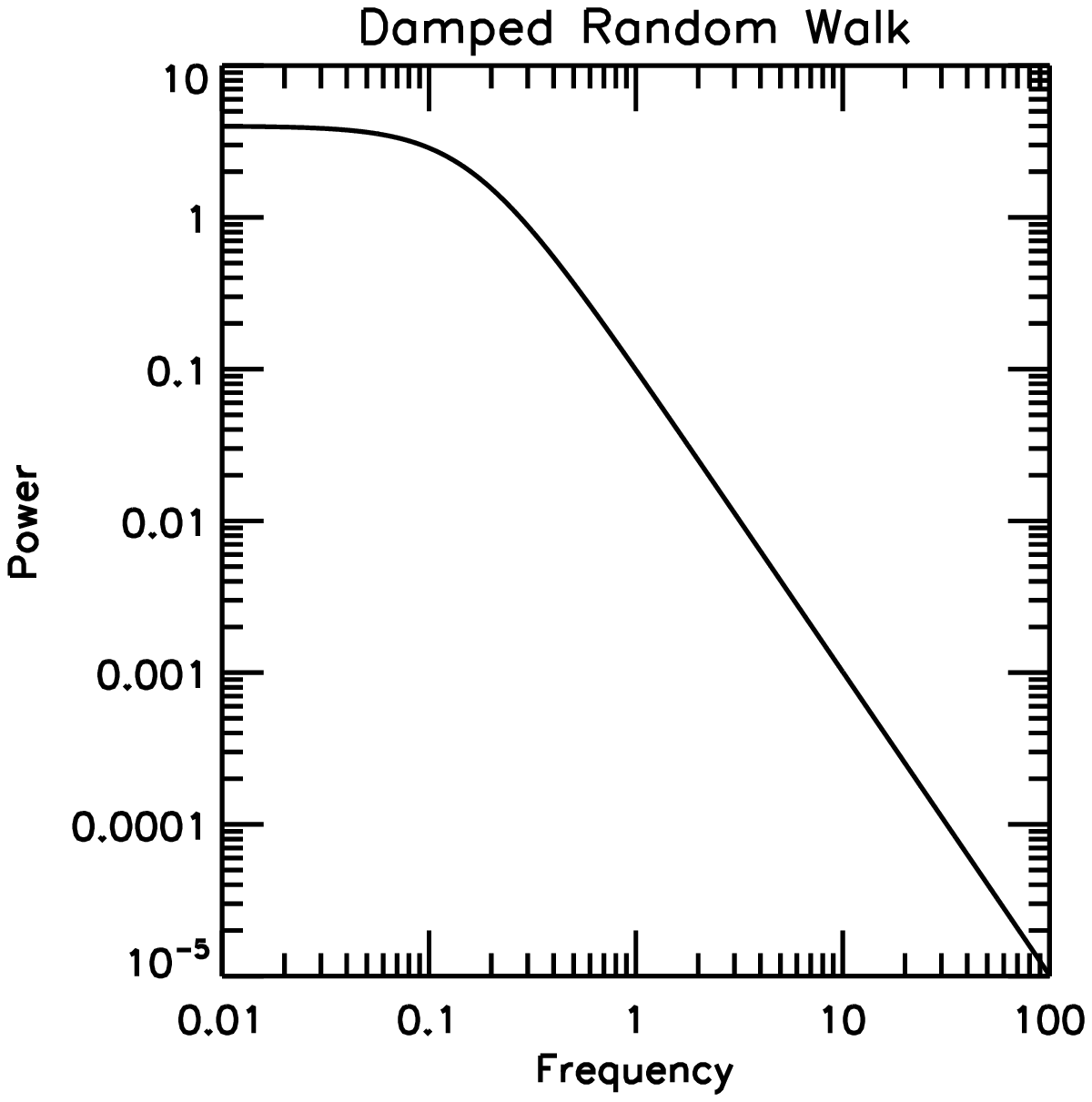}
} {
\centering
\parbox[b][0.48\textwidth][c]{0.4\textwidth}{\large Random walk power spectrum undefined}
} {
\caption{Theoretical power spectra for the light curve models discussed in this section, in arbitrary units. A frequency of 1~unit corresponds to a time separation of 1~unit in Figure~\ref{fig_analytic_lc_demos}. Light curve model parameters are as in Figure~\ref{fig_analytic_lc_demos}.} \label{fig_analytic_powerspec}
\label{fig_analytic_powerspec_sine}\label{fig_analytic_powerspec_aatau}\label{fig_analytic_powerspec_gp1}\label{fig_analytic_powerspec_gp2}\label{fig_analytic_powerspec_drw}
}

\subsection{Sinusoid}\label{defn_lc_sine}

As the archetypal periodic function, a sinusoidal signal serves as a good reference point for comparison to aperiodic functions. This signal is defined as
\begin{equation}
m(t) = m_0 + A \sin{(\omega (t - t_0))} \label{eqn_analytic_lc_sine}
\end{equation}
Note that in this convention, $A$ is the half-amplitude, not the peak-to-peak amplitude.
The difference between the 5th and 95th percentiles of magnitude (hereafter the ``5-95\% amplitude'') of a sine is $2 A \cos{\left(\frac{\pi}{2} (0.10)\right)} = 1.975 A$. The RMS amplitude is $A \sqrt{\pi} = 1.772 A$.

The power spectral density of a sinusoid, following the Fourier transform convention of \citet{RandomWalk}, is the well-known result 
\begin{displaymath}
S(\nu) = \frac{1}{2} A^2 \left(\delta(\nu-\frac{1}{P})+\delta(\nu+\frac{1}{P})\right)
\end{displaymath}
where $\delta$ denotes the Dirac delta function. By definition, a sinusoid has only a single frequency component.

\subsection{Squared Exponential Gaussian Process}

The squared exponential Gaussian process is a probabilistic model where magnitudes are drawn from a Gaussian distribution, and the correlation between magnitudes at any two times follows a Gaussian function of the time difference between them. It has the form 
\begin{equation}
\begin{array}{rcl}
E(m(t)) & = & m_0 \\
V(m(t)) & = & \sigma_m^2 \\
\covar{(m(t_i), m(t_j))} & = & \sigma_m^2 e^{-(t_i-t_j)^2/2\tau^2}
\end{array} \label{eqn_analytic_lc_gp}
\end{equation}
where $E(X)$ denotes the expected value of random variable $X$, $V(X)$ the variance of $X$, and $\covar(X, Y)$ the covariance of random variables $X$ and $Y$. $\tau$ is the correlation time and $\sigma_m$ is the RMS amplitude of the squared exponential Gaussian process.
The 5-95\% amplitude for a squared exponential Gaussian process is $2 \sigma_m \sqrt{2} \ \inverf{0.90}$ = $3.291 \sigma_m$.

The power spectral density of a squared exponential Gaussian process, following the Fourier transform convention of \citet{RandomWalk}, is 
\begin{displaymath}
S(\nu) = 2 \sqrt{2 \pi} \tau \sigma_m^2 e^{-(2\pi \tau \nu)^2/2}
\end{displaymath}
The power spectrum is flat for $\nu \lesssim 1/\tau$, but has almost no power at higher frequencies, as shown in Figure~\ref{fig_analytic_powerspec_gp1}.

\subsection{Damped Random Walk}

A damped random walk, or more formally an Ornstein-Uhlenbeck process \citep{DampedRandomWalk_First}, is a stochastic model in which a variable undergoes random Gaussian perturbations but also experiences a restoring force. Although a damped random walk is normally defined in terms of a stochastic differential equation, it is also a Gaussian process \citep{RandomWalk_GP}, though with weaker short-timescale correlations than the squared exponential Gaussian process introduced previously. If $(m(t) - m_0)$ follows an Ornstein-Uhlenbeck process (we subtract $m_0$ because we follow the formalism of \citet{RandomWalk}, who defines the process to have mean zero), then 
\begin{equation}
\begin{array}{rcl}
E(m(t)) & = & m_0 \\
V(m(t)) & = & \frac{D \tau}{2} \\
\covar{(m(t_i), m(t_j))} 
 & = & \frac{D \tau}{2} e^{-|t_i-t_j|/\tau}
\end{array} \label{eqn_analytic_lc_drw}
\end{equation}
where $D$ is the diffusion constant for the random walk and $\tau$ is the damping time. 

The 5-95\% amplitude for a damped random walk is $2 \sqrt{D \tau} \ \inverf{0.90}$ = $2.327 \sqrt{D \tau}$. The RMS amplitude is $\sqrt{\frac{D \tau}{2}}$.

The power spectral density of a damped random walk, following \citet{RandomWalk}, is 
\begin{displaymath}
S(\nu) = \frac{2 D \tau^2}{1 + (2\pi \tau \nu)^2}
\end{displaymath}
The power spectral density is flat for $\nu \lesssim 1/\tau$, but decays as $1/\nu^2$ at higher frequencies (Figure~\ref{fig_analytic_powerspec_drw}).

\subsection{Undamped Random Walk}

The random walk, or, more formally, a Wiener process \citep{RandomWalk_First}, is the limit of a damped random walk when the damping time $\tau$ becomes infinite. It is more commonly thought of as a model in which a variable is changed only through random Gaussian perturbations. Since a random walk has no characteristic timescale, it serves as a test case for how timescale metrics perform when there is no well-defined quantity to measure.

Following \citet{RandomWalk}, the mean and variance are 
\begin{equation}
\begin{array}{rcl}
E(m(t)) & = & m_0 \\
V(m(t)) & = & D (t-t_0) \\
\covar(m_i, m_j) & = & \lim_{\tau \rightarrow \infty} \frac{D \tau}{2} e^{-(t_2-t_1)/\tau} \left(1 - e^{-2(t_1-t_0)/\tau}\right)
\end{array} \label{eqn_analytic_lc_rw1}
\end{equation}
where $D$ is the diffusion constant, $t_0$ is the starting point of the random walk, $t_1$ is the earlier of $t_i$ and $t_j$, and $t_2$ is the later of $t_i$ and $t_j$. 

\subsection{White Noise}

A Gaussian white noise process is a probabilistic model where magnitudes are drawn from a Gaussian distribution, and are independent of the magnitudes at all other times. Formally, 
\begin{equation}
\begin{array}{rcl}
E(m(t)) & = & m_0 \\
V(m(t)) & = & \sigma_m^2 \\
\covar{(m(t_i), m(t_j))} & = & \sigma_m^2 \delta(t_i, t_j)
\end{array} \label{eqn_analytic_lc_wn}
\end{equation}
where $\delta$ denotes the Kronecker delta, not the Dirac delta. $\sigma_m$ can be interpreted as the RMS amplitude of the white noise process.
The 5-95\% amplitude for a white noise process is $2 \sigma_m \sqrt{2} \ \inverf{0.90}$ = $3.291 \sigma_m$.

The power spectral density of a white noise process, following \citet{RandomWalk}, is 
\begin{displaymath}
S(\nu) = 2\sigma_m^2
\end{displaymath}
By definition, a white noise process has a uniform power spectrum.

\section{Simulations}\label{siminfo}

In this section, we describe the software, input parameters, and analysis methods we used to test timescale metrics on simulated light curves. The metrics themselves are presented in detail in the next section.

\subsection{LightcurveMC: an extensible light curve simulation program}

We have developed a program to generate random light curves and to perform automated statistical analysis of each light curve. LightcurveMC is designed to be highly modular, allowing new light curve types or new analysis tools to be introduced without excessive development overhead. The statistical tools are completely agnostic to how the light curve data is generated, and the light curve generators are completely agnostic to how the data will be analyzed. The use of fixed random seeds throughout guarantees that the program generates consistent results from run to run.

All figures and results in this work were generated using LightcurveMC~\lcmcversion. It is available, with documentation, from the Astrophysics Source Code Library as ascl:1408.012 \citep{LightcurveMC}. For the simulation runs presented here, the program was built using GCC 4.4.7-3 on Red Hat Enterprise Linux 6.4 and linked against kpfutils\footnote{\url{https://github.com/kfindeisen/kpfutils}}~\kpflibversion, Timescales\footnote{\url{https://github.com/kfindeisen/Timescales}}~\timescversion, Boost\footnote{\url{http://www.boost.org/}} 1.41.0, GSL\footnote{\url{http://www.gnu.org/software/gsl/}} 1.10, TCLAP\footnote{\url{http://tclap.sourceforge.net/}} 1.2.1, R\footnote{\url{http://www.r-project.org/}} 2.15.1, Rcpp\footnote{\url{http://dirk.eddelbuettel.com/code/rcpp.html}} 0.10.3, RInside\footnote{\url{http://dirk.eddelbuettel.com/code/rinside.html}} 0.2.10, gptk\footnote{\url{http://cran.r-project.org/web/packages/gptk/}} 1.06, and numDeriv\footnote{\url{http://cran.r-project.org/web/packages/numDeriv/}} 2012.9-1. The random generation of squared exponential and two-timescale Gaussian processes is sensitive to the version of GSL used, but otherwise the program output reported here should be reproducible regardless of compiler, interpreter, or library version.

\subsection{Input Cadences}

{
\def\pairwise{$\{\Delta t\}$}

We tested timescale metrics on four observing cadences, representing different young star monitoring programs recently carried out at Caltech and the Infrared Processing and Analysis Center. The cadences, presented in Table~\ref{tbl_lcmc_cadence} and illustrated in Figure~\ref{cadencedemo}, were selected to probe different observing regimes. The PTF-NAN Full cadence is the cadence at which we observed the North America Nebula with the Palomar Transient Factory from 2009 August to 2012 December \citep[Findeisen et al., in prep.]{FlareDip, KpfThesis}, with a cadence varying between hourly and biweekly over the course of the survey. The PTF-NAN 2010 cadence is the subset of these observations that were taken in 2010, when the cadence was consistently every 1-3 days. The YSOVAR 2010 cadence represents a month of Spitzer observations for the Young Stellar Object Variability survey \citep{Ysovar1, YsovarData} to represent a higher-cadence and shorter-base line monitoring program than either of the two PTF-NAN cadences. Finally, the CoRoT cadence represents 40~days of high-frequency observations taken by \citet{AmcCorot} to probe the extremes of both very high cadence and very large number of points.

\begin{table}[hbt]
\begin{tabular}{l|c c c c}
Cadence      & Number of Points & Base line & Char. Cadence & Longest Gap \\
             &      		& (days)   & (days)  & (days) \\
\hline
PTF-NAN Full &  910 		& 1,224.9   & 0.21    & 179.3  \\
PTF-NAN 2010 &  126 		&  252.7   & 1.98    &  17.0  \\
YSOVAR 2010  &   39 		&   35.7   & 1.26    &   2.5  \\
CoRoT        & 6,307 		&   38.7   & 0.012   &   0.78 \\
\end{tabular}
\caption{Key properties of the observing cadences considered in this work. Characteristic cadence is a measure of the ``typical'' spacing between observations and is defined in the text. The longest gap is the maximum interval, within the light curve, containing no observations.
} \label{tbl_lcmc_cadence}
\end{table}

\plotfour{btp}{
\includegraphics[width=0.50\textwidth]{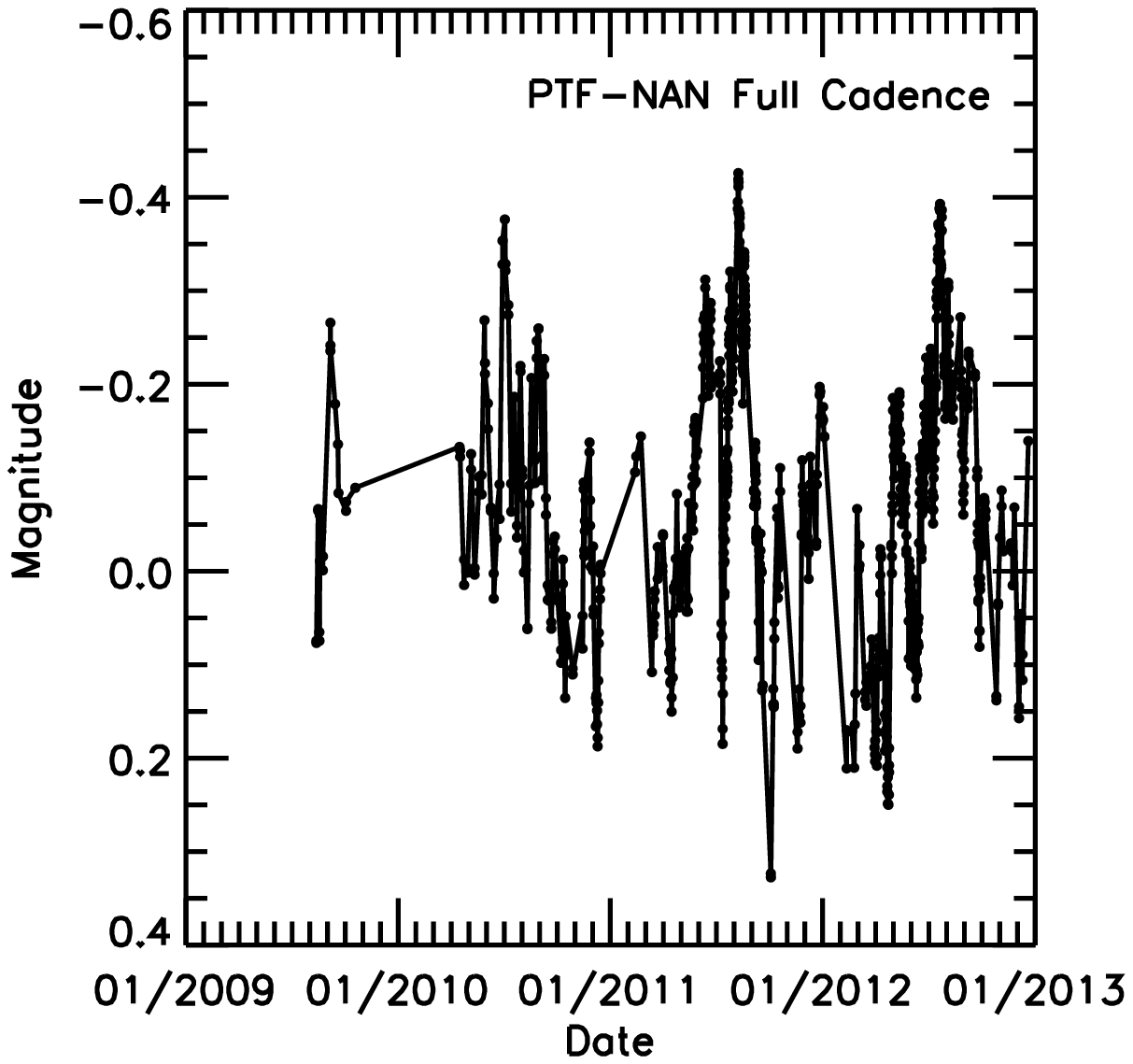}
} {
\includegraphics[width=0.50\textwidth]{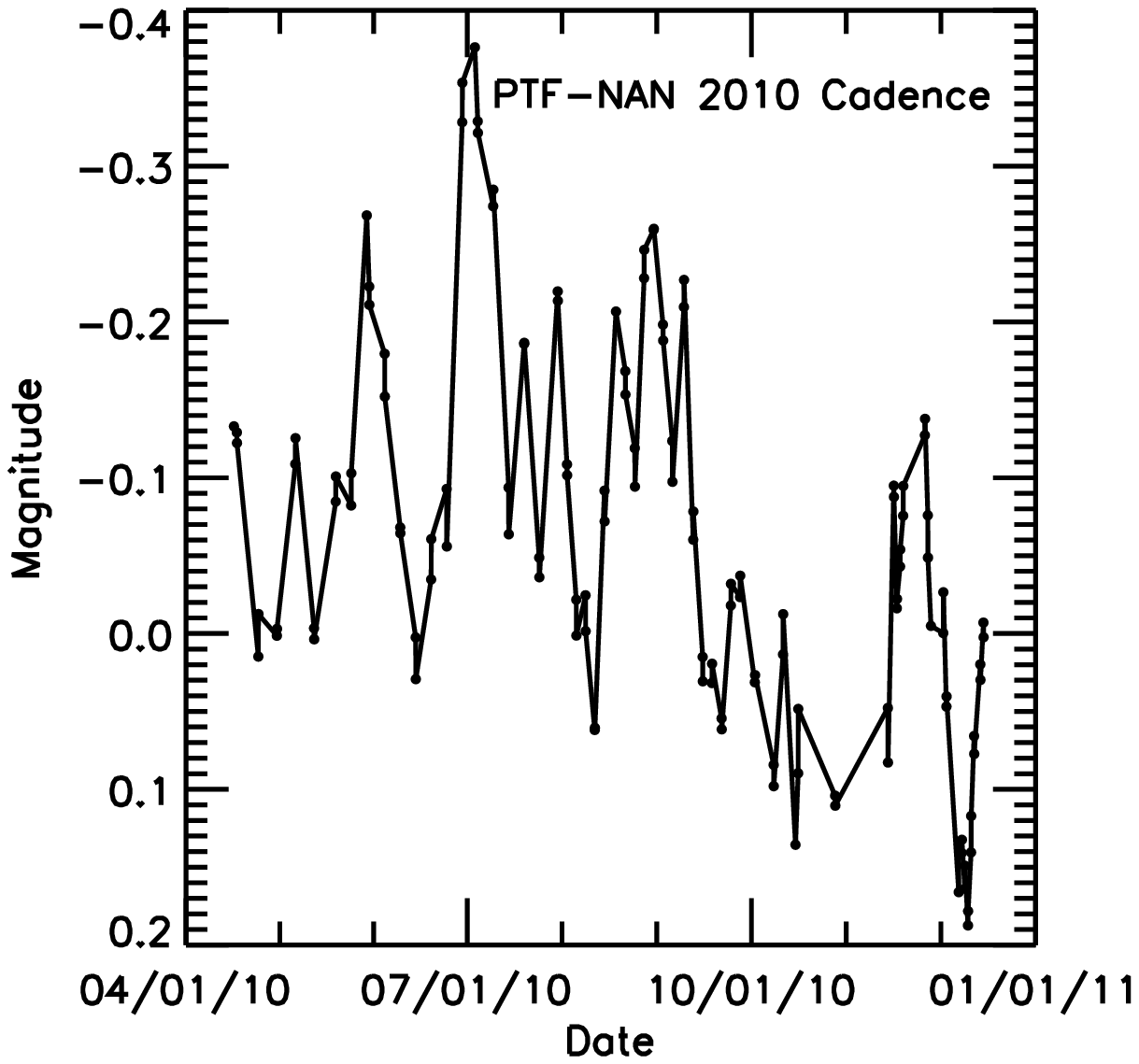}
} {
\includegraphics[width=0.50\textwidth]{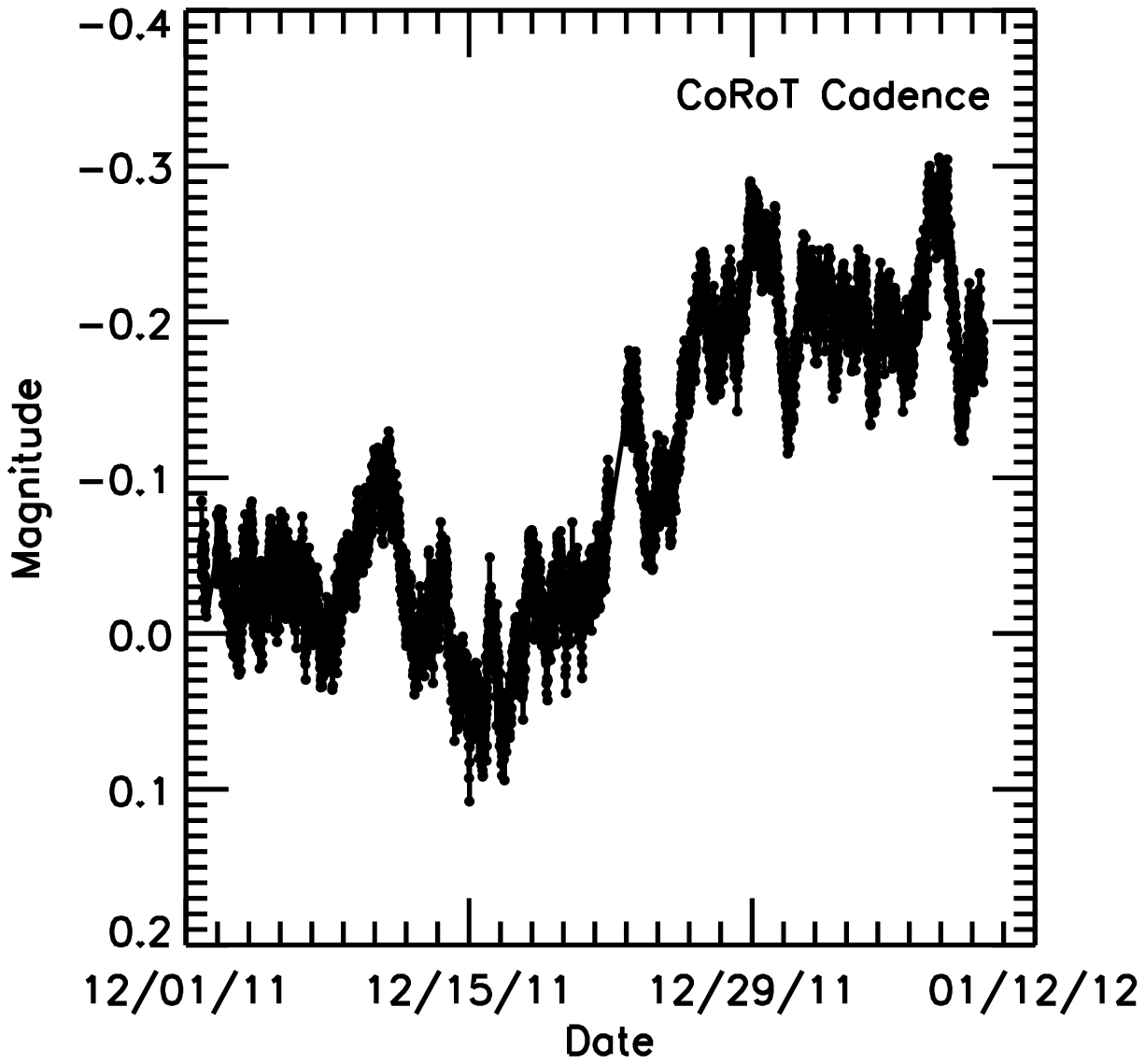}
} {
\includegraphics[width=0.50\textwidth]{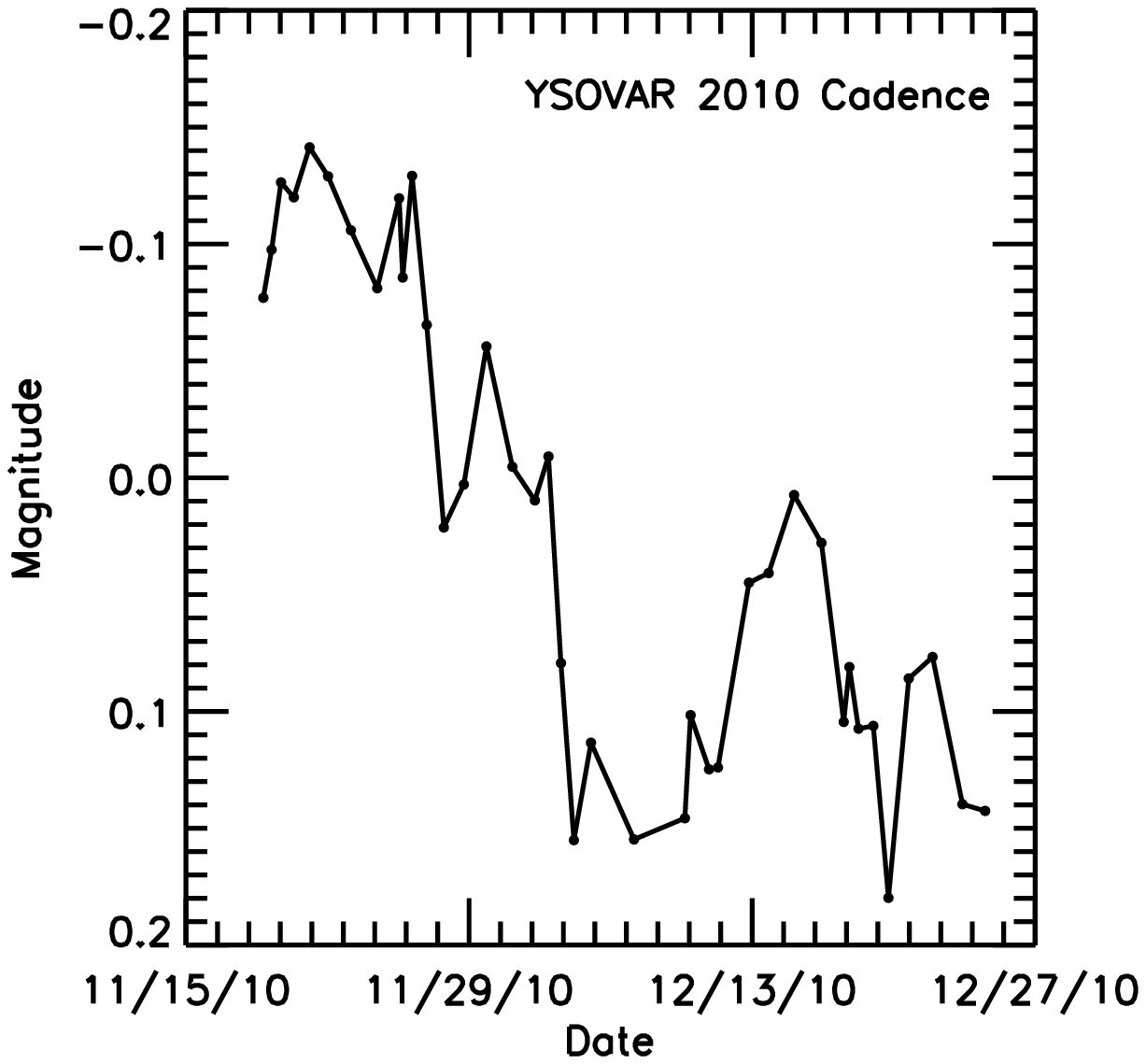}
} {
\caption{A simulated damped random walk light curve with a damping time of 16 days, sampled at each of the cadences presented in Table~\ref{tbl_lcmc_cadence}. All four panels are different subsets of the same light curve. Note that the width of the plot varies from 35 days (YSOVAR 2010) to four years (PTF-NAN Full). The CoRoT light curve shows the smallest-scale variability well, while the full PTF light curve shows long-term trends. The PTF-NAN 2010 and YSOVAR 2010 cadences provide a much sparser sampling, but comparing the CoRoT and YSOVAR 2010 figures one can see similar structures in both; the sparser coverage of the YSOVAR 2010 cadence does not preclude a characterization of its major variability component(s).} \label{cadencedemo}
}

\label{lcmc_cadence_stats}
We characterize each cadence by its base line, by its characteristic cadence, and 
by the length of the longest gap in the data.

To define a characteristic cadence for an irregularly sampled time series, we note that a regular time series of $N$ observations spaced by $\delta t$ probes timescales from $\delta t$ to $(N-1) \delta t$. Specifically, if one finds the set of separations between any two points in the light curve, \pairwise, there are $N-1$ pairs of points separated by $\delta t$, $N-2$ pairs separated by $2 \delta t$, and so on, for a total of $N(N-1)/2$ pairs of points. Since the first $N-1$ of these $N(N-1)/2$ pairs are equal to the cadence, the cadence is the $2/N$th quantile of the set \pairwise. Reasoning that the set \pairwise\ is a complete characterization of which timescales get probed with what degree of redundancy by a data set, we define the characteristic cadence of \emph{any} time series to be the $2/N$th quantile of \pairwise. 
}

\subsection{Parameter Grid}\label{grid}

For the simulations we generated light curves from sinusoidal, squared exponential Gaussian process, damped random walk, and random walk models over a grid of light curve parameters. These parameters included the choice of cadence, the choice of light curve, the light curve amplitude, the light curve timescale, and the signal-to-noise of the data. Since the simulations were motivated by our study of optical young stellar variability, we selected model parameters reflecting our expectations for young stellar variability. 

Figure~\ref{fig_analytic_lc_demos} shows a sample of the program output, one panel per light curve type, for a regular test cadence.

We tested amplitudes of 1~mag, 0.5~mag, 0.25~mag, and 0.1~mag, values typical for young stars \citep[e.g.,][]{HHG1994, RotorCtts}, measured between the 5th and 95th percentiles (``5-95\% amplitude''). The lower limit of 0.1~mag amplitude was set by the limited photometric precision of our data. The formulas from section~\ref{analytic_testlclist} were used to transform these amplitudes into model parameters. For example, since the 5-95\% amplitude for a sine wave is $1.975 A$ from section~\ref{defn_lc_sine}, we selected $A = 1/1.975 = 0.506$~mag to generate a 1~mag sinusoidal signal.

We added Gaussian white noise to each light curve, in flux space, at signal-to-noise ratios of 20, 10, and 4, measured with respect to the theoretical median flux of the light curve. The amplitude-to-noise ratio depends on both the signal-to-noise ratio and the selected amplitude. We also had a corresponding run with no noise (for testing Gaussian process models, one of the timescale metrics, we instead adopted a signal-to-noise ratio of 300 because the fits could not converge if there was exactly zero noise). Points that had negative flux after adding noise were counted as nondetections and removed from the analysis, but we did not simulate detection limits explicitly.

We tested the PTF-NAN Full cadence, the PTF-NAN 2010 cadence, and the YSOVAR 2010 cadence from Table~\ref{tbl_lcmc_cadence}. To keep computation times down, we did not test all 48 combinations of amplitude, signal-to-noise, and cadence; instead, we tested each amplitude at a signal-to-noise of 20 and PTF-NAN Full cadence (values typical for much of the PTF-NAN data), each signal-to-noise at an amplitude of 0.5~mag and PTF-NAN Full cadence, and each cadence at a signal-to-noise of 20 and amplitude of 0.5~mag, for a total of 9 combinations.

For simulations using either PTF-NAN cadence, we tested light curve timescales (denoted $\tau$ or $P$ in section~\ref{analytic_testlclist}) of 0.5~days, 2~days, 5~days, 16~days, 64~days, and 256~days at all 9 combinations of amplitude, signal-to-noise, and cadence. This timescale grid was chosen because we wanted to determine our sensitivity to variability across the entire time range probed by our cadence. The lower limit was set to several hours, both because we had little data probing intranight variability, and because previous work showed little variation on such short timescales \citep{ShortVarLimit,CodyMost}. The upper limit was set to of order a year, so that we could still sample the characteristic timescale three or more times in our multi-year survey. For simulations using the YSOVAR 2010 cadence, we instead tested timescales of 0.1~days, 1~day, 2~days, 5~days, 20~days, and 40~days, applying analogous arguments to the cadence and time baseline of the YSOVAR 2010 cadence.

In addition to the primary parameter grid described above, we ran simulations of only one of the timescale metrics, peak-finding, on the cadence used by CSI~2264. These separate simulations were run at a signal-to-noise ratio of 100, a value typical of the light curves studied by \citeauthor{AmcCorot}. We used the same amplitude and timescale grids as for the YSOVAR 2010 cadence. 

In either grid, at each grid point we generated 1,000 light curves of each type, giving them the appropriate amplitude and timescale. Since a random walk does not have a well-defined amplitude or timescale, we assigned it a diffusion constant $D = \frac{2\sigma^2}{\tau}$, where $\sigma$ and $\tau$ were the amplitude and timescale parameters adopted for the damped random walk. This convention for $D$ gave the damped and undamped random walks at the same grid point equal diffusion constants for ease of comparison (cf. Equation~\ref{eqn_analytic_lc_drw}). We also generated 1,000 white noise light curves at each combination of amplitude, signal-to-noise, and cadence to compare to the structured light curves. The final tally was 213,000 simulated light curves. Each timescale metric was tested on each of the light curves, with one exception. To keep running times down, the much slower Gaussian process fitting, presented in section~\ref{lcmc_gp}, was tested on only the first 30 light curves of each set of 1,000.

\subsection{Timescale Characterization}

The simulation setup described in sections~\ref{analytic_testlclist} and \ref{siminfo} is generic, and can be applied to a variety of problems in observational time-domain astronomy, including accurate amplitude or shape parameter estimation. Here, we are primarily interested in identifying the characteristic timescales of observed light curves. The speed at which a source varies is an important physical probe of a variety of physical mechanisms, as described in section~\ref{intro_whytime}. 

Characterizing timescales in aperiodic, irregularly sampled signals is a difficult problem. For a metric to be useful in real applications, we must show that it gives meaningful and precise results, and that it can do so regardless of the (unknown a priori) type of light curve. We would also prefer a metric that is not affected by either physical details of a source, such as the light curve amplitude, or observational details of the data, such as signal-to-noise ratio or cadence.

The grid presented in section~\ref{grid} probes a broad range of both ``physical'' and observational parameters. In section~\ref{perform}, we plot metric performance against light curve shape, input timescale, signal-to-noise ratio, and cadence. While the effects of light curve amplitude and signal-to-noise ratio are not perfectly degenerate (i.e., a 1~mag variable at a signal-to-noise ratio of 10 is not equivalent to a 0.1~mag variable at a signal-to-noise ratio of 100, due to the nonlinear relationship between magnitude and flux changes), the difference is subtle enough that, for brevity, we only present how timescale metrics depend on signal to noise at fixed amplitude, but not how they depend on amplitude at fixed signal to noise.

\section{Timescale Recovery Metrics}

\subsection{\dmdt\ Plots}\label{lcmc_dmdt}

The \dmdt\ plot is a nonparametric representation of a light curve that describes the frequency with which a particular degree of variability is observed on a particular timescale. In some ways it resembles the self-correlation analysis of \citet{SelfCorr_First,SelfCorr_TTS2}, although it preserves more information about the light curve's behavior and thus allows a broader range of analysis techniques. It is defined by pairing up all observations $m_i(t_i)$ of a light curve, and recording only the time and magnitude differences:
\begin{displaymath}
\begin{array}{rcll}
\Delta m_{ij} & = & |m_i - m_j| & (i > j) \\
\Delta t_{ij} & = & |t_i - t_j| & (i > j)
\end{array}
\end{displaymath}
where the restriction $i > j$ is to ensure each pair is considered only once. These differences may then be represented as a scatter plot of $\Delta m$ as a function of $\Delta t$, presenting directly how much variability is observed on which time base lines. If the original light curve had $N$ data points, the corresponding \dmdt\ plot has $N(N-1)/2$ pairs of $(\Delta t, \Delta m)$ values. The \dmdt\ plot is closely related to the structure function $SF$, as $SF(\Delta t) = E({\Delta m}^2)$.

\begin{figure}[htb]
\plottwo{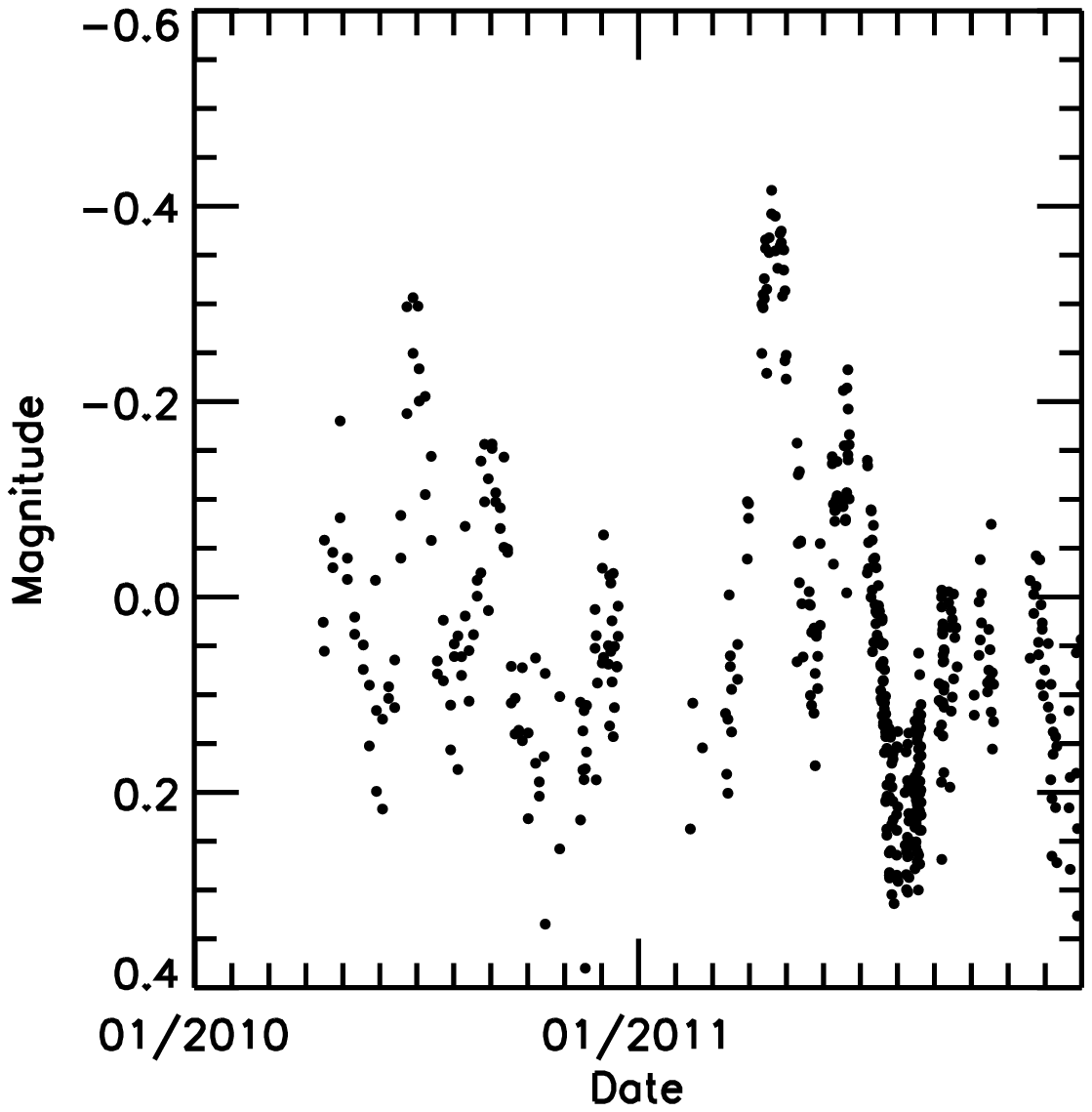}{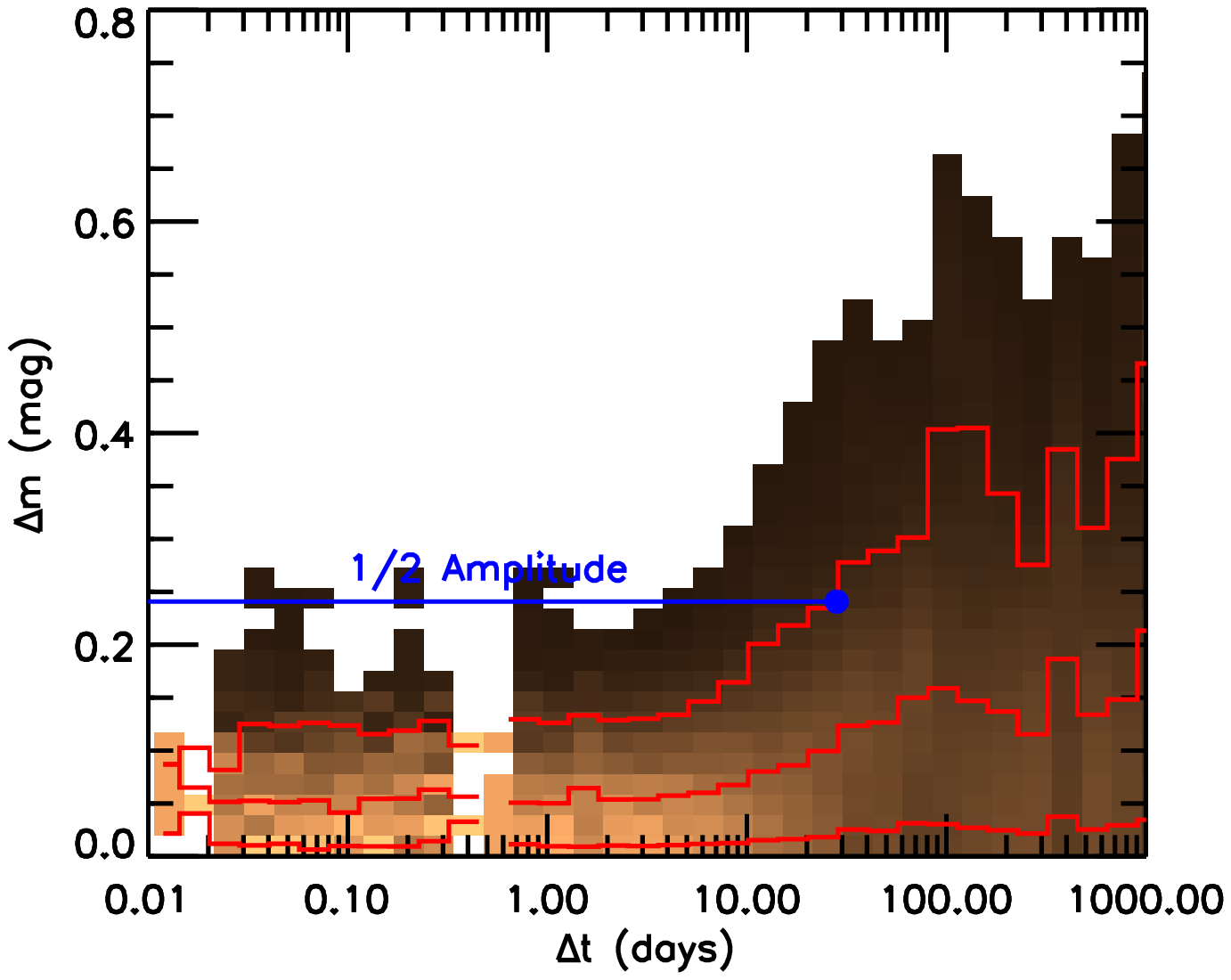}
\caption{An example of a simulated damped random walk light curve with a damping time of 16~days (left) and a binned version of the corresponding \dmdt\ plot (right). Light shades denote a high density of points, while dark shades denote a low density. The blue line illustrates the exercise of defining the timescale as the point at which the quantiles of $\Delta m$ (red lines) cross a threshold (half the 5-95\% amplitude, in this example).} \label{lcmc_demo_dmdt}
\end{figure}

For light curves with hundreds or thousands of epochs, \dmdt\ plots may have thousands to millions of unique pairs of points. A plot with this many points is difficult to interpret, as nearly all the points simply blend together. Therefore, another layer of abstraction, such as a histogram or a density estimator, may be used to present a \dmdt\ plot. An example of a histogram-based \dmdt\ plot is shown in Figure~\ref{lcmc_demo_dmdt}.

In general, different timescales will be sampled to different degrees by a \dmdt\ plot. For example, consider a time series consisting of N points uniformly spaced by an interval $\delta t$. The allowed values of $\Delta t$ will have the form $\Delta t = n \delta t$, with $1 \le n \le N-1$, and the number of pairs with each value will be $N-n$. The shortest timescales will be by far the best-sampled, with the median value of $\Delta t$ being approximately $0.3 N \delta t$ in the limit $N \gg 1$. Data gaps and other complexities can lead to other bias patterns.

Any method of analyzing \dmdt\ plots, whether qualitative or formal, must correct for the differing number of pairs at different timescales, because the relative number of pairs at different timescales is a property of the experimental setup rather than of the source(s) being studied. At present, we are binning the pairs in $\log \Delta t$, and describing the \dmdt\ plots in terms of summary statistics on $\Delta m$ within each time bin. This representation removes the biases associated with variable sampling, though (as with all histograms) the results will tend to converge slowly to the true distribution as the number of observations increases, and the results may be biased by changes in the $\Delta m$ distribution across the width of an individual bin. In addition, the shortest timescale bins, being the narrowest, are poorly populated, introducing sampling noise into the results.

We convert a \dmdt\ plot into a scalar timescale by finding the first $\Delta t$ at which the 90\% quantile exceeds one half of the 5-95\% amplitude. Both thresholds were chosen on the basis of intuition: the 90\% quantile should represent the ``upper envelope'' of variability that only occurs $\sim 10\%$ of the time, allowing us to probe intermittent variability, while one half the total amplitude is a reasonable definition of a significant magnitude change. Discussion of alternative measures, both other choices of quantile and other choices of threshold, can be found in \citet{KpfThesis}. 

In general, the 90\% quantile increases monotonically with timescale. This behavior is easy to understand if the star is varying incoherently: the amount by which the star changes brightness in, for example, 100 days cannot be less than the amount by which it changes brightness in 10 days because it has had the opportunity to undergo a 10-day change within the 100-day period. In particular, if the star has no variability mechanisms operating on timescales longer than 10 days, then the 100-day brightness change will simply be the net result of 10 uncorrelated (or nearly uncorrelated) 10-day changes. A flattening in the 90\% quantile curve therefore means that all the variability occurs on shorter timescales.

For the simulations presented in this work, the first bin edge was set at $10^{-1.97}$~days, with subsequent bins at increments of 0.15~dex up to the full light curve base line. This choice of bins allowed all possible timescales to be probed, even in the highest-cadence portions of the PTF-NAN, YSOVAR, and CoRoT observing patterns, while the nightly observing gaps in the PTF-NAN data would only deplete points from two bins, from $10^{-0.47}$~days to $10^{-0.32}$~days and from $10^{-0.32}$~days to $10^{-0.17}$~days. These precise bin edges led to the cleanest separation between well-populated and poorly populated bins.

In practice, the variance in timescales across different realizations of the same light curve generator usually exceeds the $\sim 40\%$ error introduced by quantizing the output to 0.15~dex bins. Therefore, finer bins would not improve performance, and are likely to worsen it by increasing the scatter in each bin.

The use of $\Delta t$ bins in analyzing the \dmdt\ plot forces the output timescale onto a grid set by the bin edges. While the scatter in most simulation runs exceeds the bin width, in some runs enough of the observations fall into the same bin to give much less scatter than would be present if we used finer bins or an unbinned analysis.
To give a more representative measure of the intrinsic precision of \dmdt\ analysis, we add, in quadrature, $\frac{1}{\sqrt{12}}$ of the bin width to the standard deviation of all \dmdt-based timescales. The factor of $\frac{1}{\sqrt{12}}$ comes from the fact that the standard deviation of a uniformly distributed random variable is $\frac{1}{\sqrt{12}}$ the length of the interval in which the variable may be found. With no constraint on where within a bin the unbinned timescale would have lied, its position is effectively a uniform random variable in log $\Delta t$ space.

\subsection{Peak-Finding}\label{lcmc_pf}

Peak-finding is a timescale metric developed by \citet{AmcCorot} for well-sampled aperiodic light curves. A direct generalization of periods, peak-finding computes a timescale based on the spacing between local minima and local maxima in a light curve. The more regularly spaced these minima and maxima are, the more precise the metric.

\begin{figure}[htb]
\plottwo{f4a.eps}{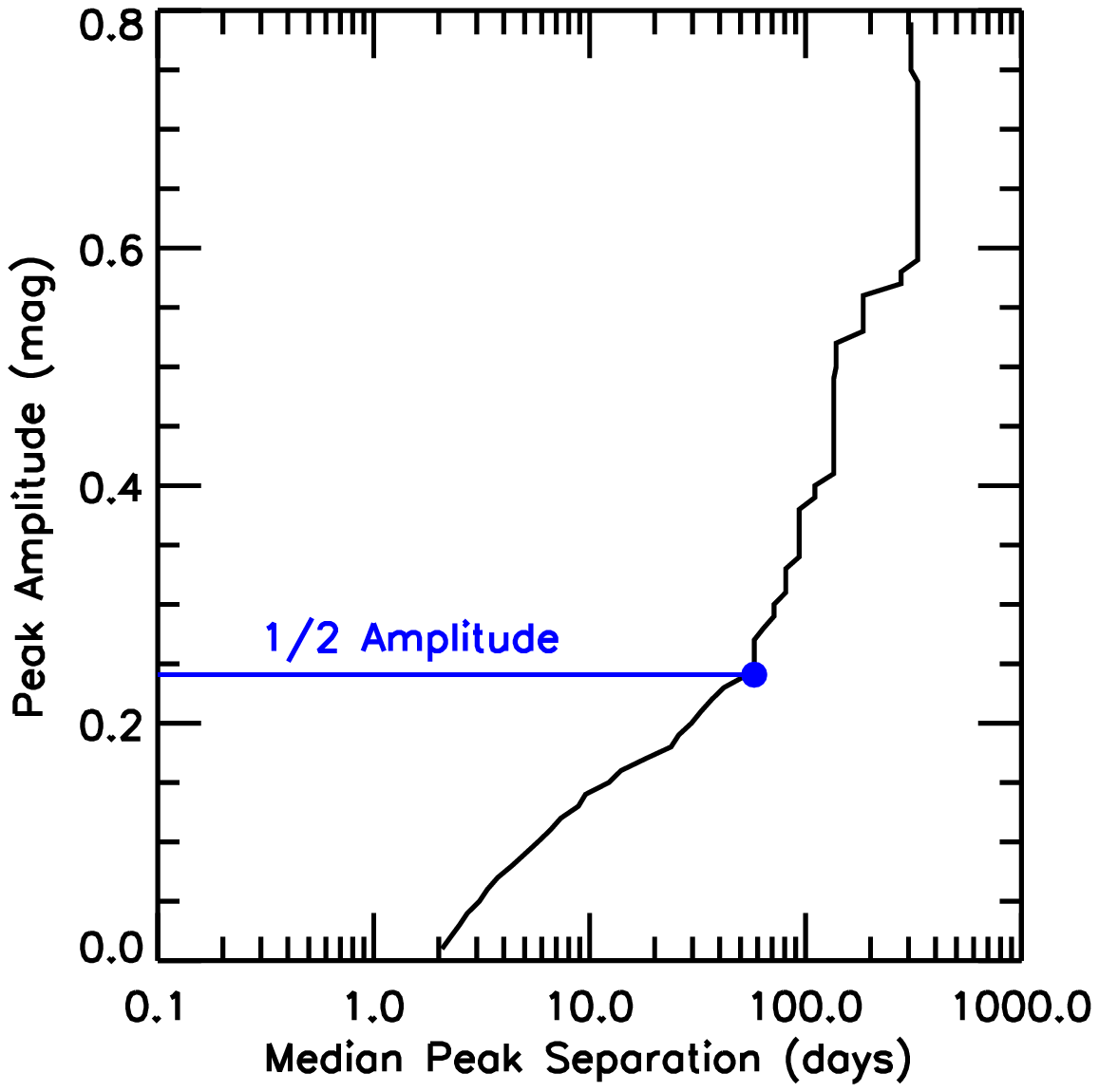}
\caption{An example of a simulated damped random walk light curve with a damping time of 16~days (left) and a peak-finding plot derived from it (right). The blue line illustrates the exercise of defining the timescale as the point at which the peak-finding curve crosses a threshold (half the 5-95\% amplitude, in this example).} \label{fig_lcmc_demo_peak}
\end{figure}

We begin with the first point on the light curve, then identify the first local minimum or maximum that differs from the first point by a predetermined magnitude threshold. After each local minimum we find the first local maximum differing from it by the threshold, and vice versa. In this way, the method builds up a list of alternating minima and maxima, discarding low-amplitude fluctuations. The mean time between minima and maxima separated by a given amplitude threshold, including the end points, is a measure of the speed of fluctuations of that amplitude. By repeating the process for a variety of thresholds, one builds up a plot (Figure~\ref{fig_lcmc_demo_peak}) of timescale as a function of magnitude scale, from the level of measurement noise up to the full light curve amplitude.

The procedure described above is a slightly older version of the method from that presented by \citeauthor{AmcCorot}, which begins from the global maximum of the light curve, working outward in both directions, rather than from the first point. 
In addition, we have altered the algorithm of \citet{AmcCorot} by considering the median separation between fluctuations, rather than the mean. The mean separation would be biased high by the large seasonal gaps in the two PTF-NAN observing cadences. The median separation is much more robust to coverage gaps provided that most of the variability is on timescales shorter than the length of an observing season, because then the many minima and maxima within each season's coverage dominate the median. 
Use of the median instead of the mean does not have a significant effect for simulations using the YSOVAR 2010 or CoRoT cadences, which have much more uniform coverage.

\label{peakfind_cut_choice}
For the simulations, the median peak separation was calculated at each multiple of 0.01~mag, up to the full light curve amplitude. The strategy with which we converted the peak-finding plot into a single timescale varied between the two sets of parameter grids:
\begin{itemize}
\item For simulations using the CoRoT cadence, we identified the highest magnitude threshold at which at least one minimum and one maximum were found, and adopted the median separation at 80\% of that highest threshold as our timescale. This definition was chosen to allow direct comparison of the simulation results to the work of \citet{AmcCorot}.
\item For simulations using the YSOVAR 2010 cadence or either PTF-NAN cadence, we identified the median separation between peaks differing by at least half the light curve's 5-95\% amplitude. We preferred this timescale definition because other simulations, presented in \citet{KpfThesis}, showed that this definition of timescale had considerably less scatter but slightly higher systematics than the 80\% criterion used by \citet{AmcCorot}.
\end{itemize}

\subsection{Gaussian Process Regression}\label{lcmc_gp}

Gaussian process regression is an increasingly popular analysis tool for modeling aperiodic time series that are assumed to consist of a smooth (but unknown) function plus noise. Since a Gaussian process has no specific functional form, a good fit is instead characterized by a high likelihood that the data were drawn from a Gaussian distribution with a particular covariance matrix.

\begin{figure}[htb]
\plottwo{f4a.eps}{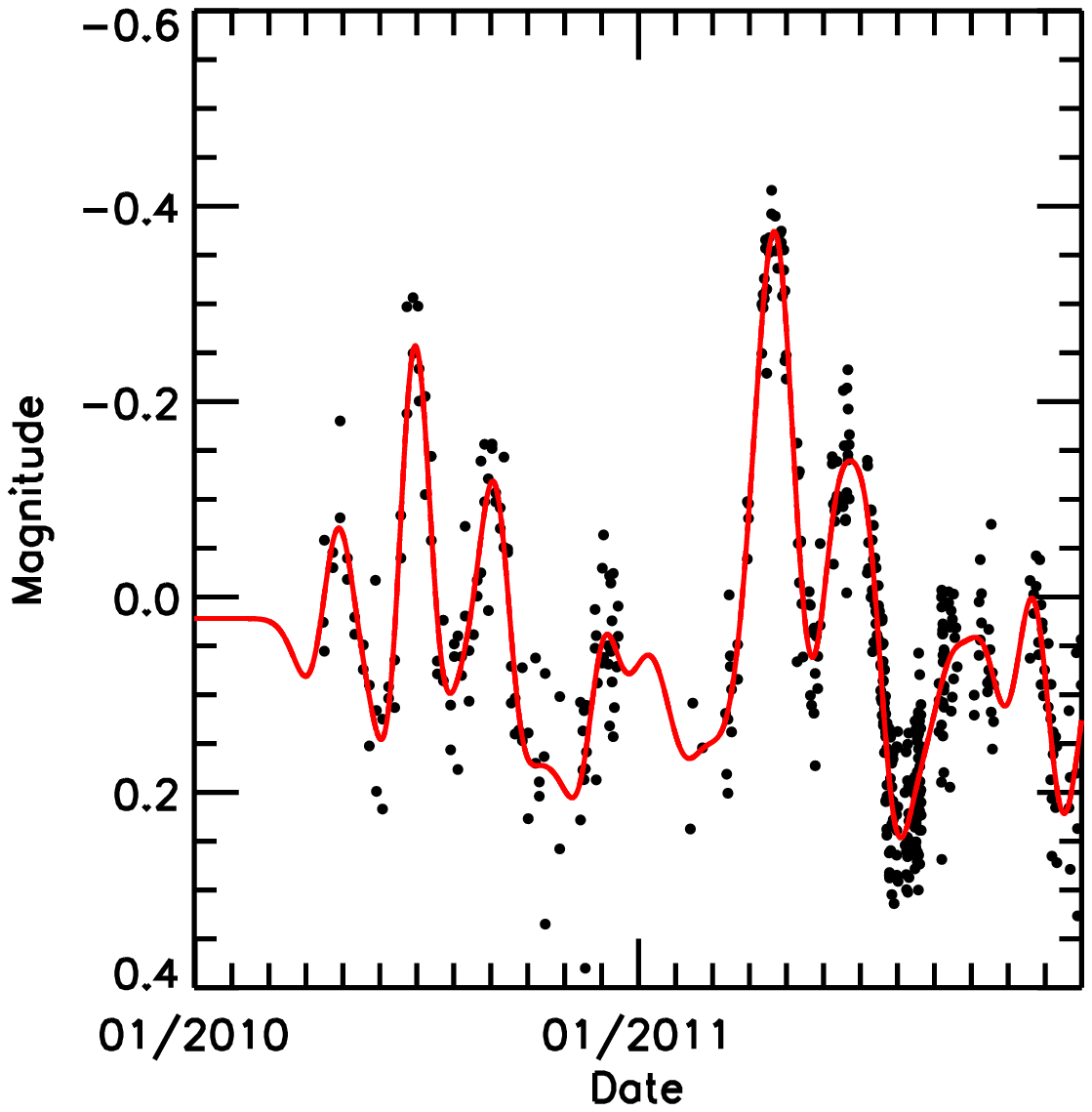}
\caption{An example of a simulated damped random walk light curve with a damping time of 16~days (left) and the best GP fit (right). Formal errors in the model are far smaller than the scatter of actual data points. The timescale is one of the model parameters, but cannot be directly read from the plot.} \label{fig_lcmc_demo_gpfit}
\end{figure}

In the most common case, the one we use here, the covariance matrix $K$ of measurements taken at $t_i$ and $t_j$ is assumed to be the sum of a squared exponential Gaussian process, characterized by an amplitude $\sigma$ and a coherence time $\tau$, and a white noise process, characterized by an amplitude $\sigma_n$:
\begin{displaymath}
K_{ij} = K(t_i, t_j) = \sigma^2 \exp{\left(-\frac{(t_i-t_j)^2}{2 \tau^2}\right)} + \sigma_n^2 \delta(t_i, t_j)
\end{displaymath}
where $\delta$ denotes the Kronecker delta. For the purposes of this study, the two amplitudes $\sigma$ and $\sigma_n$ are nuisance parameters, and only the best fit coherence time $\tau$ is reported. The fitting package we used (\texttt{gptk}) used conjugate gradient descent to maximize the likelihood
\begin{displaymath}
L(\sigma, \tau, \sigma_n|\vec{m}) = - \frac{1}{2} \transp{\vec{m}} K^{-1} \vec{m} - \frac{1}{2} \log{|K|} - \frac{n}{2} \log{2\pi}
\end{displaymath}
given the vector of observed magnitudes $\vec{m}$.

Because the likelihood function for Gaussian process regression involves the inverse of the $N \times N$ covariance matrix $K$, where $N$ is the number of data points, computing the likelihood function is a cubic operation in $N$. Since conjugate gradient descent may require up to $N$ iterations to converge, the overall task of fitting a light curve is \emph{quartic} in $N$. To keep running times reasonable, we only attempted Gaussian process regression on light curves simulated on the YSOVAR 2010 cadence ($N = 39$), and only generated 30 light curves per grid point rather than the usual 1,000. The light curve parameter grid was the same as described in section~\ref{grid}, except that the noise-free case was replaced with a signal-to-noise ratio of 300. Since the amount of noise in the data is one of the free parameters, and since \texttt{gptk} fits the parameters in log space, attempting to fit a noise-free squared exponential Gaussian process ($\ln{\sigma_n} = -\infty$) with a noisy squared exponential Gaussian process would never converge.

\section{Timescale Metric Examples}\label{timescale_examples}

In this section we present average \dmdt\ and peak-finding plots for the simulated light curves, concentrating on sinusoidal and damped random walk models. We also present the scatter of each plot around its mean. All plots are computed for the PTF-NAN Full cadence. These plots illustrate many of the metrics' properties described in the previous section. In addition, they help place into context the results of section~\ref{perform}, where the plots from multiple runs are distilled into a single number, then averaged together.

\subsection{\dmdt\ Plots}

\subsubsection{Sinusoid}

\plotthree{ptb}{
\includegraphics[width=0.32\textwidth]{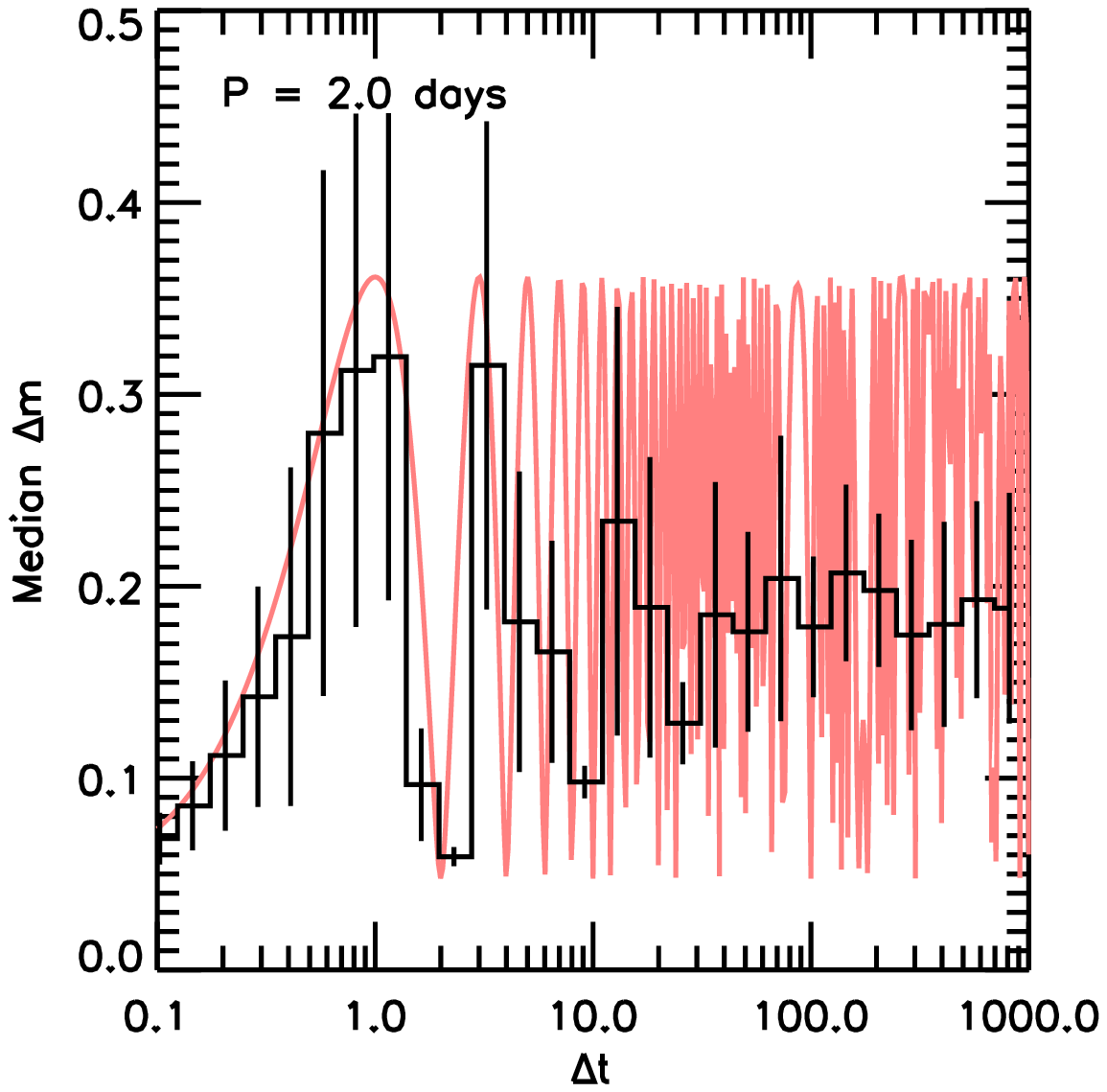}
} {
\includegraphics[width=0.32\textwidth]{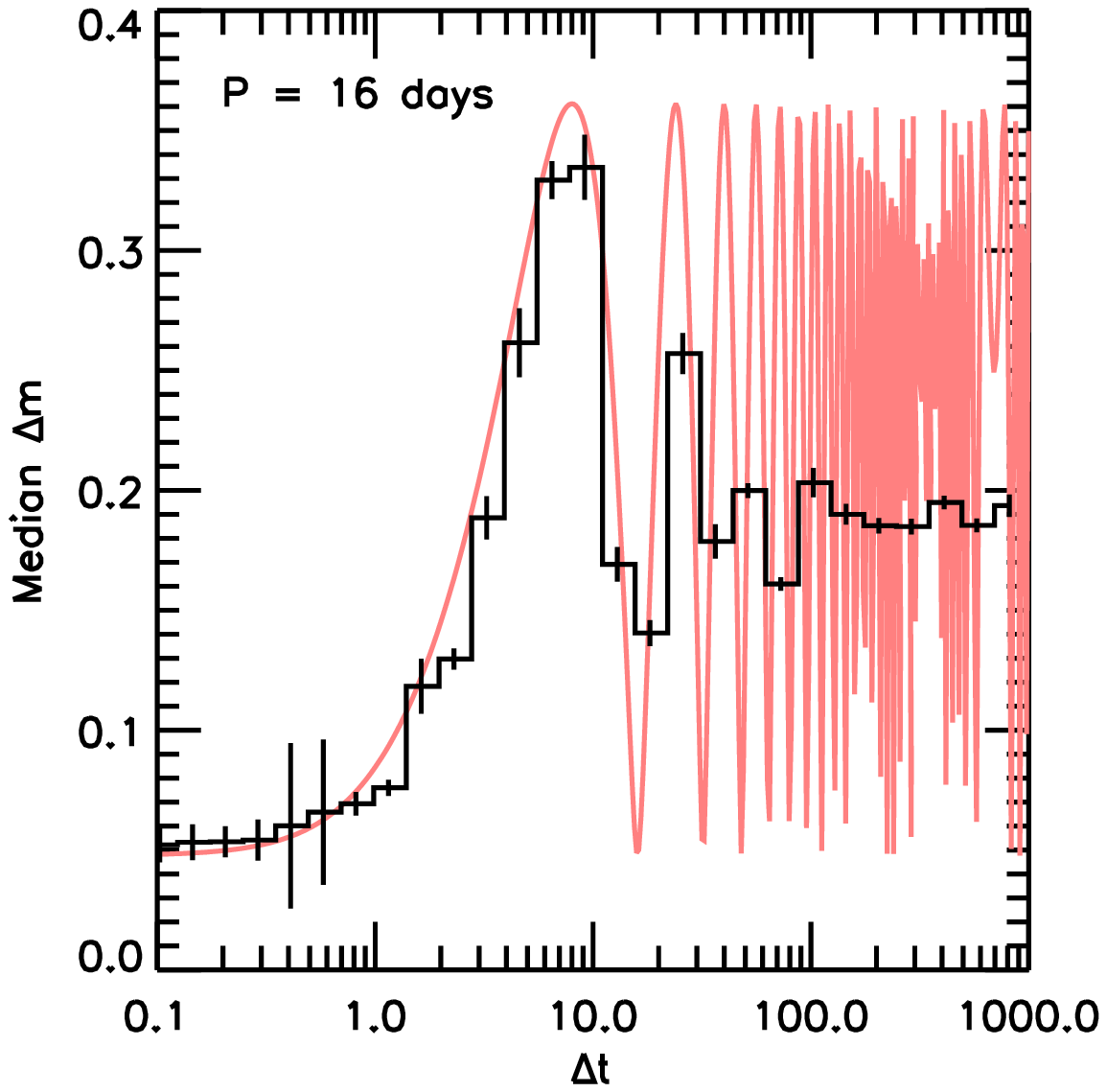}
} {
\includegraphics[width=0.32\textwidth]{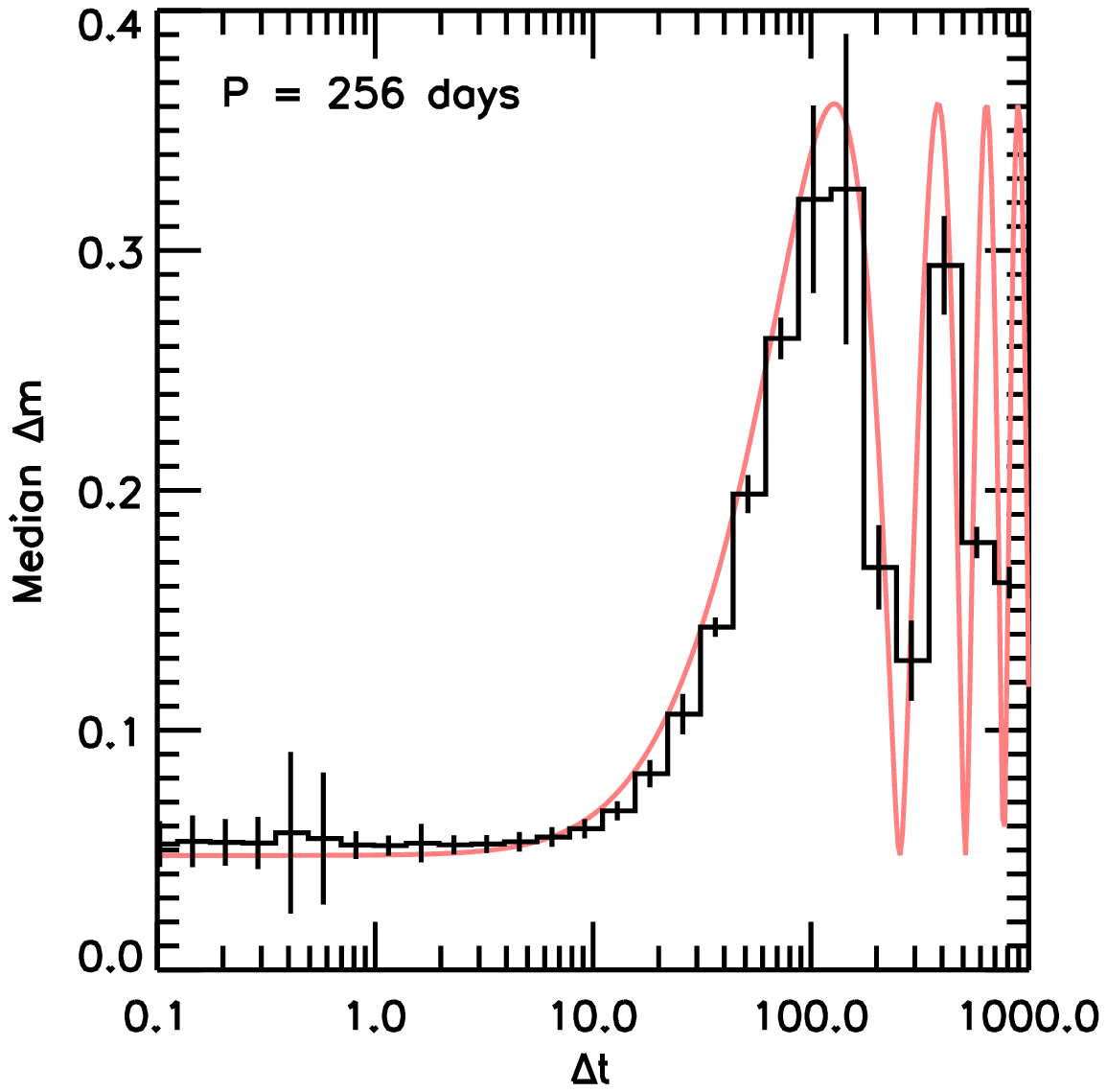}
} {
\caption{The average value of the $\Delta m$ median from 1,000 simulations of a sine wave at several representative periods and a 5-95\% amplitude of 0.5~mag. Error bars represent the standard deviation of the $\Delta m$ median in each $\Delta t$ bin; note the large uncertainties in the two sparsely populated bins at $\Delta t \sim 0.5$~days. The red curve shows the $\Delta m$ median predicted theoretically by \citet{KpfThesis}.} \label{fig_lcmc_sine_dmdt_ptf}
}

We present in Figure~\ref{fig_lcmc_sine_dmdt_ptf} the average value of the $\Delta m$ median, as a function of the time interval between time observations, for a set of 1,000 simulations of a sinusoidal light curve with different periods. All light curves are observed using the PTF-NAN Full cadence. The most dramatic change with the sine period is that the first peak in $\Delta m$, and the rise leading up to it, shift so that the first peak always appears at half the period. At very long timescales, the wide $\Delta t$ bins provide little information, as they do not resolve the periodic variability.

In \citet{KpfThesis}, we presented theoretical expectations for \dmdt\ plots assuming infinite cadence and observing base line. These predictions are shown as the red curve on the plots. The values of $\Delta m$ in each bin generally follow the predictions, although in most bins $\Delta m$ is averaged over multiple cycles. As a result, while in principle the $\Delta m$ values oscillate with the period of the sinusoidal signal, the oscillations can only be discerned for one or at most two periods of the measured $\Delta m$ curve.

\subsubsection{Damped Random Walk}

\plotthree{ptb}{
\includegraphics[width=0.32\textwidth]{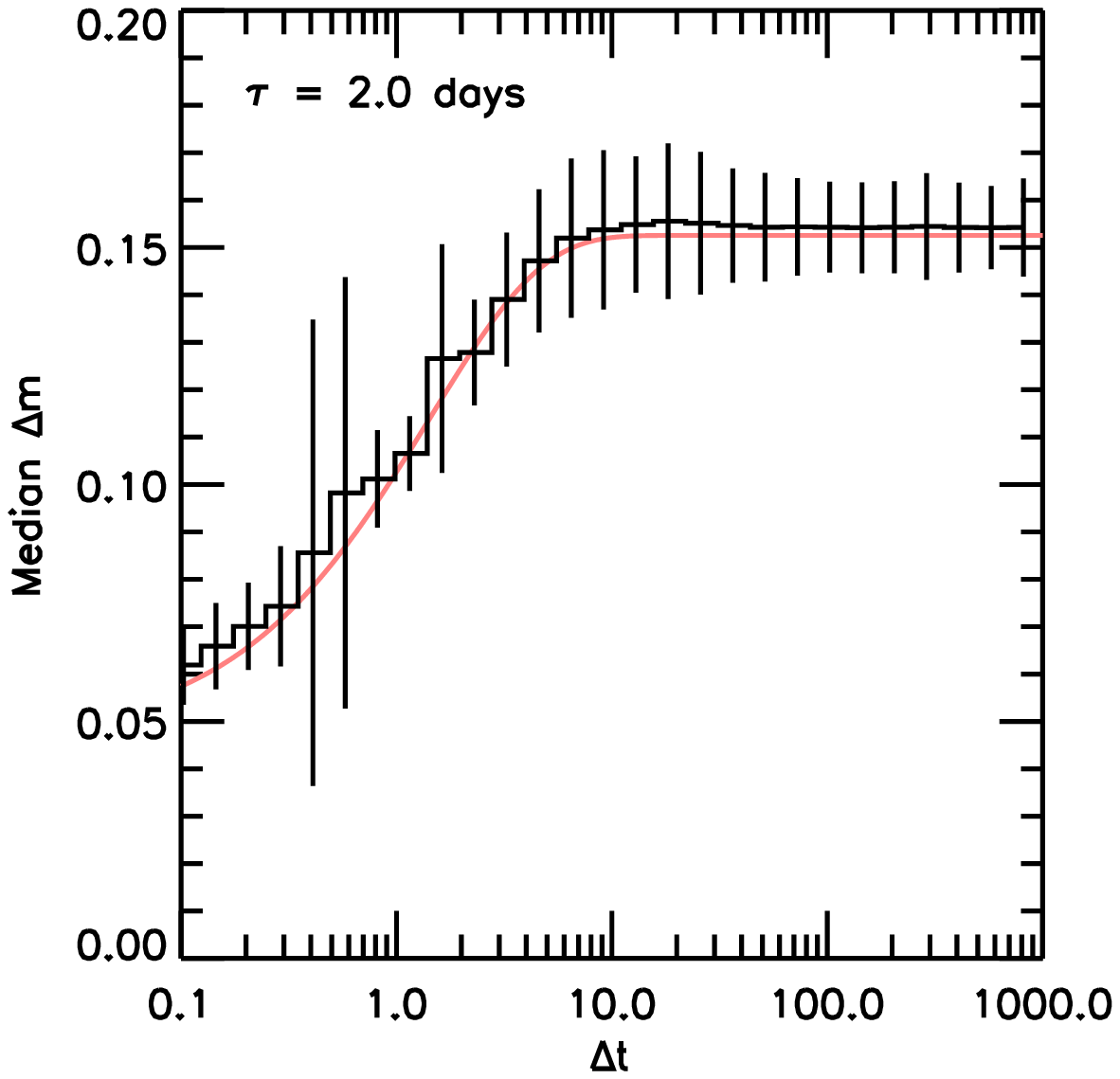}
} {
\includegraphics[width=0.32\textwidth]{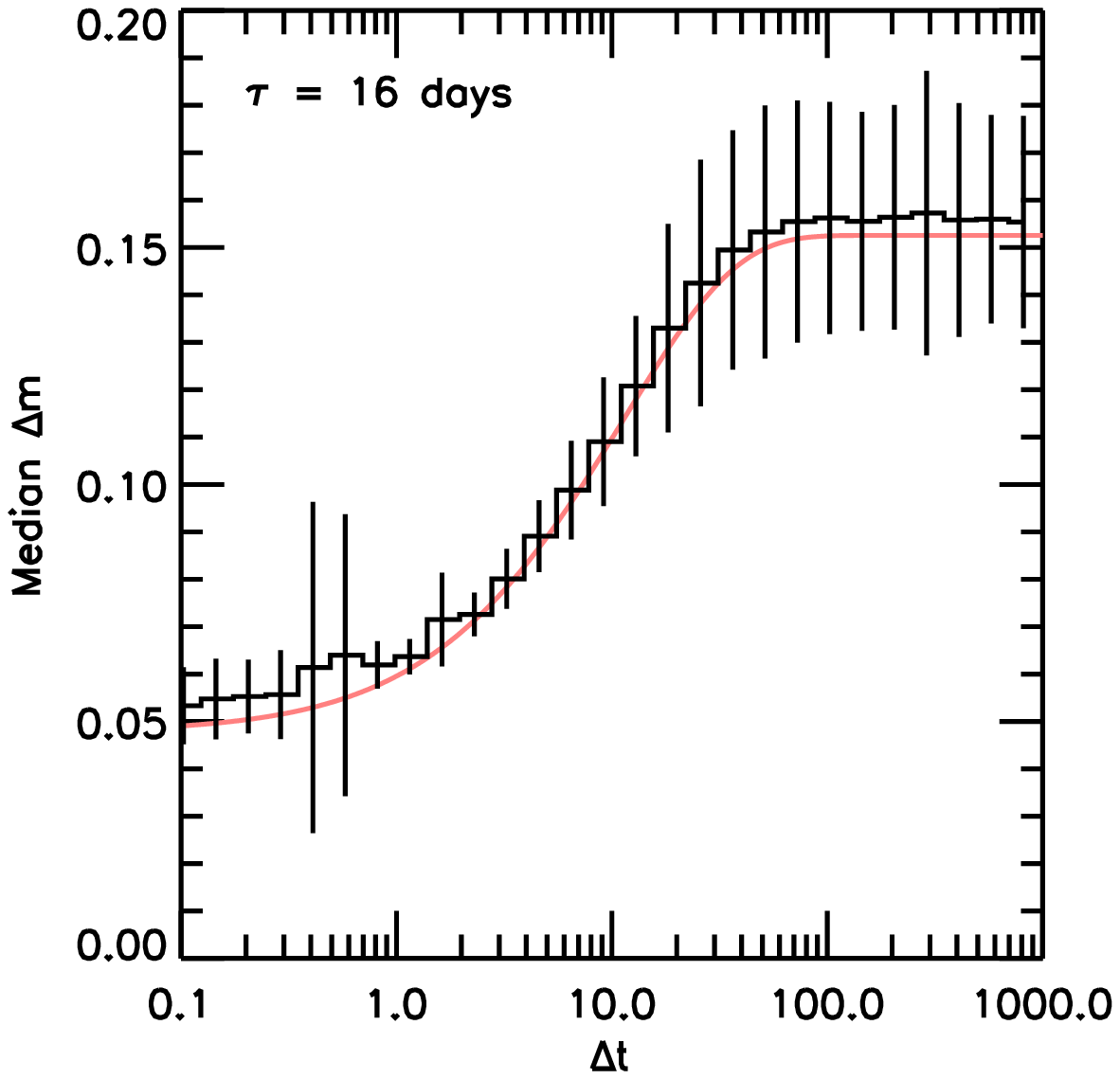}
} {
\includegraphics[width=0.32\textwidth]{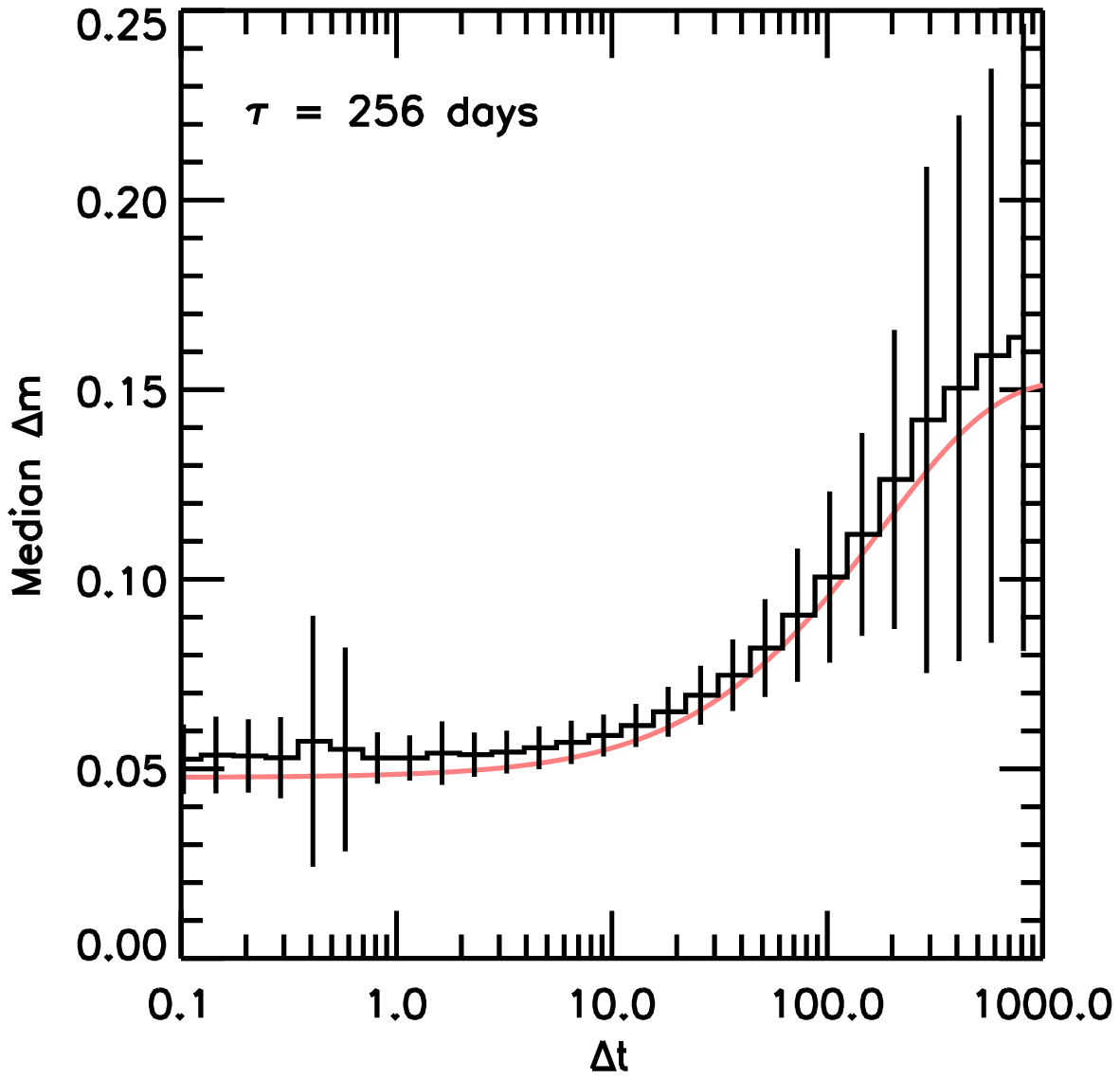}
} {
\caption{The average value of the $\Delta m$ median from 1,000 simulations of a damped random walk at several representative timescales and a 5-95\% amplitude of 0.5~mag. Error bars represent the standard deviation of the $\Delta m$ median in each $\Delta t$ bin. The red curve shows the $\Delta m$ median predicted theoretically by \citet{KpfThesis}.} \label{fig_lcmc_drw_dmdt_ptf}
}

We present in Figure~\ref{fig_lcmc_drw_dmdt_ptf} the average value of the $\Delta m$ median, as a function of the time interval between time observations, for a set of 1,000 simulations of a damped random walk with different correlation times. All light curves are observed using the PTF-NAN Full cadence. The median $\Delta m$ value steadily rises up to a few times the light curve's damping time, where it levels off to reflect the absence of longer-term variability. The two sparsely populated bins at $\Delta t \sim 0.5$~days show consistently large scatter, reflecting the small-number statistics used to find the median. In addition, the 64- and 256-day models show significant scatter at the longest $\Delta t$ bins probed by the observations; we believe this reflects there not being enough long base lines to probe the characteristic timescale of the variability.

In \citet{KpfThesis}, we presented theoretical expectations for \dmdt\ plots assuming infinite cadence and observing base line. These predictions are shown as the red curve on the plots. The values of $\Delta m$ in each bin generally follow the predictions, but there is a large amount of scatter from run to run.

\subsubsection{Other Light Curves}

\dmdt\ plots for a squared exponential Gaussian process qualitatively resemble those for a damped random walk, in that they have a rise and a leveling out on average but have a large degree of scatter from run to run. The \dmdt\ plots for a random walk, on the other hand, show nearly 100\% scatter for bins probing timescales longer than 200~days. The rise in scatter for a damped random walk model suggests that \dmdt\ plots cannot accurately represent variability with (damping) timescales exceeding $\sim 1/5$ of the observing base line; the light curve provides an unrepresentative snapshot of variability on longer timescales, and the \dmdt\ plot cannot present information that is not in the data. \label{lcmc_dmdt_maxtimescale}

\subsection{Peak-Finding}

\subsubsection{Sinusoid}

\plotthree{ptb}{
\includegraphics[width=0.32\textwidth]{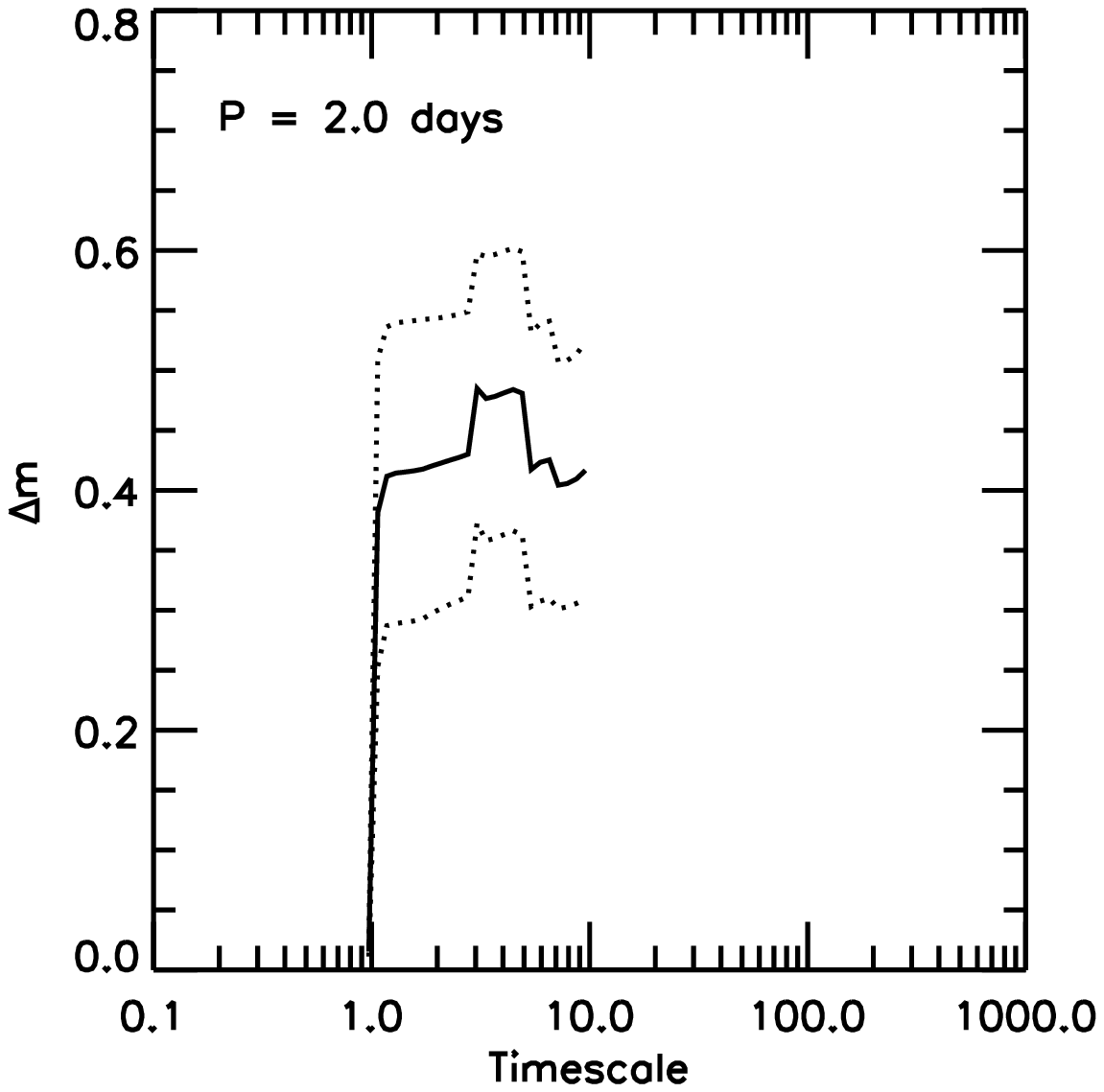}
} {
\includegraphics[width=0.32\textwidth]{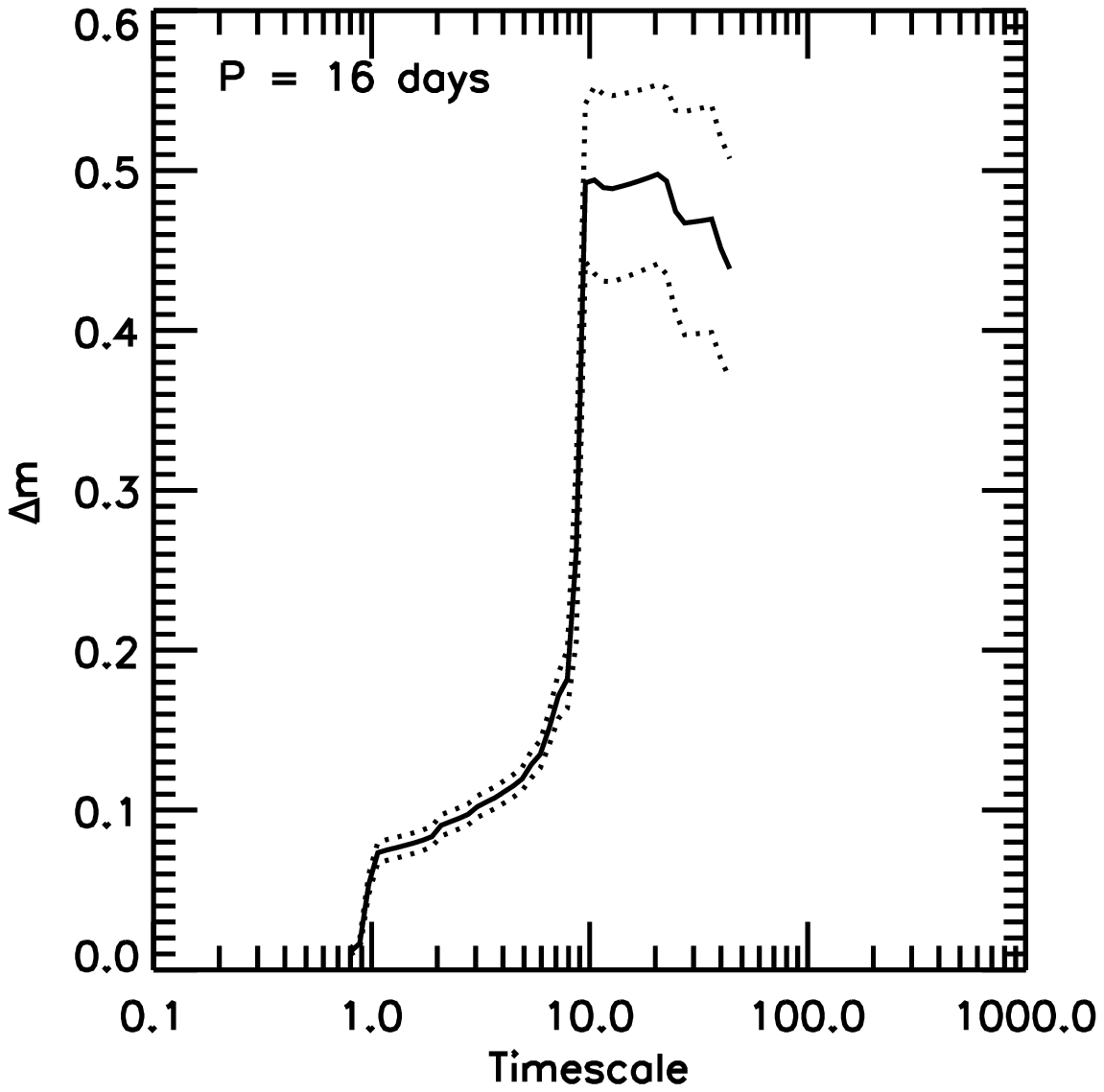}
} {
\includegraphics[width=0.32\textwidth]{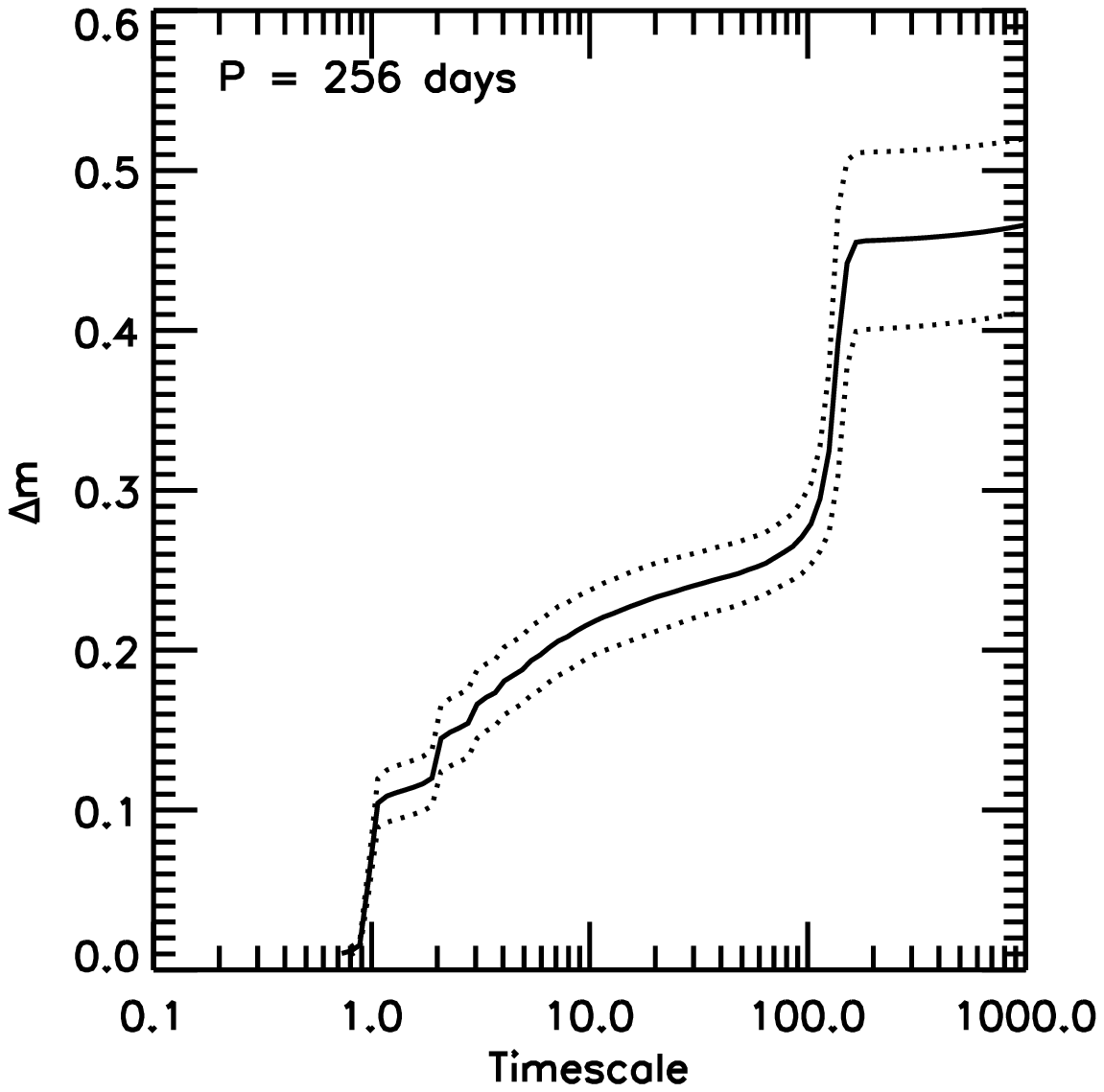}
} {
\caption{The mean peak-finding function from 1,000 simulations of a sine wave at several representative periods and a 5-95\% amplitude of 0.5~mag. Dotted lines represent the standard deviation of the peak-finding function at each time lag.} \label{fig_lcmc_sine_peak_ptf} \label{fig_lcmc_sine_peak_ptf_bump}
}

We present in Figure~\ref{fig_lcmc_sine_peak_ptf} the average value of the peak-finding curve for a set of 1,000 simulations of a sinusoidal light curve with different periods, observed using the PTF-NAN Full cadence. A common feature to all curves is a steep rise at half the period. This rise makes timescales based on the peak-finding curve highly consistent for sinusoids. The steep rise terminates close to the light curve amplitude (all simulations were run at a 5-95\% amplitude of 0.5~mag, which for a sine corresponds to a peak-to-peak amplitude of 0.51~mag). The irregular behavior after the peak (including occasional descents) is an artifact of the averaging process used to make these plots; only the simulation runs in which the peak-finding plot was slow to reach the theoretical maximum of 0.51~mag have a peak-finding curve defined at long time offsets, so only those get included in the average.

\subsubsection{Damped Random Walk}

\plotthree{ptb}{
\includegraphics[width=0.32\textwidth]{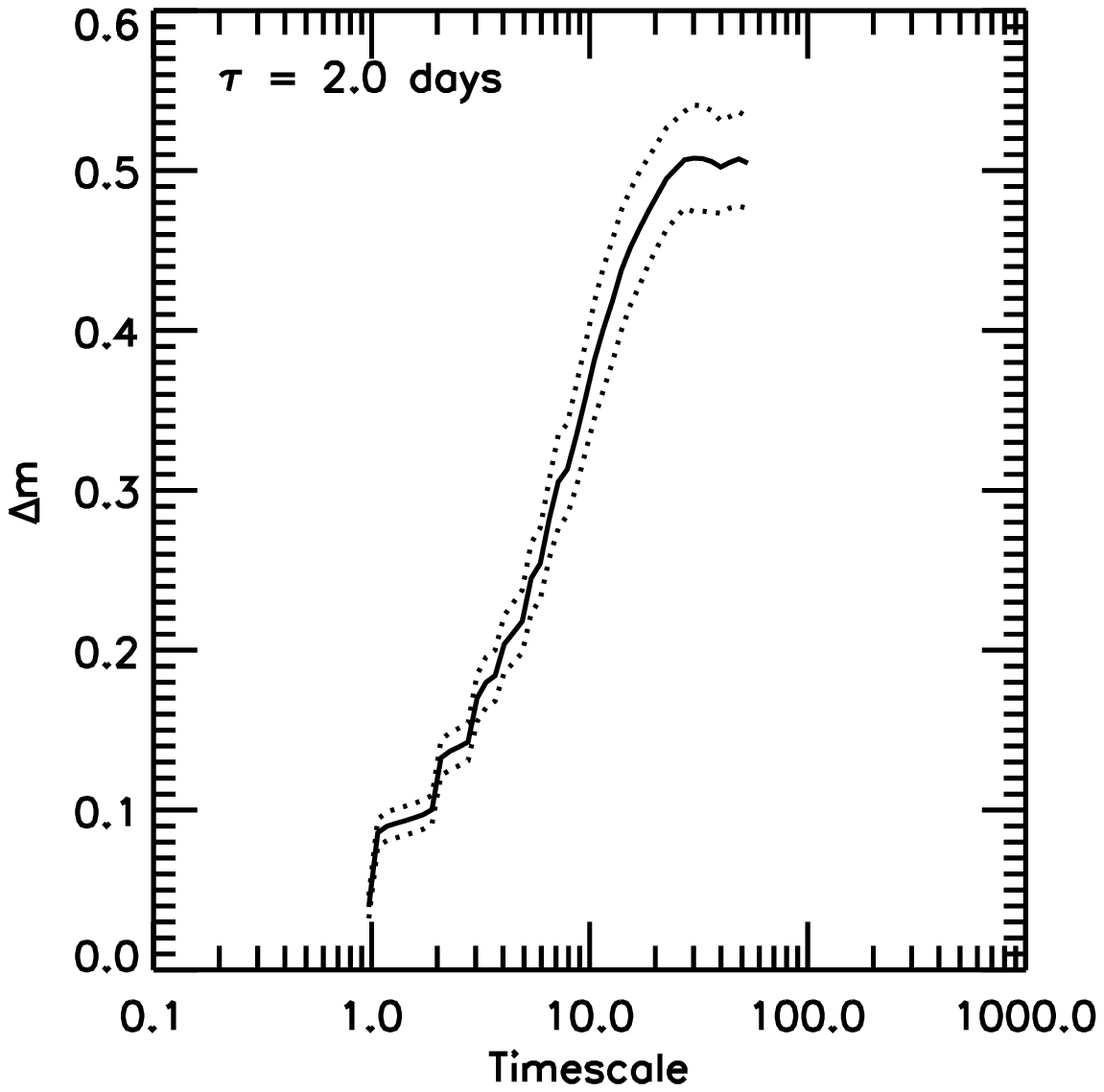}
} {
\includegraphics[width=0.32\textwidth]{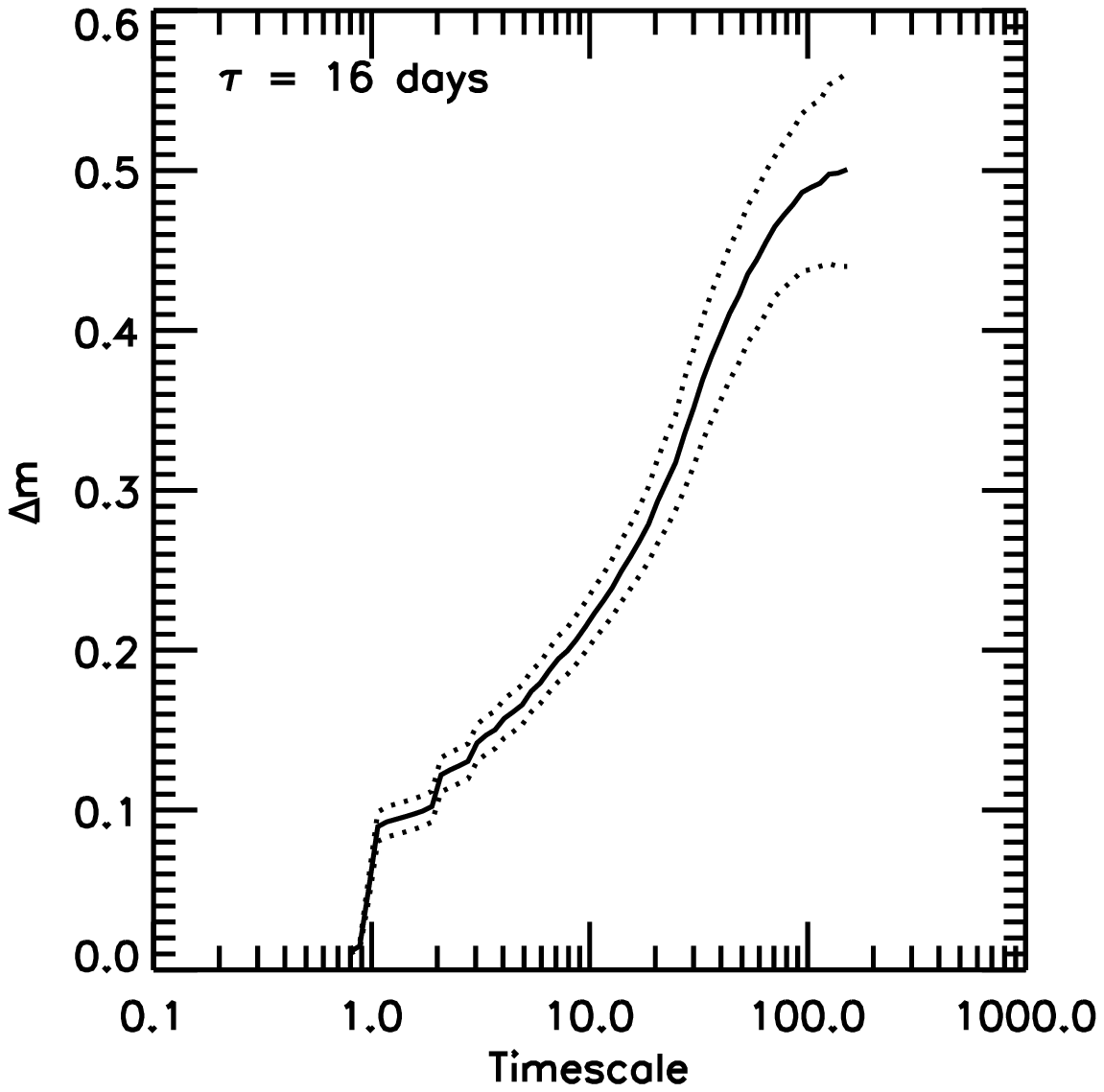}
} {
\includegraphics[width=0.32\textwidth]{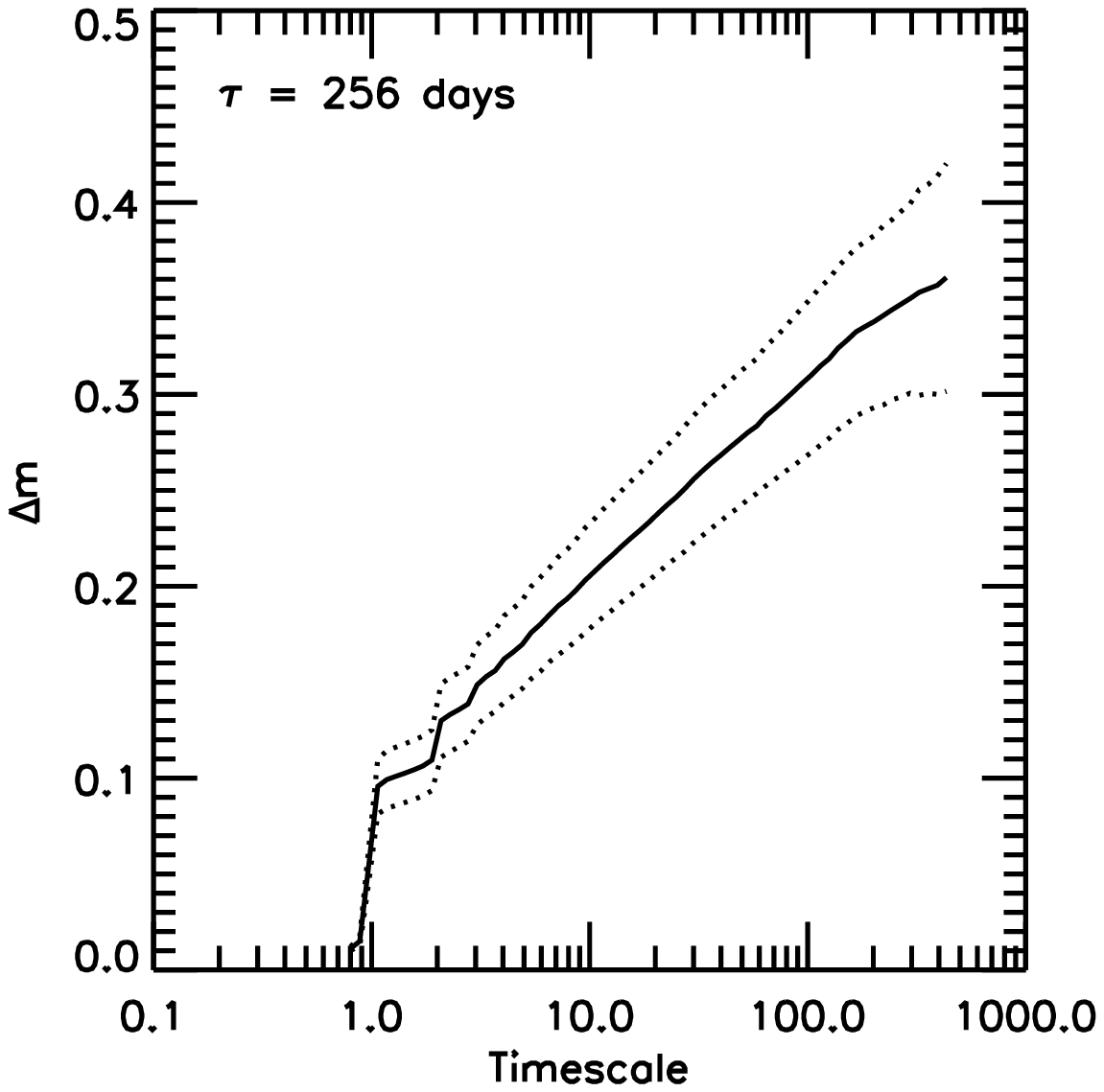}
} {
\caption{The mean peak-finding function from 1,000 simulations of a damped random walk at several representative timescales and a 5-95\% amplitude of 0.5~mag. Dotted lines represent the standard deviation of the peak-finding function at each time lag.} \label{fig_lcmc_drw_peak_ptf}
}

Figure~\ref{fig_lcmc_drw_peak_ptf} shows the average peak-finding curve for simulations of a damped random walk, observed using the PTF-NAN Full cadence. Except for a knee at 2-3 times the noise level, the peak-finding curves show no distinct features. The curves corresponding to long timescale variables rise more slowly than those corresponding to short timescale variables. As with the \dmdt~plots, the peak-finding plots corresponding to the longest-timescale damped random walks have increased scatter over long time intervals. We believe, in both cases, that the high scatter reflects the lack of a representative sample of variability at these long timescales. However, the effect described in the description of peak-finding curves for sinusoidal signals, where the curve may be undefined at long time intervals, likely contributes to the scatter here as well.

\subsubsection{Other Light Curves}

On a squared exponential Gaussian process, a peak-finding plot shows a leveling out on long timescales, indicating that high-amplitude minima and maxima rapidly become rare for these light curves. However, the peak-finding plot varies too much from light curve to light curve for this feature to serve as a useful timescale metric. 
On a random walk, which has no characteristic timescale, the peak-finding plot generally slopes upward, as it does for a damped random walk. However, the peak-finding plot for a random walk has many irregularities, including areas where it levels out, at timescales short compared to the observing base line. We have not determined the cause of this behavior.

\section{Timescale Metric Performance}\label{perform}

In this section we present the ensemble of simulated timescale metrics, and summarize their performance as a function of light curve and observational parameters. We focus on the mean and average behavior of the metrics, rather than on specific cases, in order to assess their suitability for large surveys. Since manual confirmation of automated results is impractical in such surveys, it is important to understand how reliable a timescale method is when left to its own devices.

We rank each timescale metric by the following criteria: \label{new_crit_list}
\begin{description}
\item[Precision:] characterized by the relative scatter across each set of 1,000 identical runs.
\item[Discriminatory power:] the smallest difference in underlying timescales that can be distinguished using the timescale metric, characterized by the ratio of the scatter in output timescale to the slope of the dependence of output timescale on input timescale.
\item[Sensitivity to noise:] characterized by the rate at which the precision and discrimination deteriorate as the noise level increases.
\item[Sensitivity to cadence:] characterized by the bias with respect to theoretical performance, where available, and by the range of timescales having optimal precision and discrimination at each cadence.
\item[Sensitivity to incomplete data:] characterized by the difference in output timescale between the PTF-NAN 2010 and PTF-NAN Full light curves.
\end{description}

We do \emph{not} test whether timescale metrics match the model input timescales in absolute scale, because the input timescales represent a convention for parameterizing light curves. For example, while the covariance function for a squared exponential Gaussian process was written as $\covar{(m(t_i), m(t_j))} = \sigma_m^2 e^{-(t_i-t_j)^2/2\tau^2}$ in Equation~\ref{eqn_analytic_lc_gp}, it could have been written, with equal validity, as $\covar{(m(t_i), m(t_j))} = \sigma_m^2 e^{-(t_i-t_j)^2/\tau^2}$, after the transformation $\tau \rightarrow \tau/\sqrt{2}$. Therefore, it is more important that the inferred timescale be proportional to the input timescale (a trait indirectly characterized by discriminatory power) than that it match exactly.

In this section, we use these criteria to examine candidate timescale metrics, while in section~\ref{lcmc_end}, we summarize the performance of the metrics as characterized by these criteria.

\subsection{\dmdt\ Plots}

\subsubsection{Qualitative Behavior}

In Figure~\ref{fig_lcmc_dmdt_snr_ptf}, we illustrate the performance of timescales derived from \dmdt~plots for light curves observed using the PTF-NAN Full cadence. In order to clearly present the behavior across five orders of magnitude in input timescale, we normalize the inferred timescale by the timescale theoretically expected for a \dmdt\ plot given infinite cadence, infinite observing base line, and zero noise \citep{KpfThesis}. This timescale is $P/6$ for a sinusoidal signal, $1.178\tau$ for a squared exponential Gaussian process, and $0.693\tau$ for a damped random walk. These normalized timescales are shown in the top three panels of Figure~\ref{fig_lcmc_dmdt_snr_ptf}; the dotted line across the middle of the plot represents behavior consistent with analytical theory. 

For sinusoidal signals, the calculated timescale is in general consistent with analytical results. For damped random walks, the calculated timescale is consistent with theory for damping times up to 16~days; walks with timescales of 64~days or longer see a fall-off of up to a factor of three as the \dmdt\ plot no longer has a representative sampling of the variability.

\subsubsection{Precision}

Panels d-f of Figure~\ref{fig_lcmc_dmdt_snr_ptf} show the scatter in the estimated timescale over multiple simulation runs. For sinusoidal signals, the scatter is of order 10\%. For 2-day periods the scatter is much higher because at that period the largest change in the signal is over time intervals from half a day to a day, which our nightly cadence samples very poorly. For the damped random walk, the scatter is of order 40\% for short timescale variables, and 50\% or larger at long timescales. 

\subsubsection{Discrimination}

The bottom three panels of Figure~\ref{fig_lcmc_dmdt_snr_ptf} show the smallest difference in input timescale that can be distinguished using the output timescale. Since the \dmdt\ timescale is, in general, proportional to the true timescale, the discrimination is set by the scatter in the estimated timescale. For long-timescale damped random walks, the drop-off in output timescale is reflected in a reduced ability to discriminate between timescales differing by less than a factor of two.

\subsubsection{Sensitivity to Noise}

\begin{figure}
\plotnine{
\includegraphics[width=0.32\textwidth]{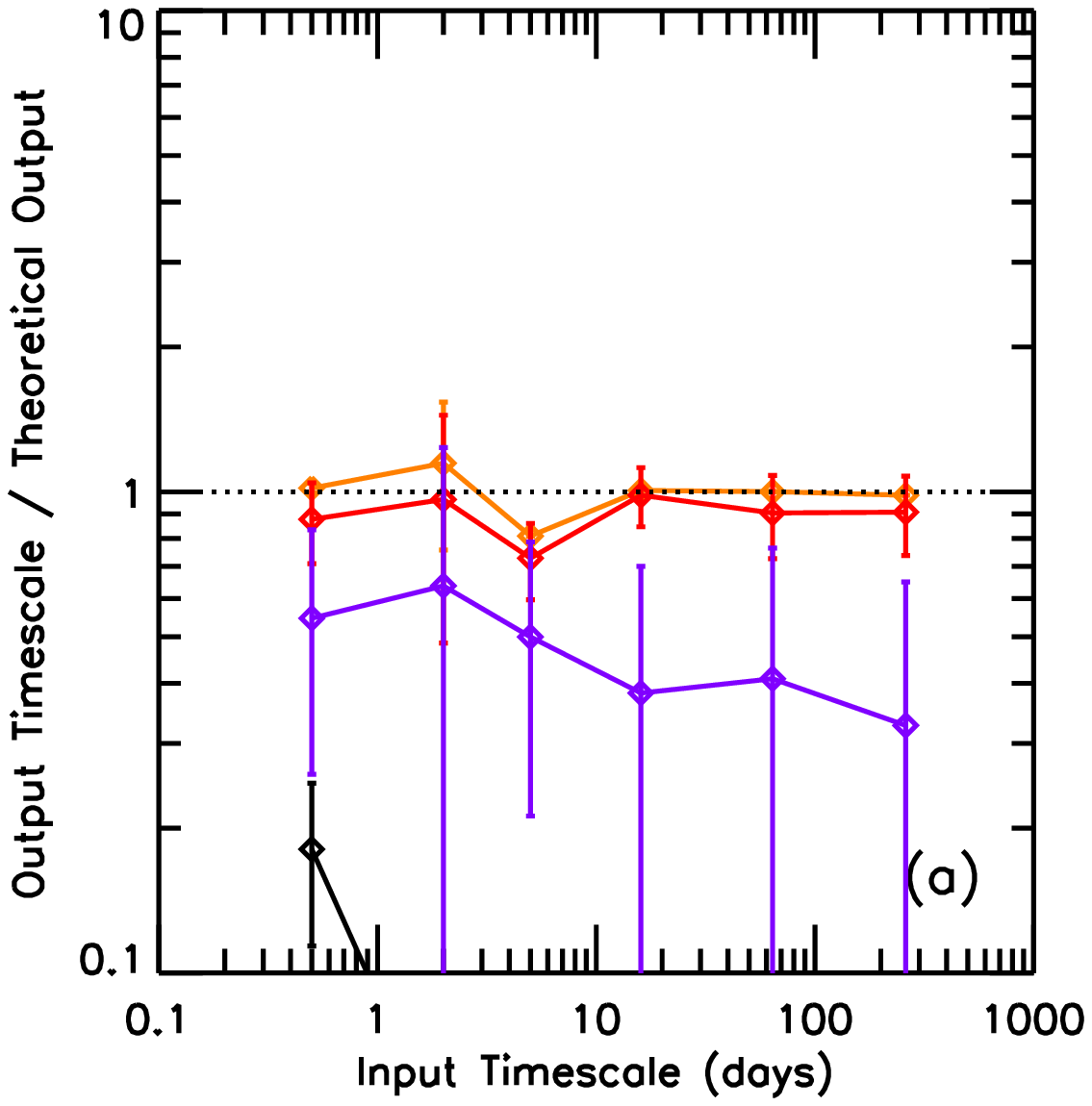}
} {
\includegraphics[width=0.32\textwidth]{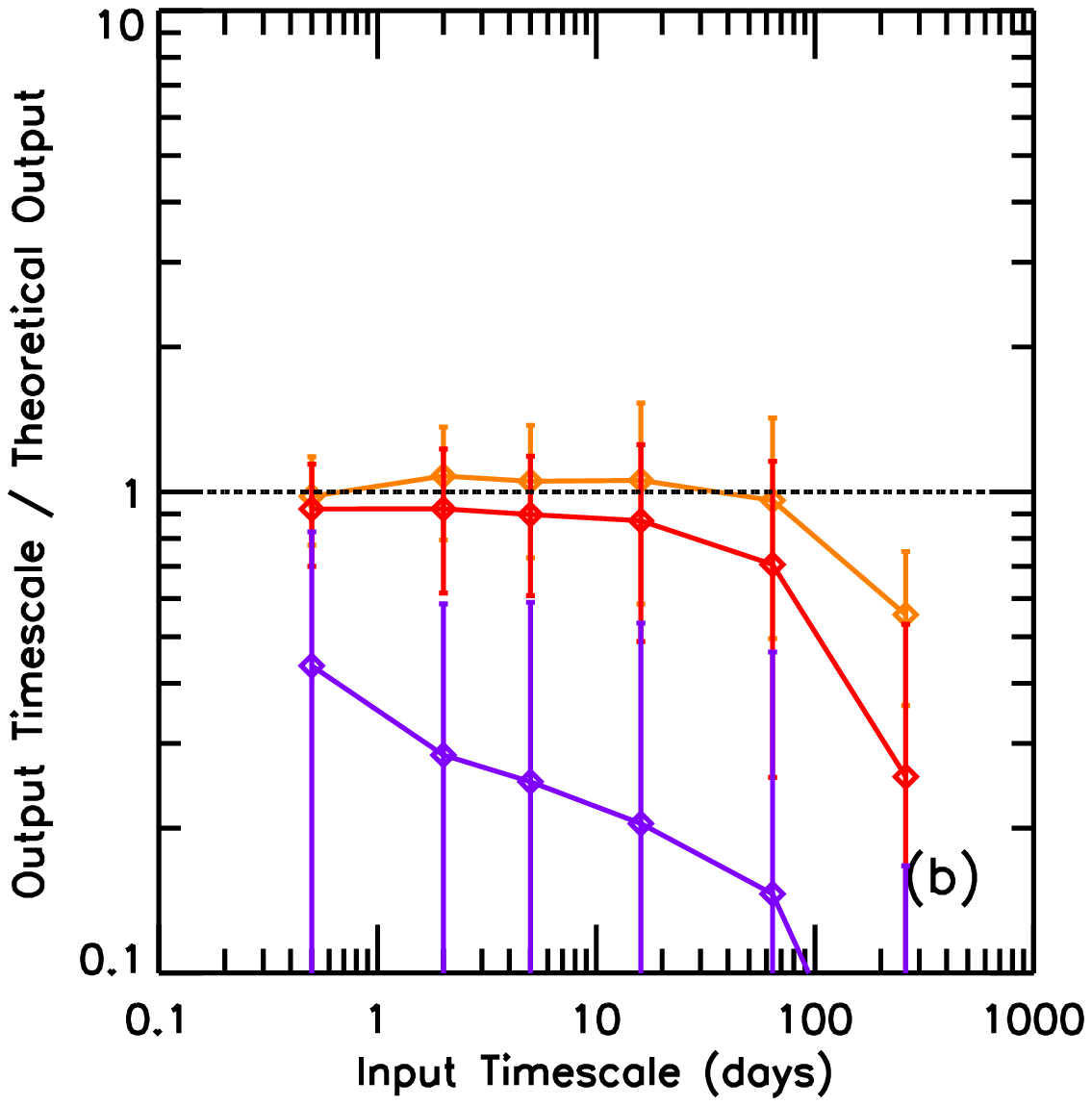}
} {
\includegraphics[width=0.32\textwidth]{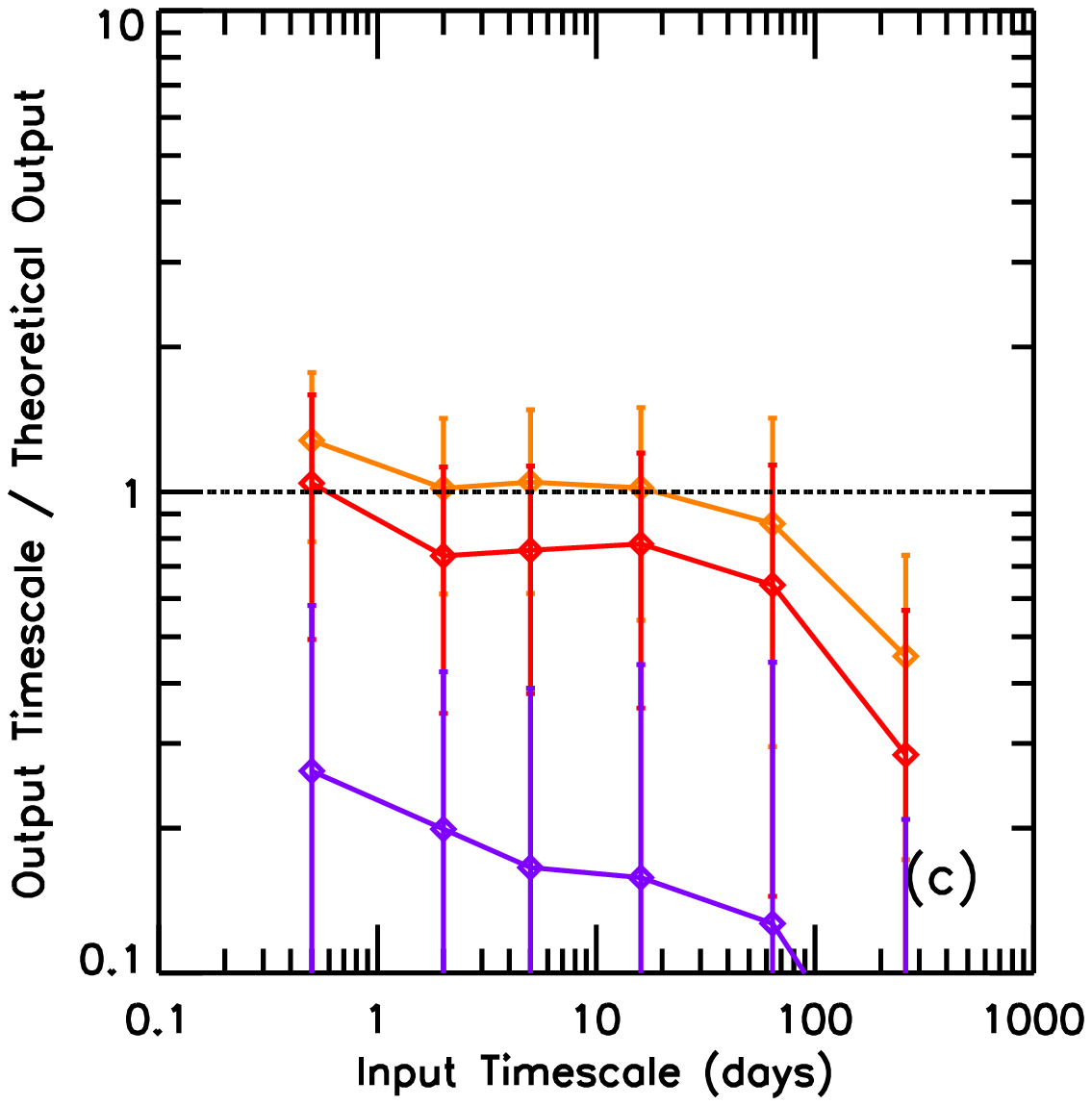}
} {
\includegraphics[width=0.32\textwidth]{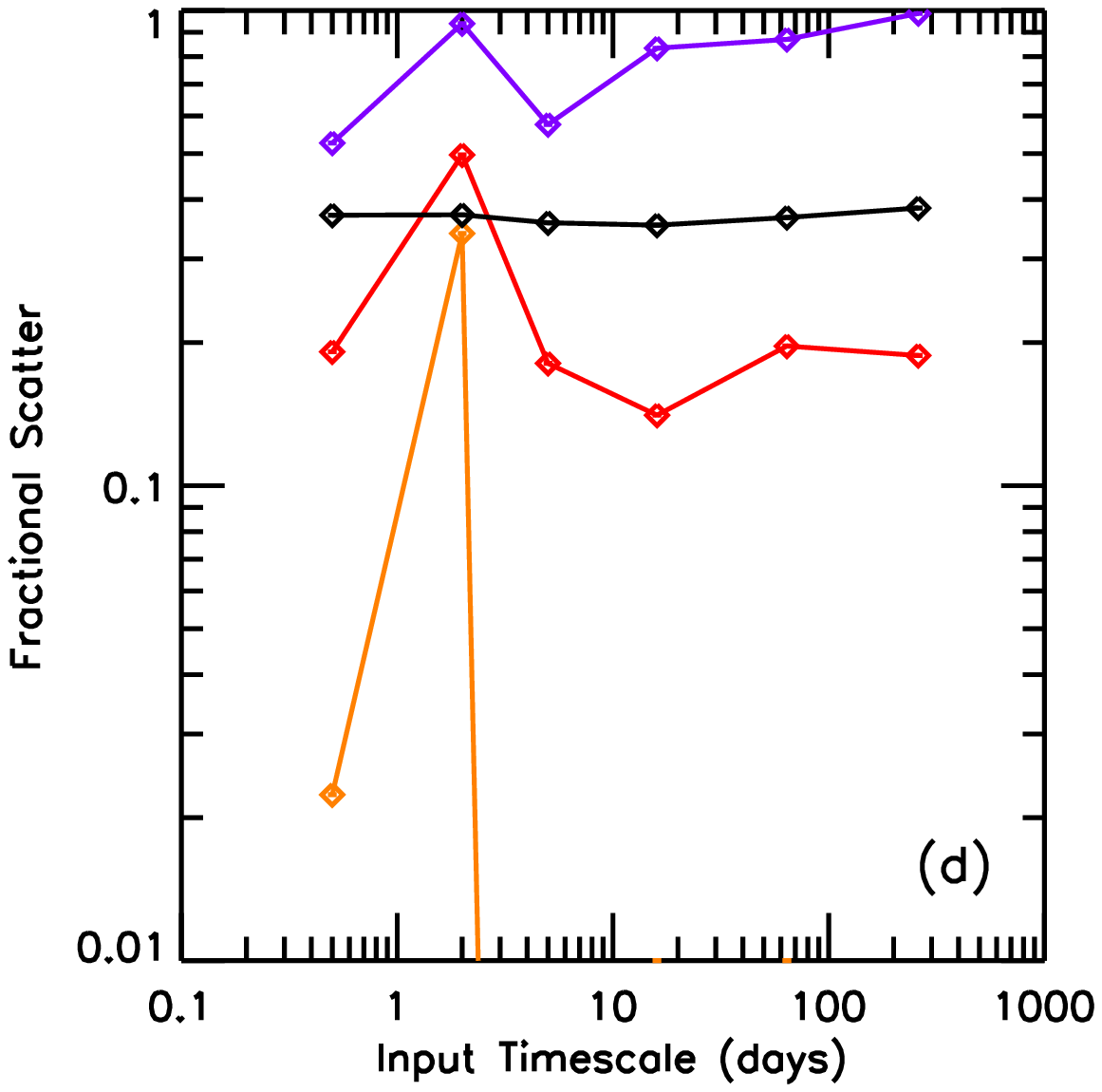}
} {
\includegraphics[width=0.32\textwidth]{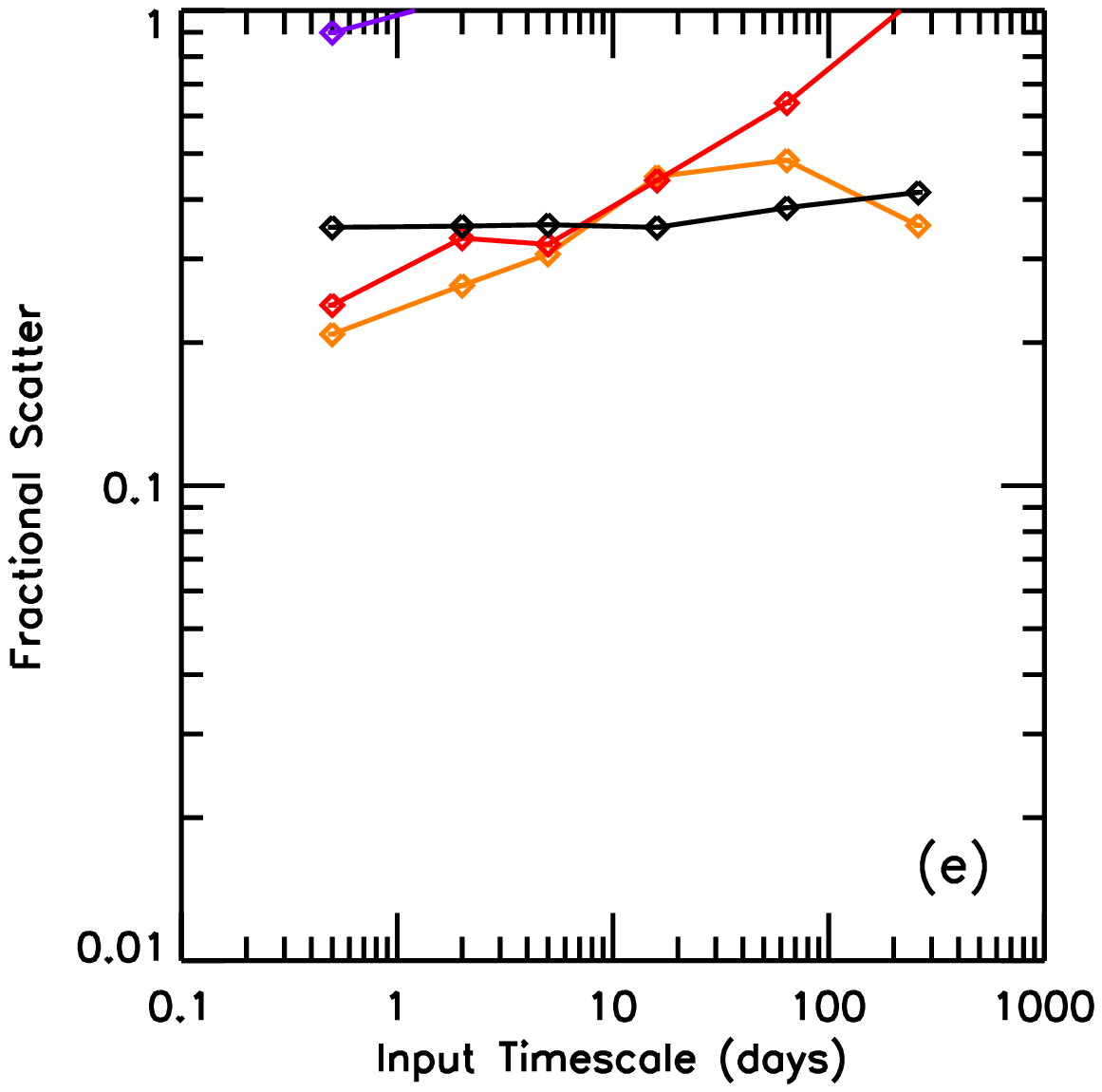}
} {
\includegraphics[width=0.32\textwidth]{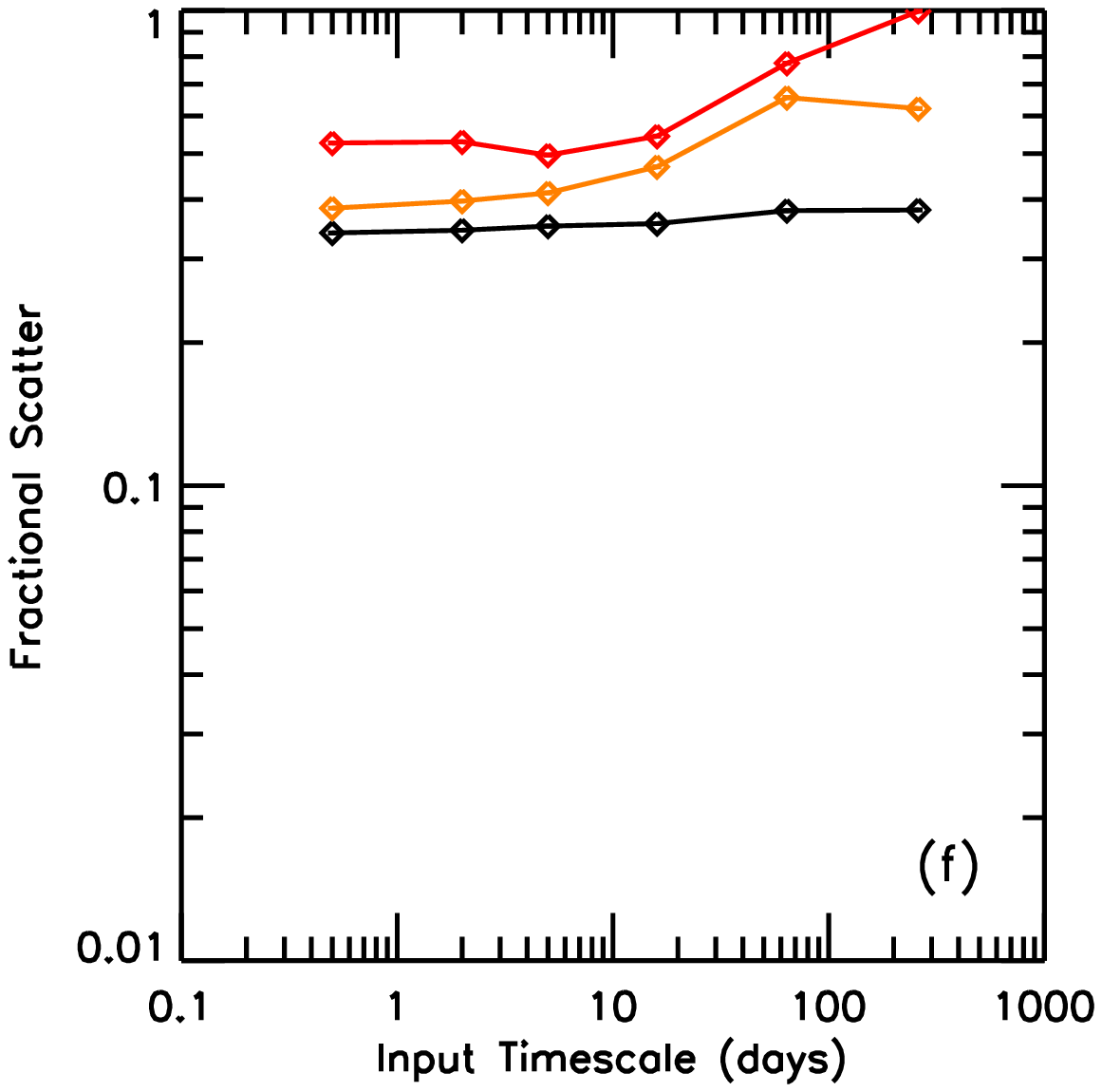}
} {
\includegraphics[width=0.32\textwidth]{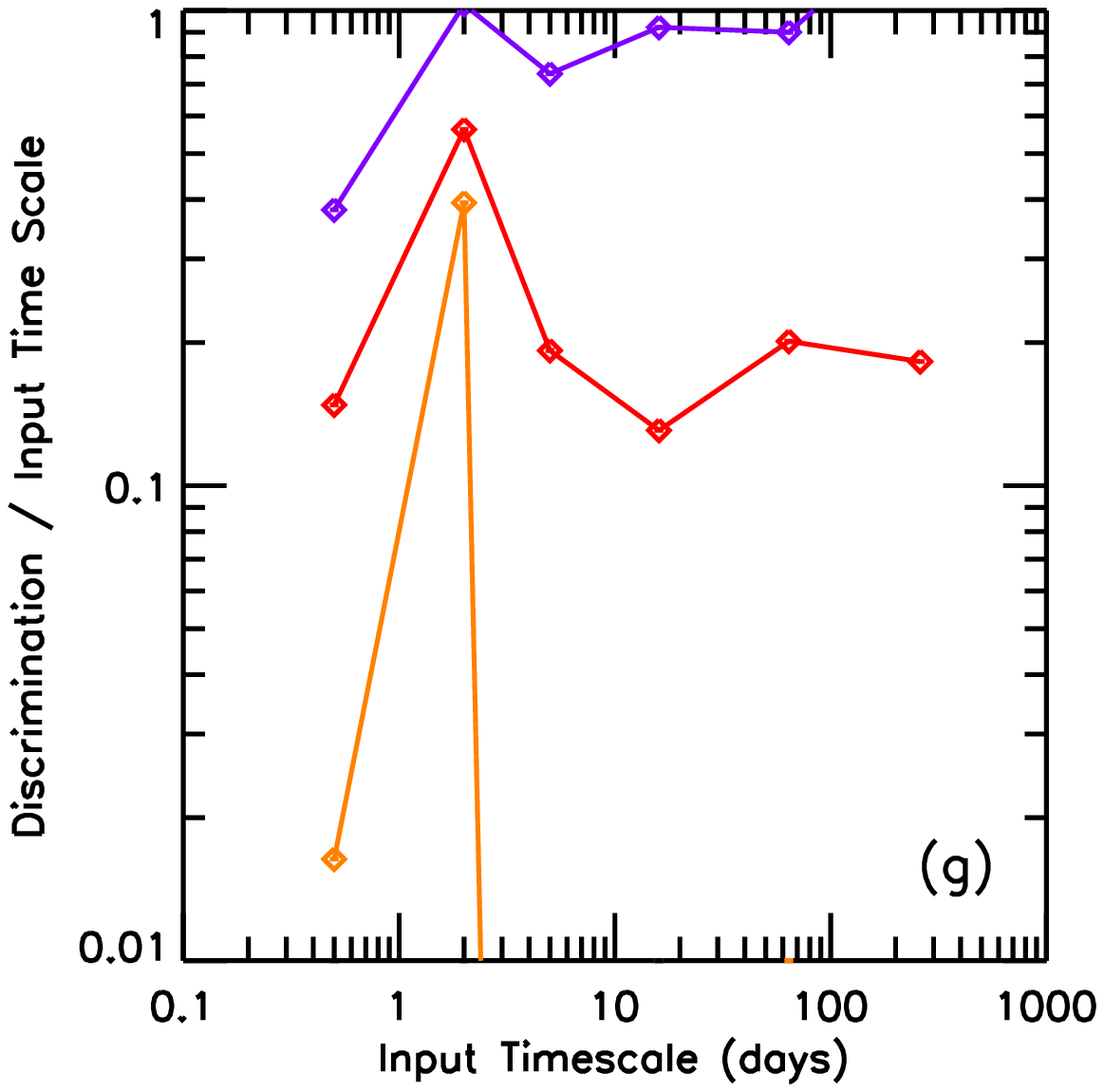}
} {
\includegraphics[width=0.32\textwidth]{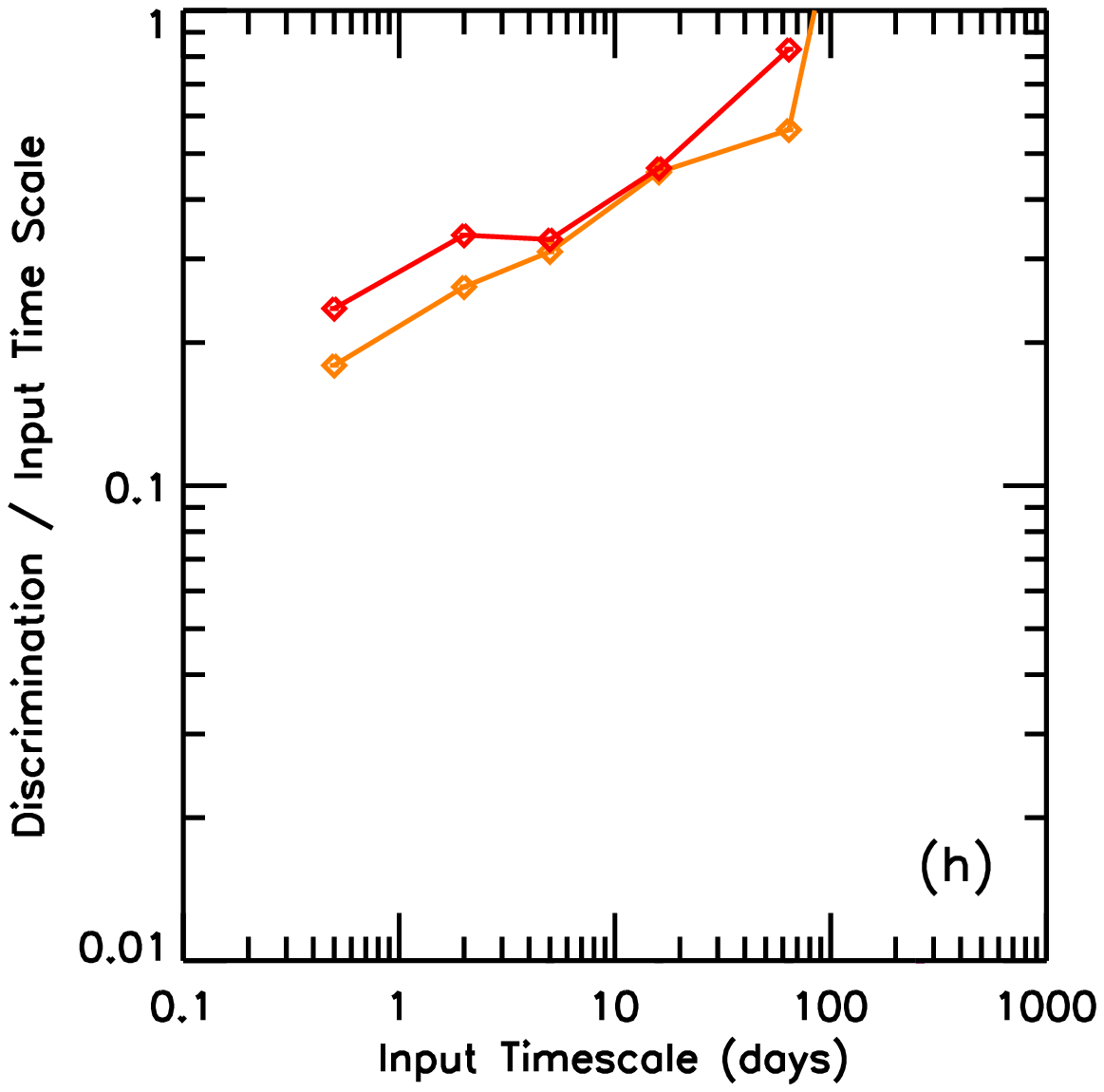}
} {
\includegraphics[width=0.32\textwidth]{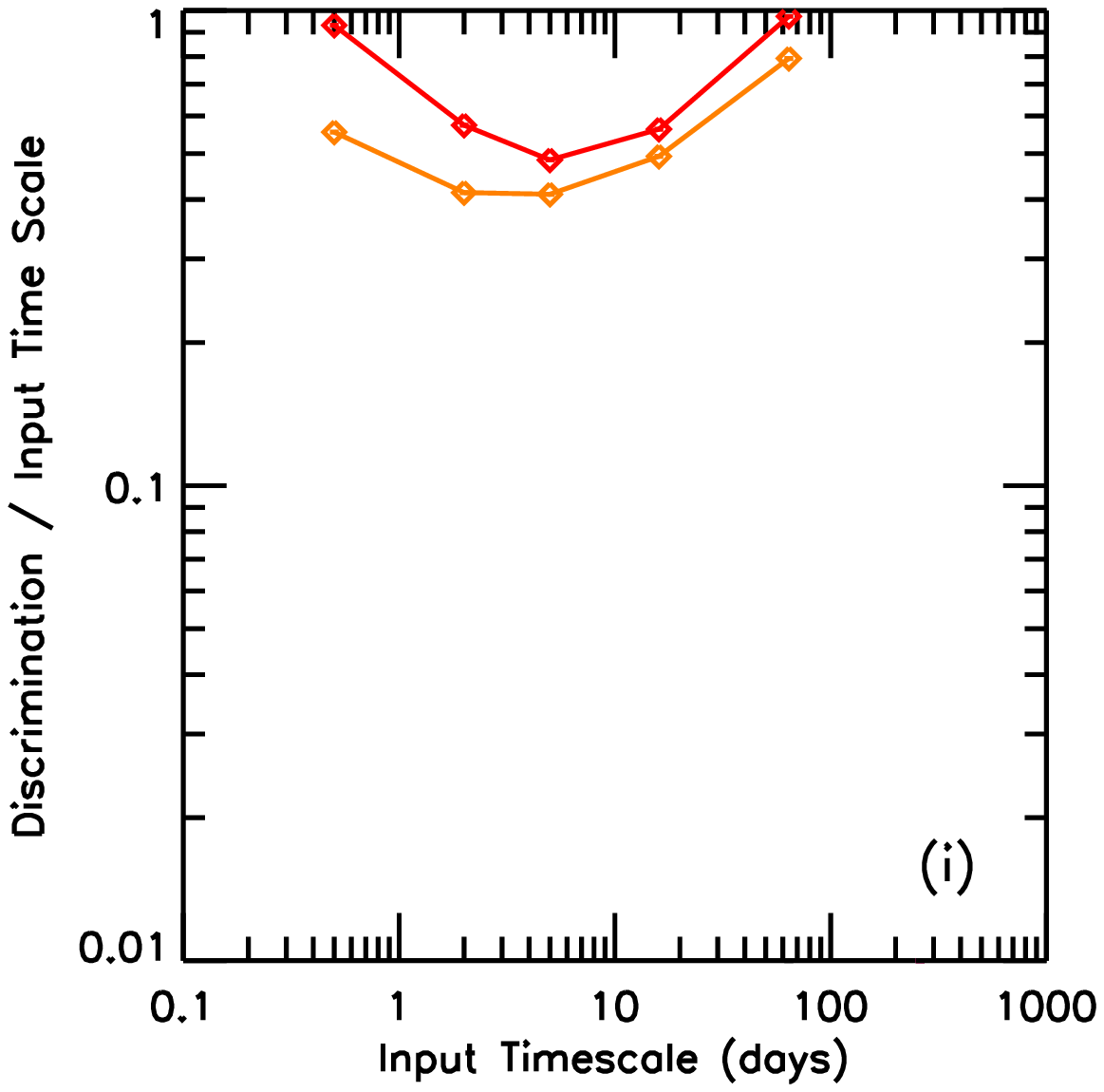}
}
\caption{The timescale calculated from a \dmdt\ plot, plotted as a function of the true underlying timescale input to the simulation. Columns are, from left to right, for a sinusoidal, squared exponential Gaussian process, and damped random walk model. Top panels show the ratio of the output timescale to the value predicted in \citet{KpfThesis}, in order to improve the dynamic range of the plot. Middle panels show the ratio of the standard deviation to the mean output timescale. Bottom panels show the degree by which the input timescale has to change to significantly affect the output timescale. In all panels, orange represents zero noise, red represents a signal-to-noise ratio of 20, purple a signal-to-noise ratio of 10, and black a signal-to-noise ratio of 4. All light curves have an expected 5-95\% amplitude of 0.5~magnitudes and are sampled using the PTF-NAN Full cadence.} \label{fig_lcmc_dmdt_snr_ptf}
\end{figure}

Figure~\ref{fig_lcmc_dmdt_snr_ptf} shows the performance of a \dmdt\ timescale as a function of the signal-to-noise ratio of the light curve. The average value of the timescale changes very little between an effectively infinite signal-to-noise and a signal-to-noise ratio of 20, particularly for a sinusoidal signal (panel a). At a signal-to-noise ratio of 4, on the other hand, the calculated timescale is always close to the smallest timescale sampled, suggesting that the \dmdt\ plot is dominated by noise.

The scatter in individual measurements (panels d-f) rises smoothly with decreasing signal-to-noise. The degree to which noise degrades measurement precision is independent of both the light curve shape and the true timescale. The discriminating power of the timescale metric (panels g-i) degrades similarly, although for a damped random walk the power is more sensitive to signal-to-noise at short timescales than at long timescales.

\subsubsection{Sensitivity to Cadence}

\begin{figure}[ptb]
\plotnine{
\includegraphics[width=0.32\textwidth]{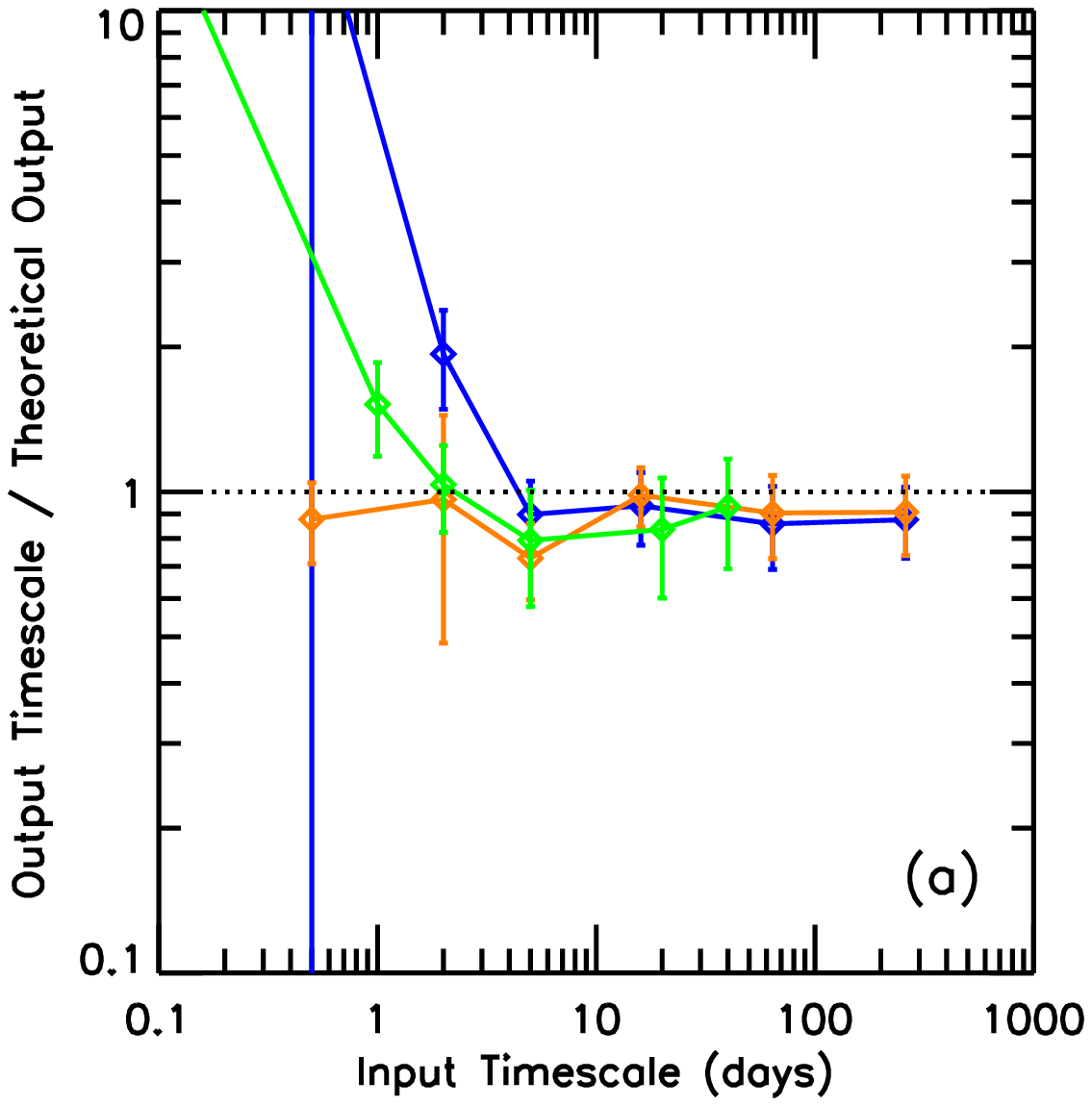}
} {
\includegraphics[width=0.32\textwidth]{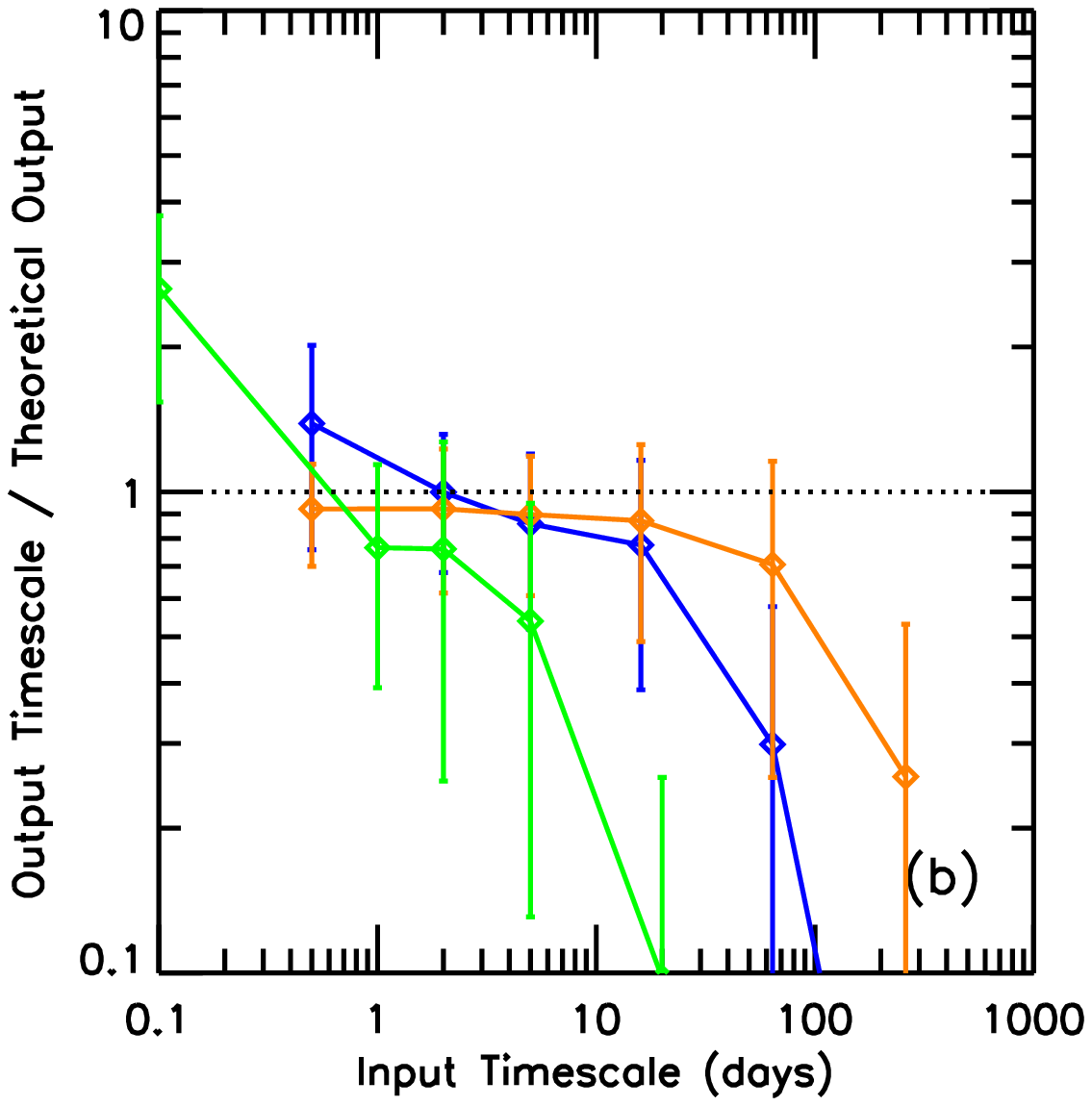}
} {
\includegraphics[width=0.32\textwidth]{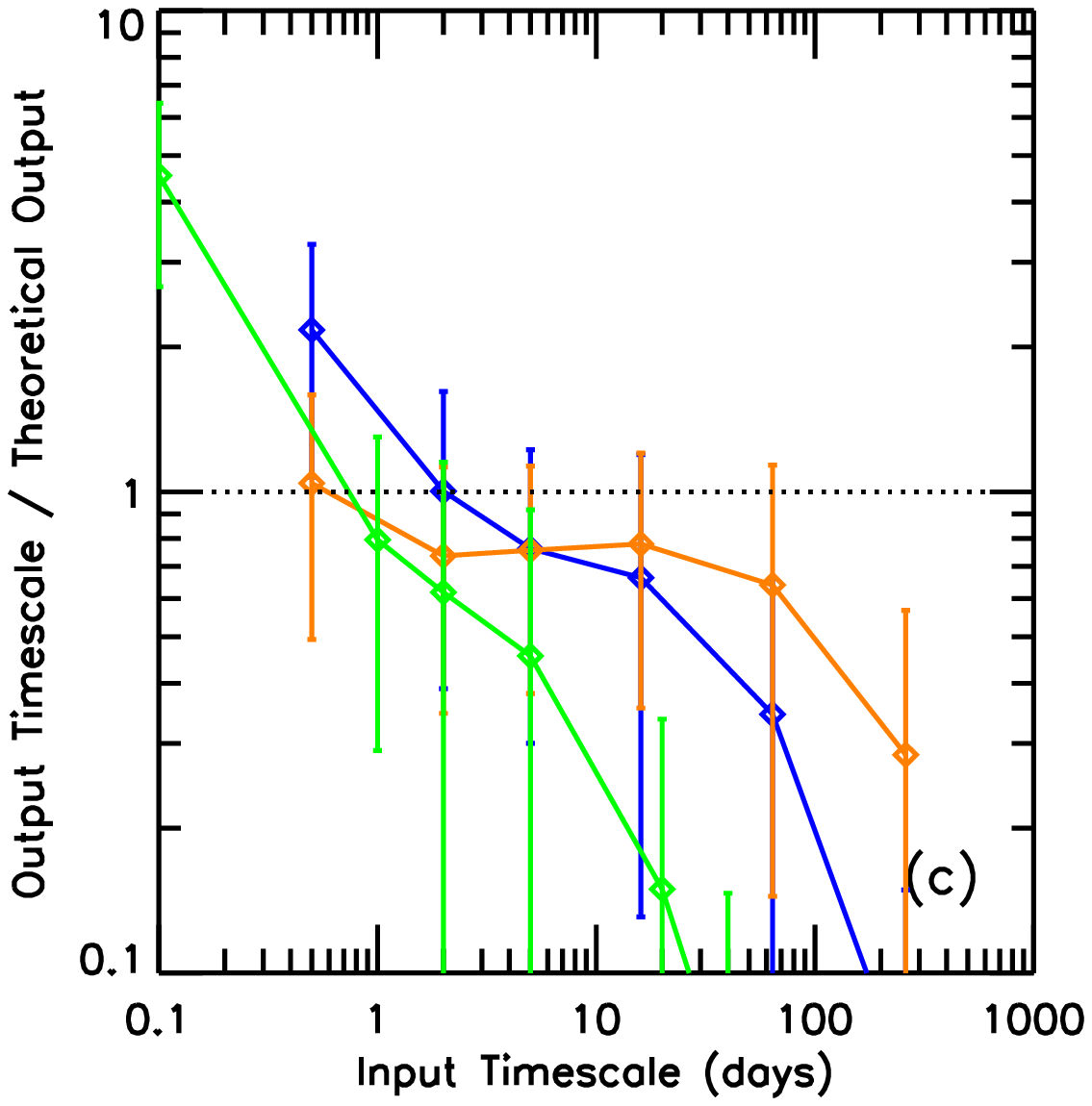}
} {
\includegraphics[width=0.32\textwidth]{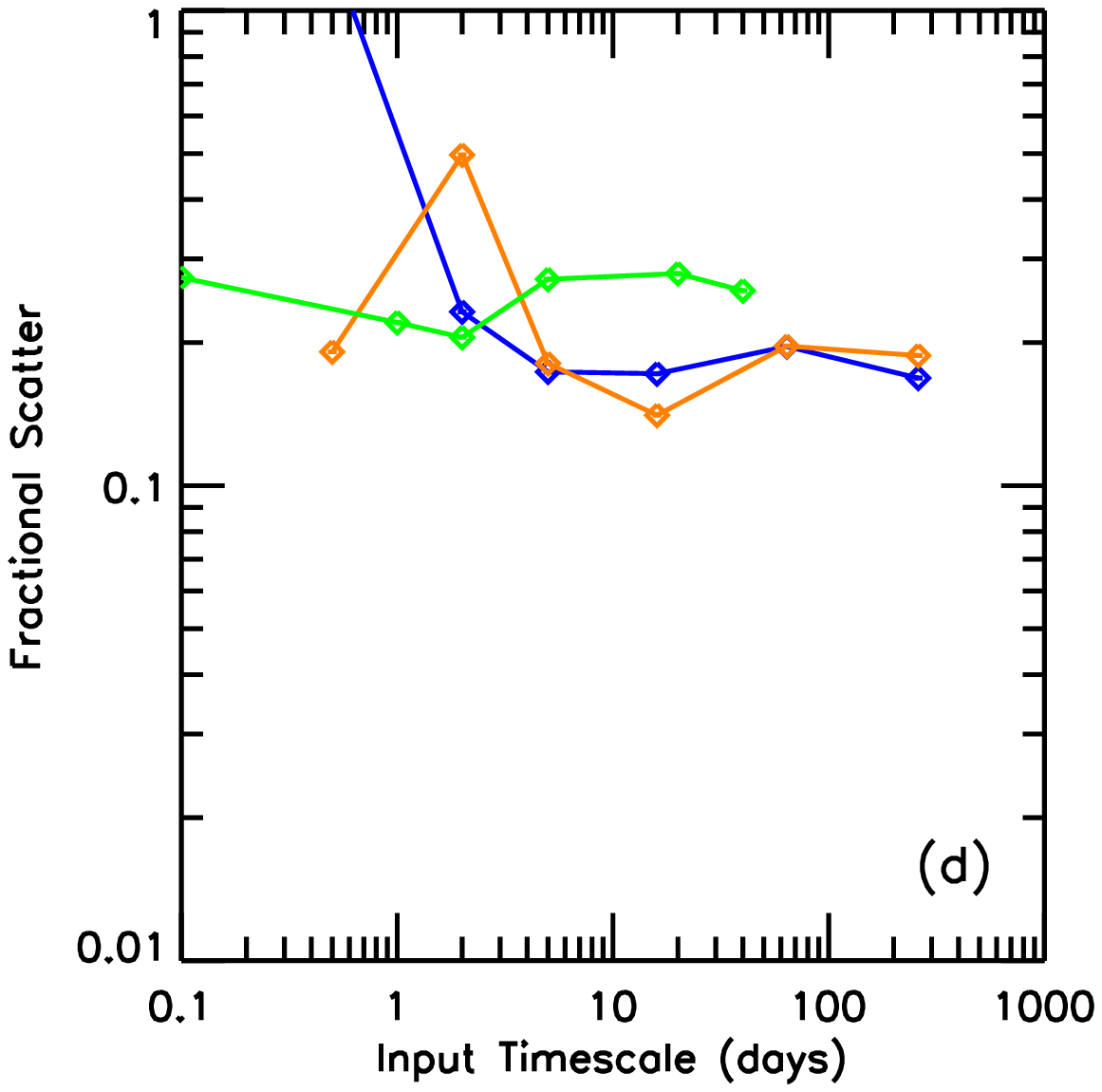}
} {
\includegraphics[width=0.32\textwidth]{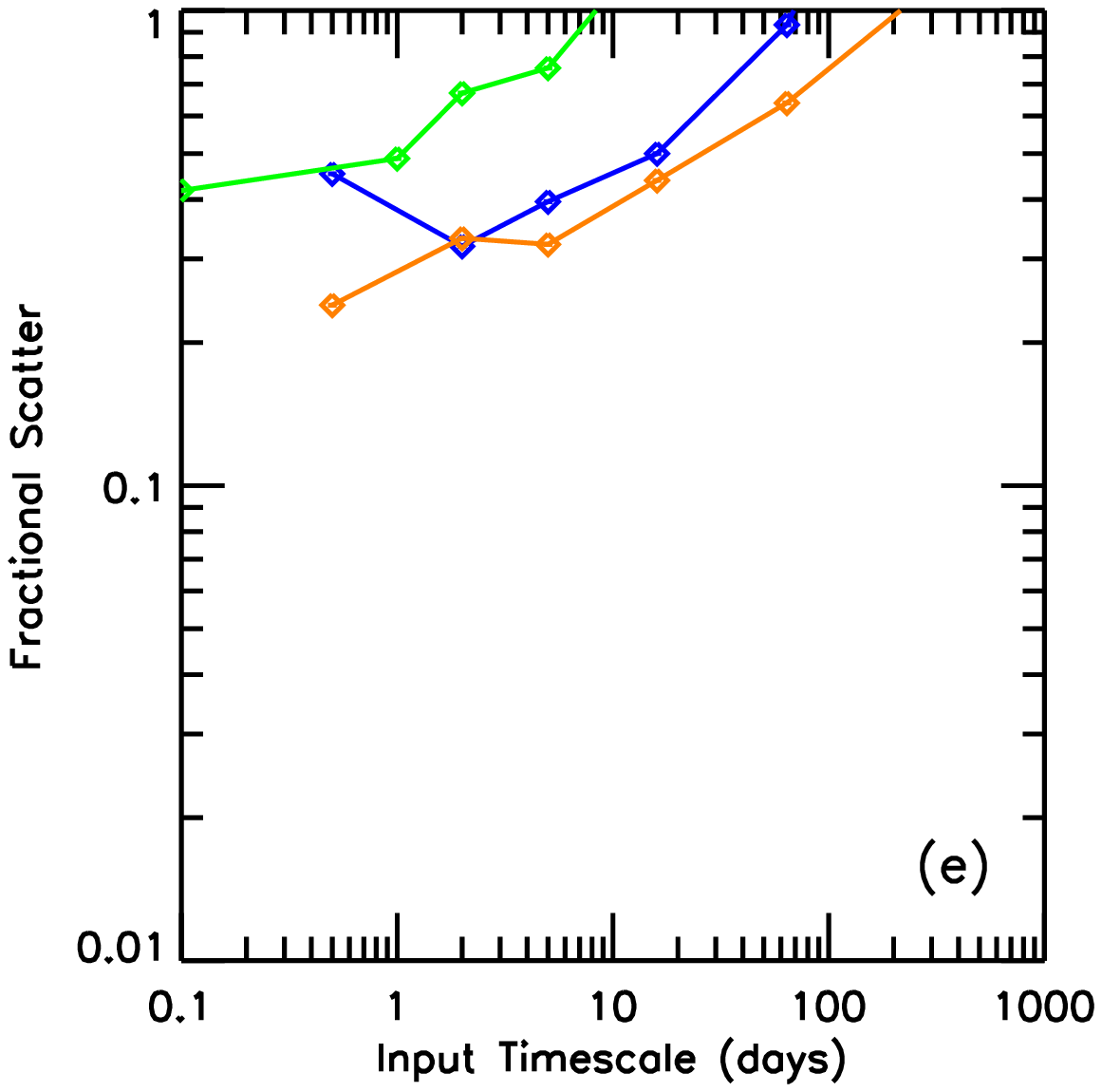}
} {
\includegraphics[width=0.32\textwidth]{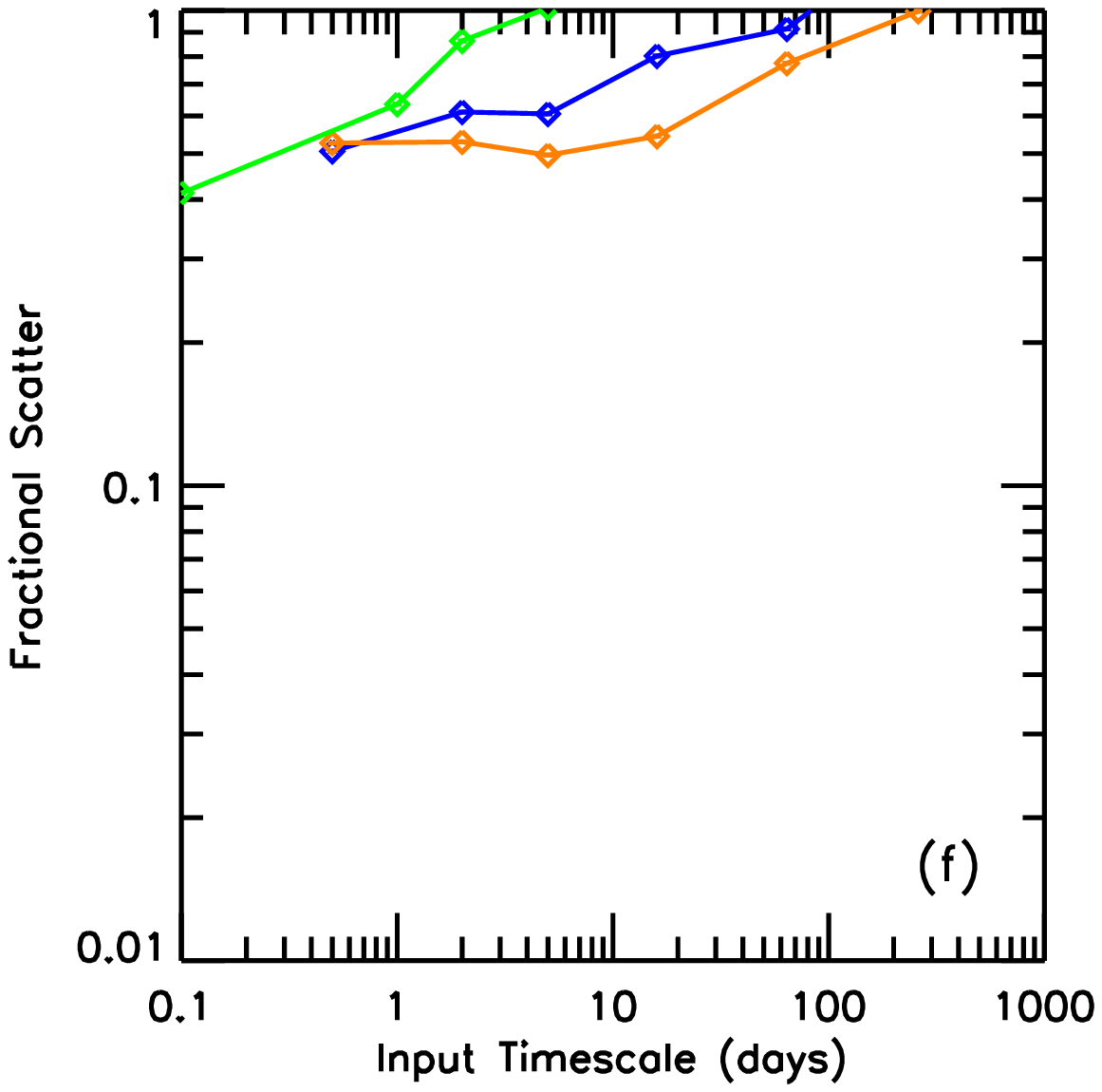}
} {
\includegraphics[width=0.32\textwidth]{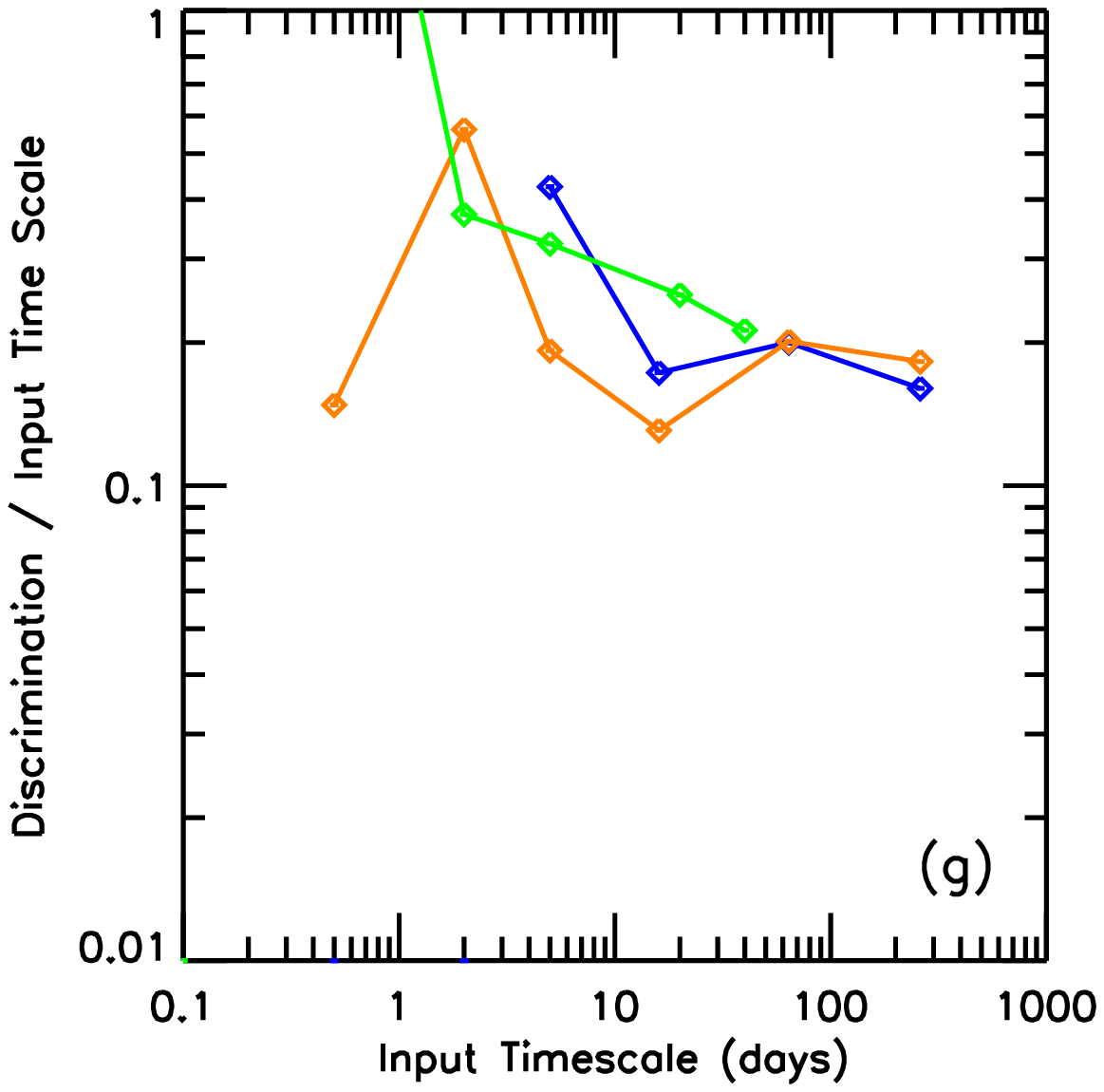}
} {
\includegraphics[width=0.32\textwidth]{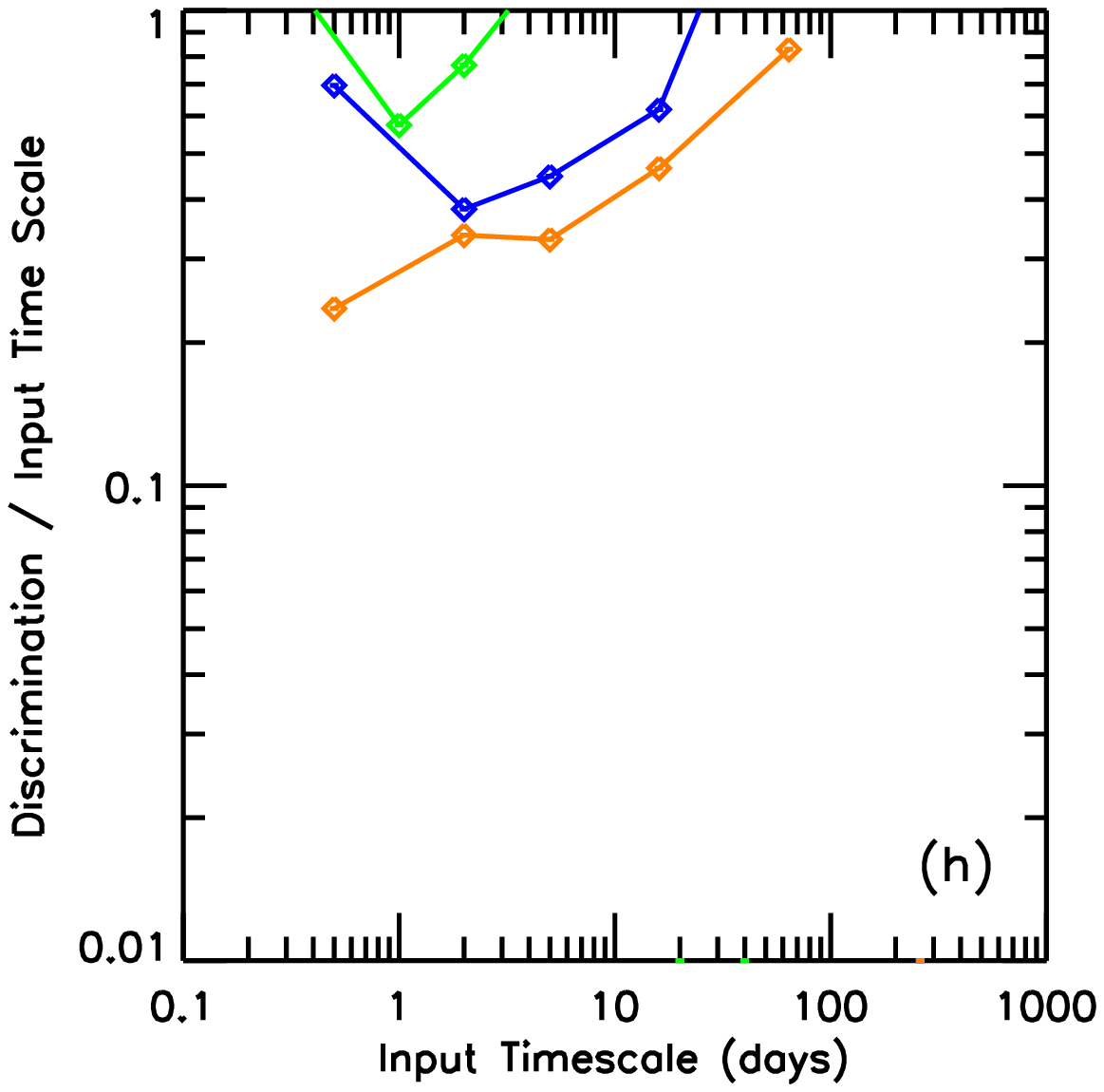}
} {
\includegraphics[width=0.32\textwidth]{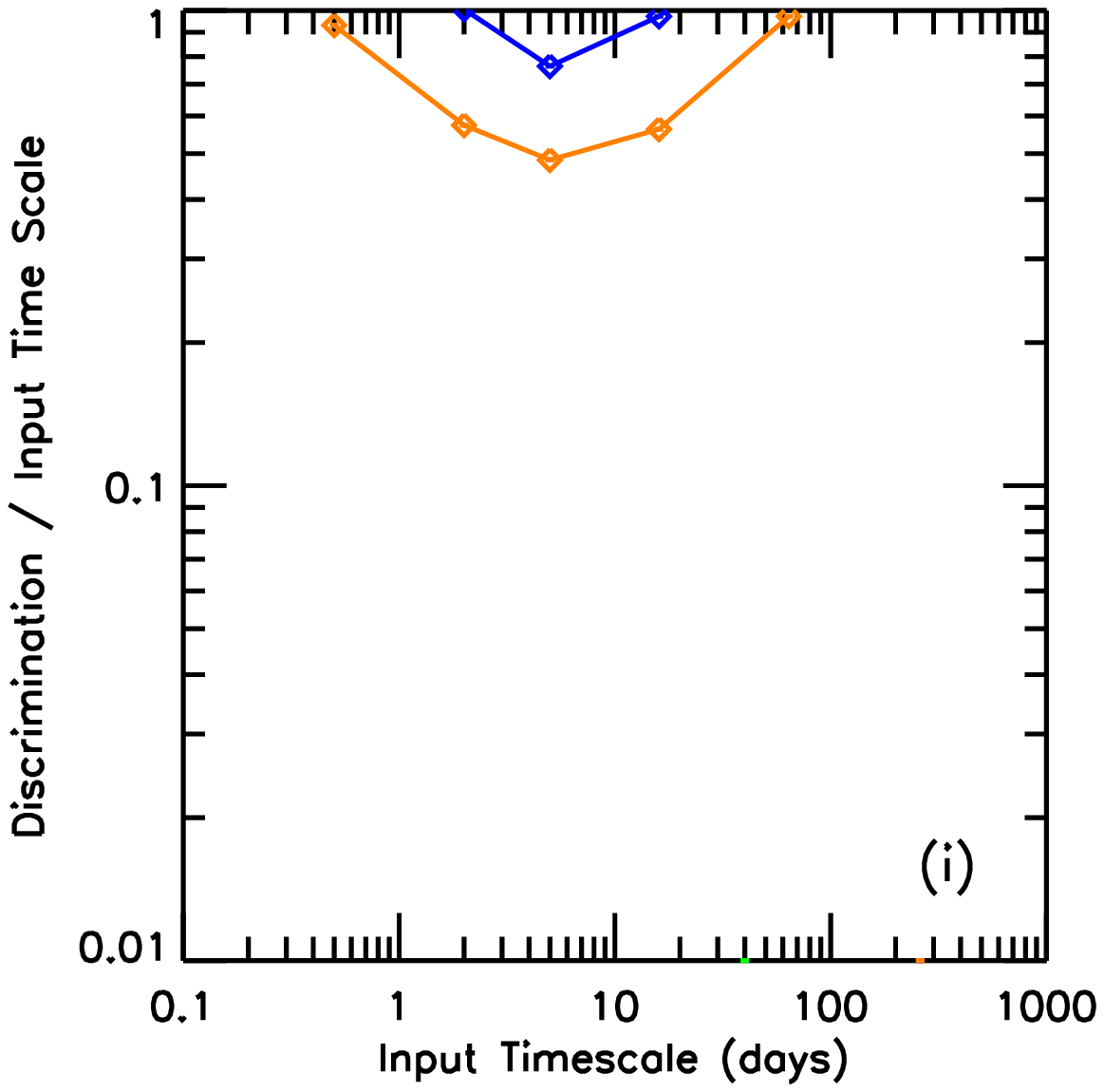}
}
\caption{As Figure~\ref{fig_lcmc_dmdt_snr_ptf}, but plotting only simulation runs with a signal-to-noise ratio of 20. Blue represents the PTF-NAN Full cadence, green the PTF-NAN 2010 cadence, and orange the YSOVAR 2010 cadence. Note that the scatter or discrimination statistic exceeds 100\% in some cases, indicating that the metric provides no useful constraint on underlying timescales in those circumstances.
The point at which the inferred timescale first becomes systematically underestimated for long-timescale sources, visible as a drop in the upper row of panels, is roughly proportional to the time base line covered by the cadence. The extent to which the inferred timescale is overestimated for short-timescale sources is roughly proportional to the spacing between observations.} \label{fig_lcmc_dmdt_cad_ptf} \label{fig_lcmc_dmdt_bordertimescale}
\end{figure}

In Figure~\ref{fig_lcmc_dmdt_cad_ptf}, we compare the behavior of a \dmdt\ timescale for the two PTF-NAN cadences and the YSOVAR 2010 cadence. For sinusoidal signals (panel a) sampled at the PTF-NAN Full cadence, the timescale is proportional to the period for periods from half a day to 256~days. At the PTF-NAN 2010 cadence, the timescale is well behaved for periods of 6~days or longer, but systematically too high for periods of 2~days or shorter. At the YSOVAR 2010 cadence, the timescale is proportional to the period for periods of two days and longer, but not for one day or shorter. For all three cases, the ``turnoff'' on the plot appears to happen at around the characteristic cadence (see section~\ref{lcmc_cadence_stats} for a definition) or by at most twice this value; the simulation grid is too coarse to make a precise assessment. This suggests that the \dmdt\ timescale cannot properly assess timescales that are not well sampled by the data.

The \dmdt\ timescale shows a somewhat different behavior for damped random walks (panel c). At the PTF-NAN Full cadence, the timescale is proportional to the damping time for timescales between half a day and 64~days. At the PTF-NAN 2010 cadence and at the YSOVAR 2010 cadence, there is no obvious interval over which the \dmdt\ timescale is proportional to the true timescale. At short timescales, the timescale is overestimated, possibly for the same reasons as for short periods in the sinusoidal case. At long timescales, the timescale is underestimated, because the observed light curve no longer probes the slowest variability in the system (cf. the \dmdt\ plots in Figure~\ref{fig_lcmc_drw_dmdt_ptf}).

When applied to a sinusoidal signal (panel d), a \dmdt\ timescale has slightly more scatter on the YSOVAR 2010 cadence ($\sim 30\%$) than on either of the PTF-NAN cadences ($\sim 20\%$). When applied to a damped random walk (panel f), the timescale has the most scatter on the YSOVAR 2010 cadence ($\sim 100\%$) and the least on the PTF-NAN Full cadence ($\sim 50\%$). The ability to discriminate input timescales (panels g-i) follows a similar pattern. This scaling is most likely driven by either the number of points or the time base line of each light curve, which is worst for the YSOVAR 2010 cadence and best for the PTF-NAN Full cadence.

\subsubsection{Summary}

We considered in this section the behavior of the \dmdt\ plot and associated timescales when applied to finite, noisy data sets. Even when given imperfect data, the \dmdt\ timescale correlates well with the true timescale of the light curve, confirming that it remains accurate even under realistic conditions. However, the behavior of individual light curves introduces a lot of scatter into the \dmdt\ timescale, with a typical standard deviation as high as 50-70\%. Therefore, \dmdt\ timescales are not precise.

The \dmdt\ timescales are moderately sensitive to the presence of noise in the light curve. Noise biases the inferred timescale downward; the effect is moderate if the noise RMS is at most one tenth the light curve amplitude, but quickly grows more severe at higher noise levels. The \dmdt\ timescale also requires a cadence with a high dynamic range of sampling; it will overestimate the timescale for sources varying faster than about $\sim 1-2$ times the characteristic cadence, and will underestimate the timescale if the true timescale is greater than about 1/15 of the monitoring base line (Figure~\ref{fig_lcmc_dmdt_bordertimescale}a-c). Since the \dmdt\ timescale is, on average, $1.178\tau$ for a squared exponential Gaussian process and $0.693\tau$ for a damped random walk, the \dmdt\ timescale is reliable so long as the output timescale is below $\sim 1/15$ the base line as well. We should note that the figure of 1/15 the base line is not as low as it sounds; for example, since the \dmdt\ timescale for a sine is typically one sixth the period, the \dmdt\ timescale can characterize periods up to 2/5 the base line.

While the \dmdt\ timescale keeps rising with input timescale past 1/15 the base line, ``saturating'' only at about 1/5 the base line (the case represented by the undamped random walk model), the discriminating power of the \dmdt\ timescale is worse than a factor of two at these long timescales.

The \dmdt\ timescale is a potentially useful timescale metric if the data are observed with a high dynamic range cadence, with a coverage window at least $\sim 30$ times the sampling interval, and if the 5-95\% amplitude of the variability is at least ten times the RMS of the noise. However, even then the timescales should be assumed to have a $1\sigma$ uncertainty of $\sim 50\%$. The \dmdt\ timescale may be more appropriate for ensemble studies than for characterizing individual light curves.

\subsection{Peak-Finding}

\subsubsection{Qualitative Behavior}\label{pf_bias}

In Figure~\ref{fig_lcmc_peak_snr_ptf}, we present the performance of peak-finding timescales for light curves sampled using the PTF-NAN Full cadence. 
For sinusoidal signals (panel a), the calculated timescale is proportional to the period for periods of two days or longer. For damped random walks (panel c), the calculated timescale increases with the damping time, but at a slower rate than a strict proportionality. In neither case do the inferred timescales ever fall below one day; the median separation between peaks is always at least one day even if the magnitude threshold is lowered far enough to probe noise.

\subsubsection{Precision}\label{pf_scatter}

The panels d-f of Figure~\ref{fig_lcmc_peak_snr_ptf} show the scatter in the estimated timescale over multiple simulation runs. For sinusoidal signals, the scatter is negligible for all but the shortest and longest periods. The scatter in the timescale is much larger for damped random walks; it is on the order of 20\% for short timescale light curves, but grows to over 100\% at longer timescales.

\subsubsection{Discrimination}

Because the peak-finding timescale is linear with the period for a sine, the discrimination (panels g-i) shows the same behavior as the scatter. Because the peak-finding timescale grows more slowly than the damping time for a damped random walk, and even levels out for long timescale signals, the discriminating power of the peak-finding plot is never better than $\sim 30\%$. 

\subsubsection{Sensitivity to Noise}

\begin{figure}[ptb]
\plotnine{
\includegraphics[width=0.32\textwidth]{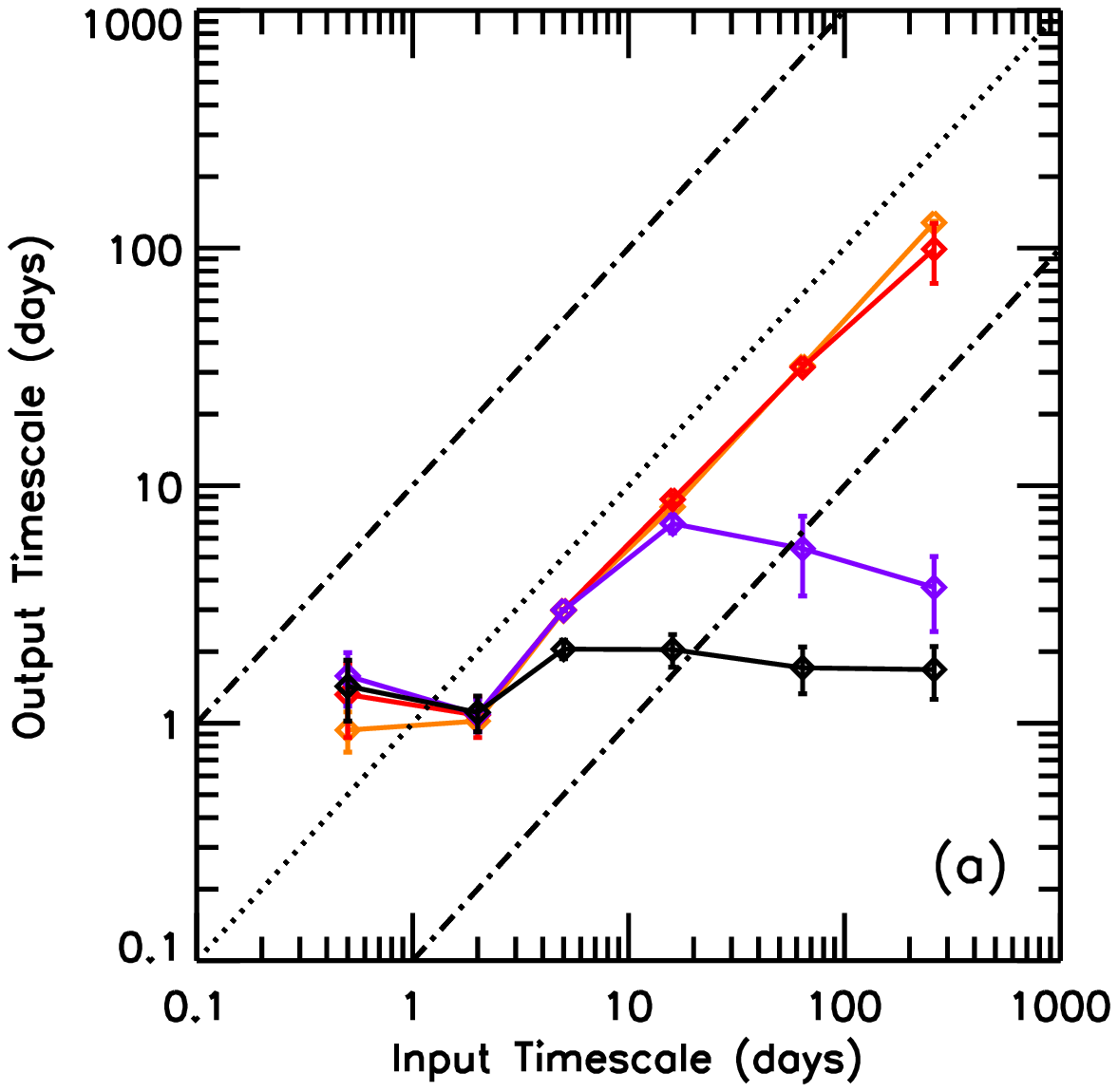}
} {
\includegraphics[width=0.32\textwidth]{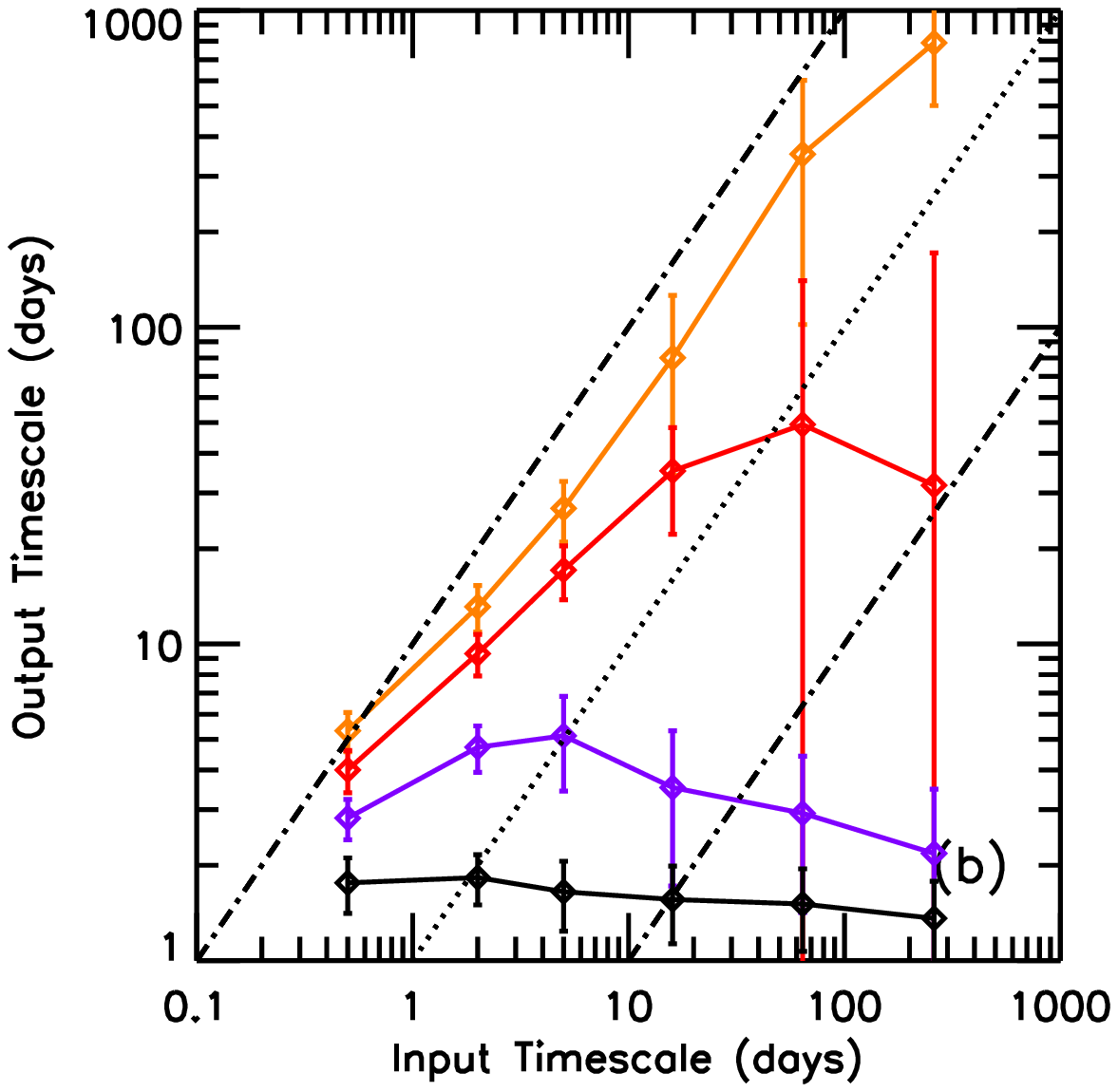}
} {
\includegraphics[width=0.32\textwidth]{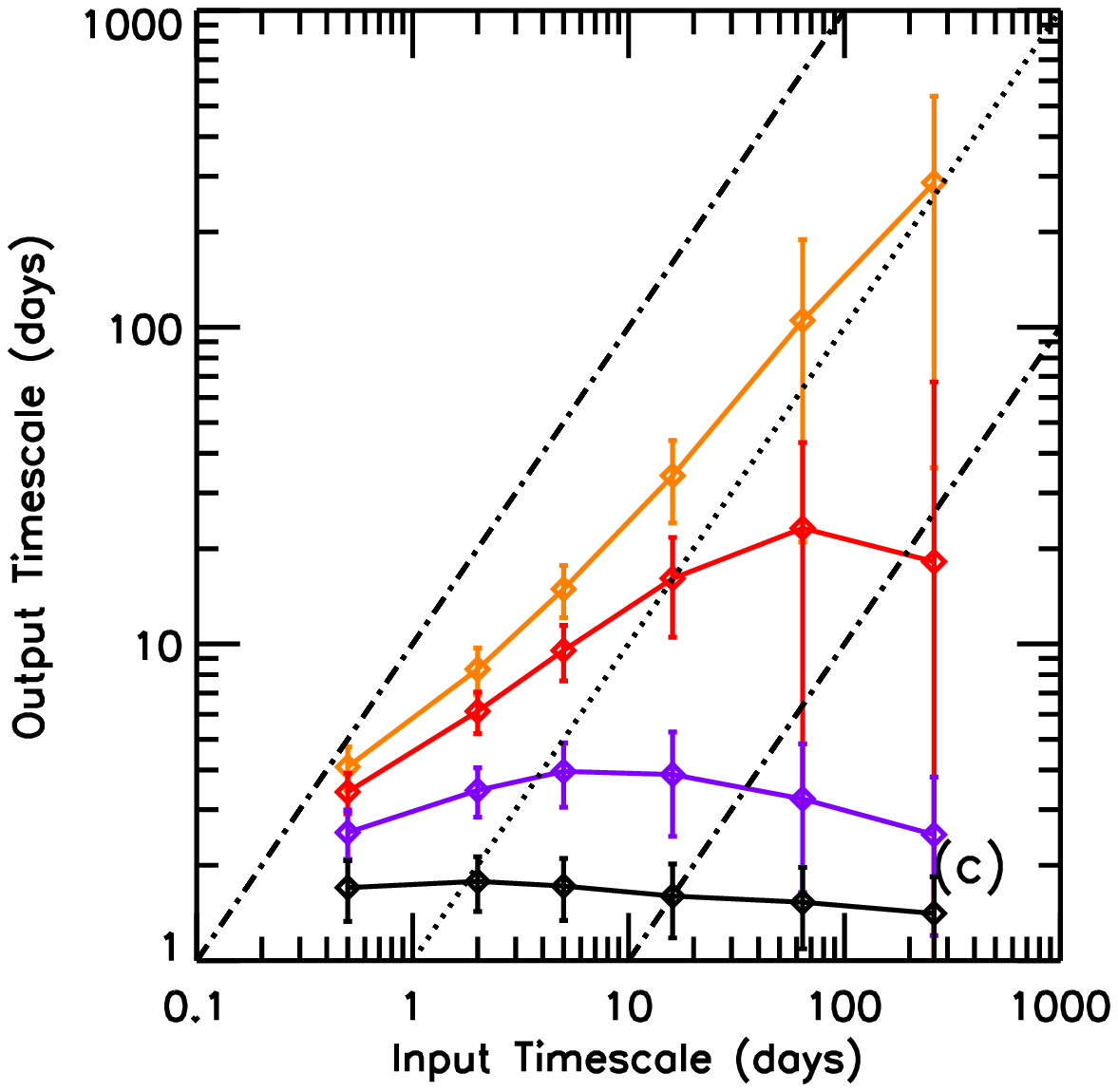}
} {
\includegraphics[width=0.32\textwidth]{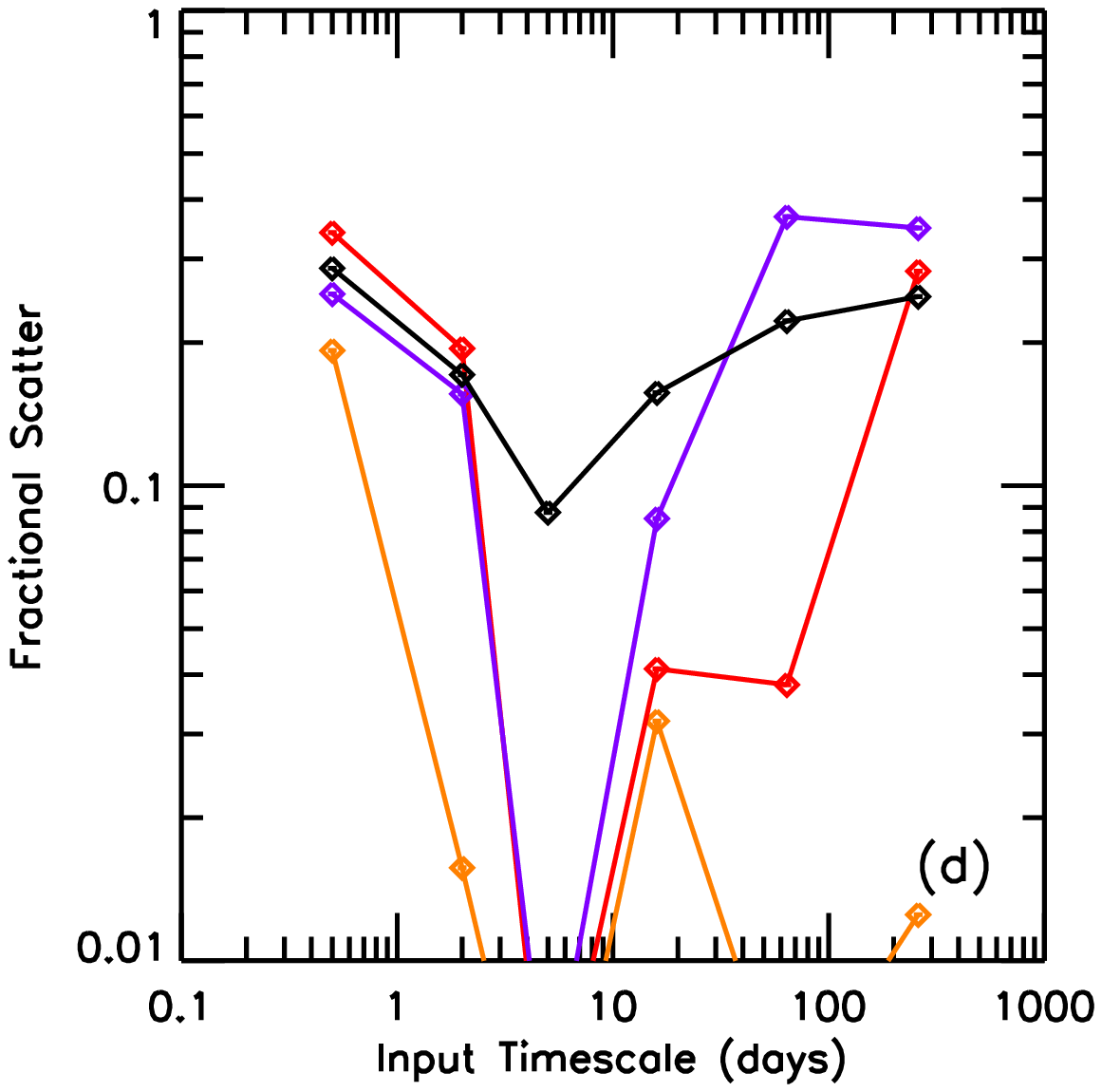}
} {
\includegraphics[width=0.32\textwidth]{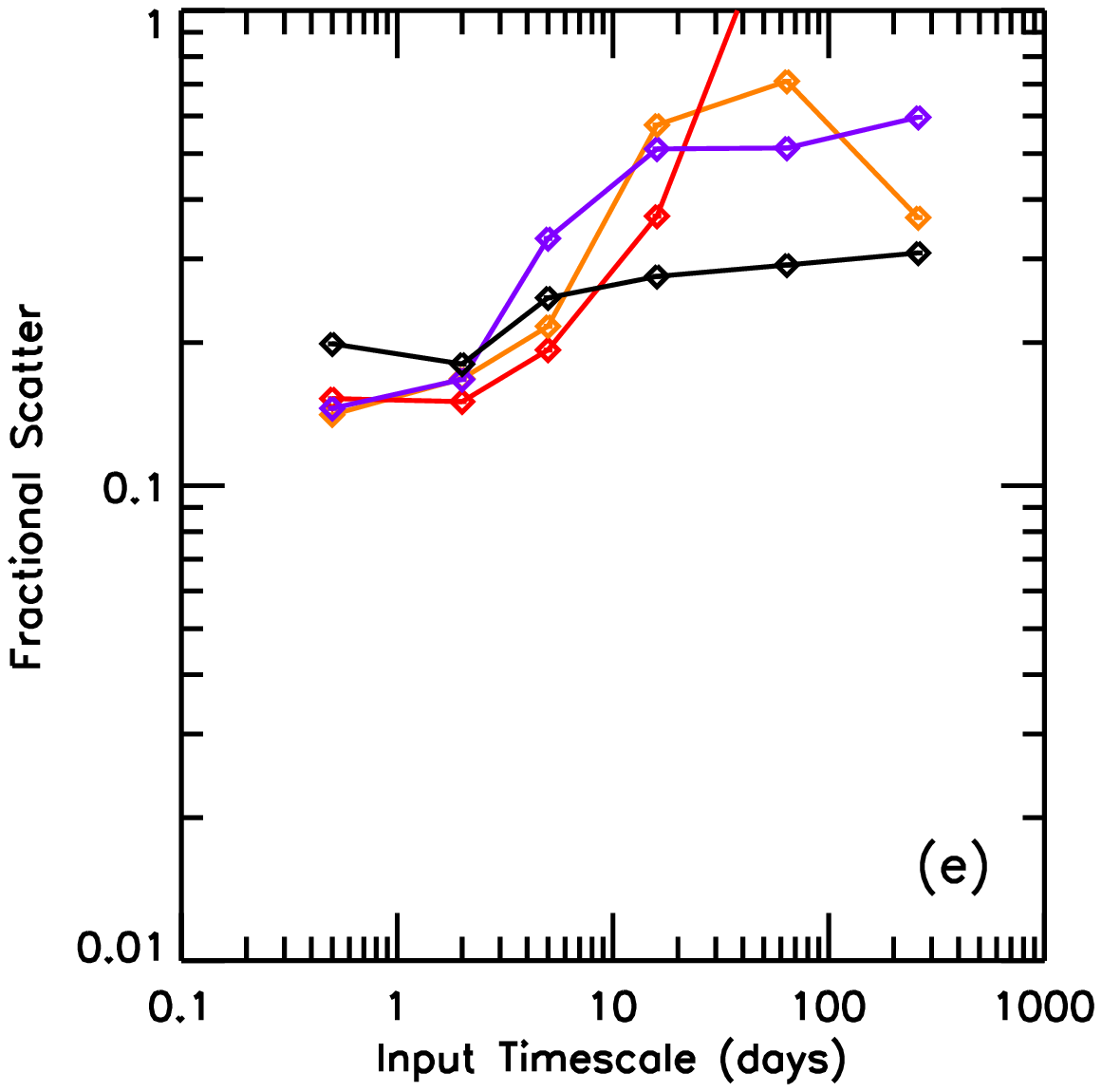}
} {
\includegraphics[width=0.32\textwidth]{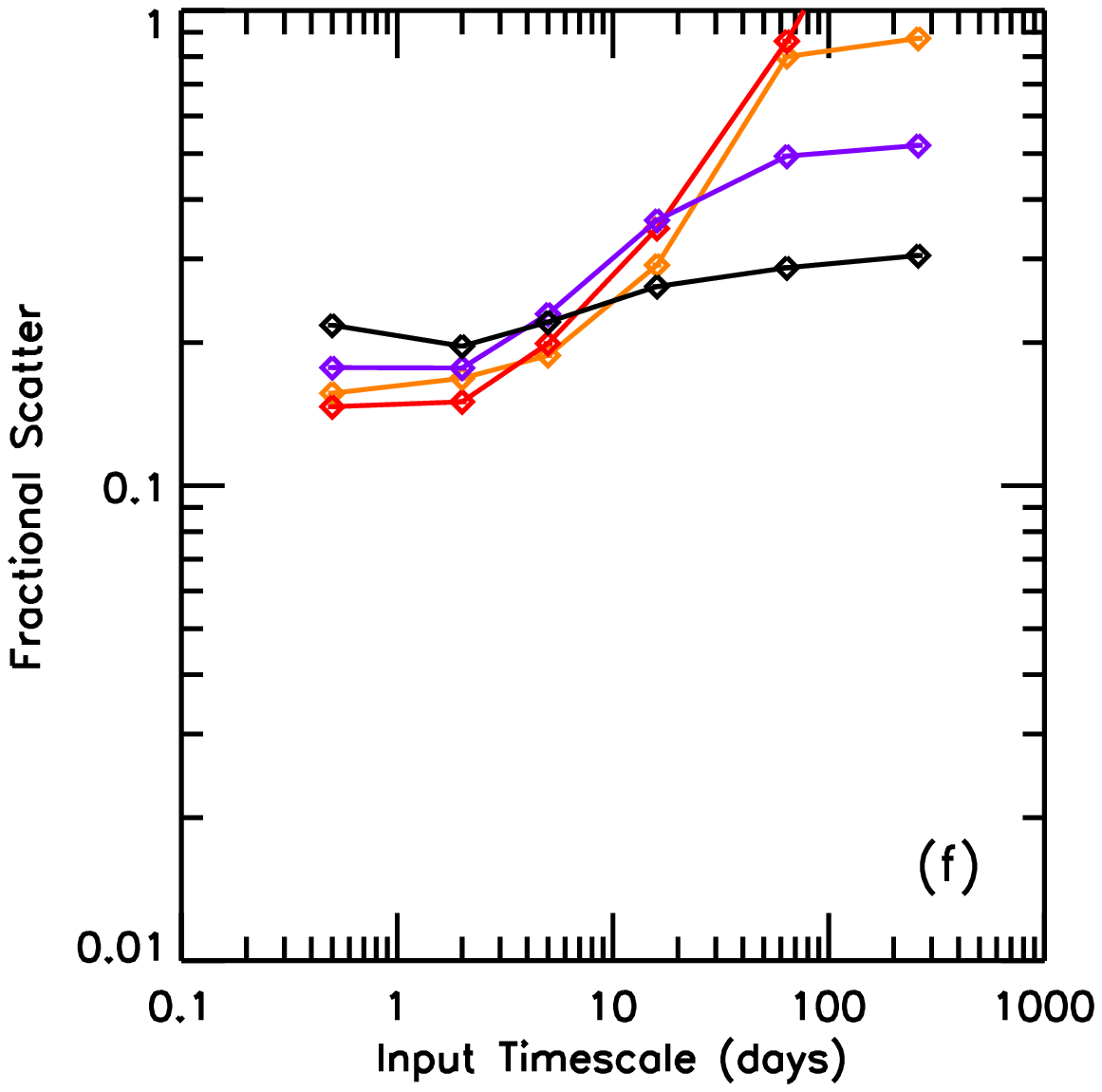}
} {
\includegraphics[width=0.32\textwidth]{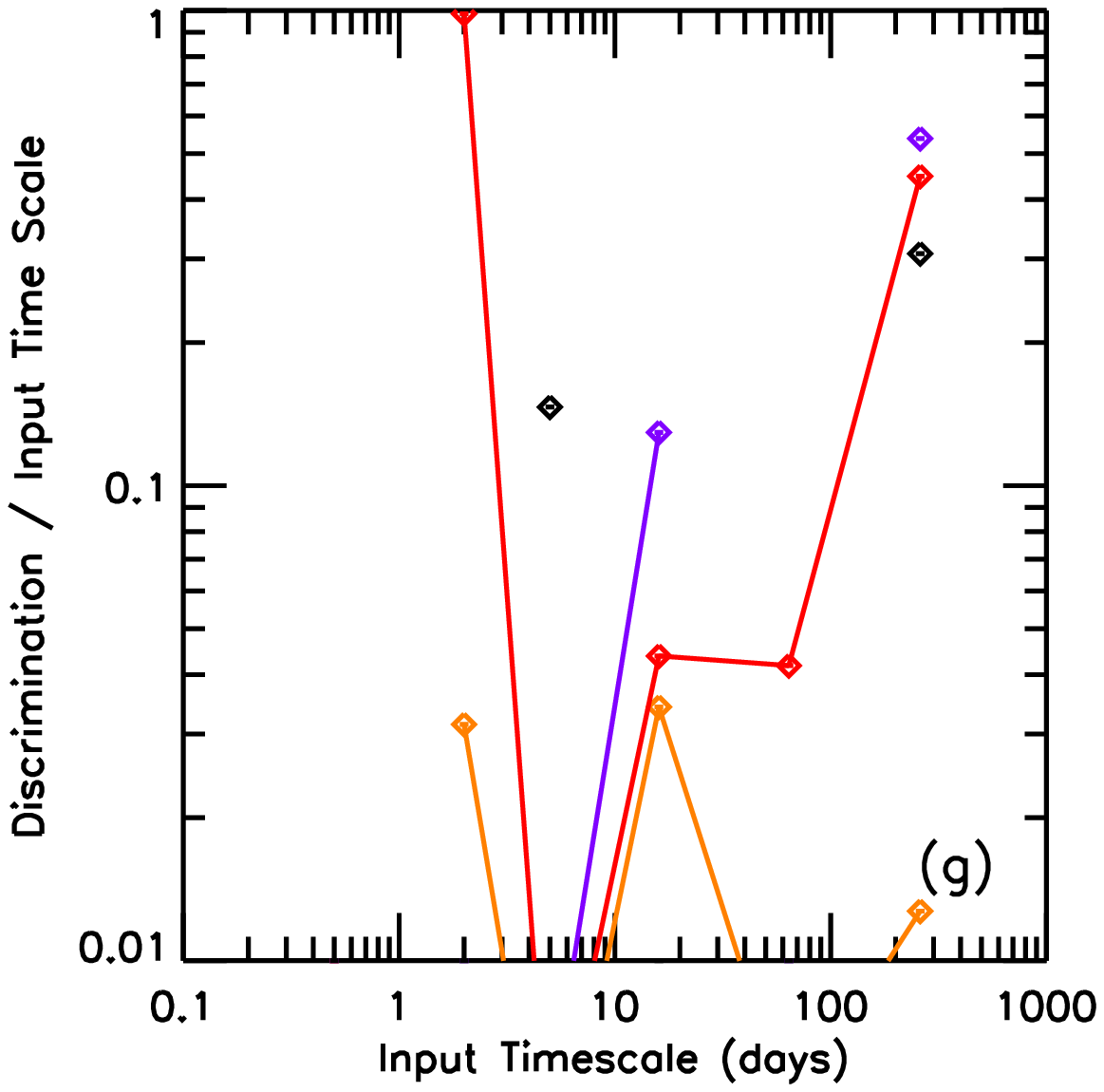}
} {
\includegraphics[width=0.32\textwidth]{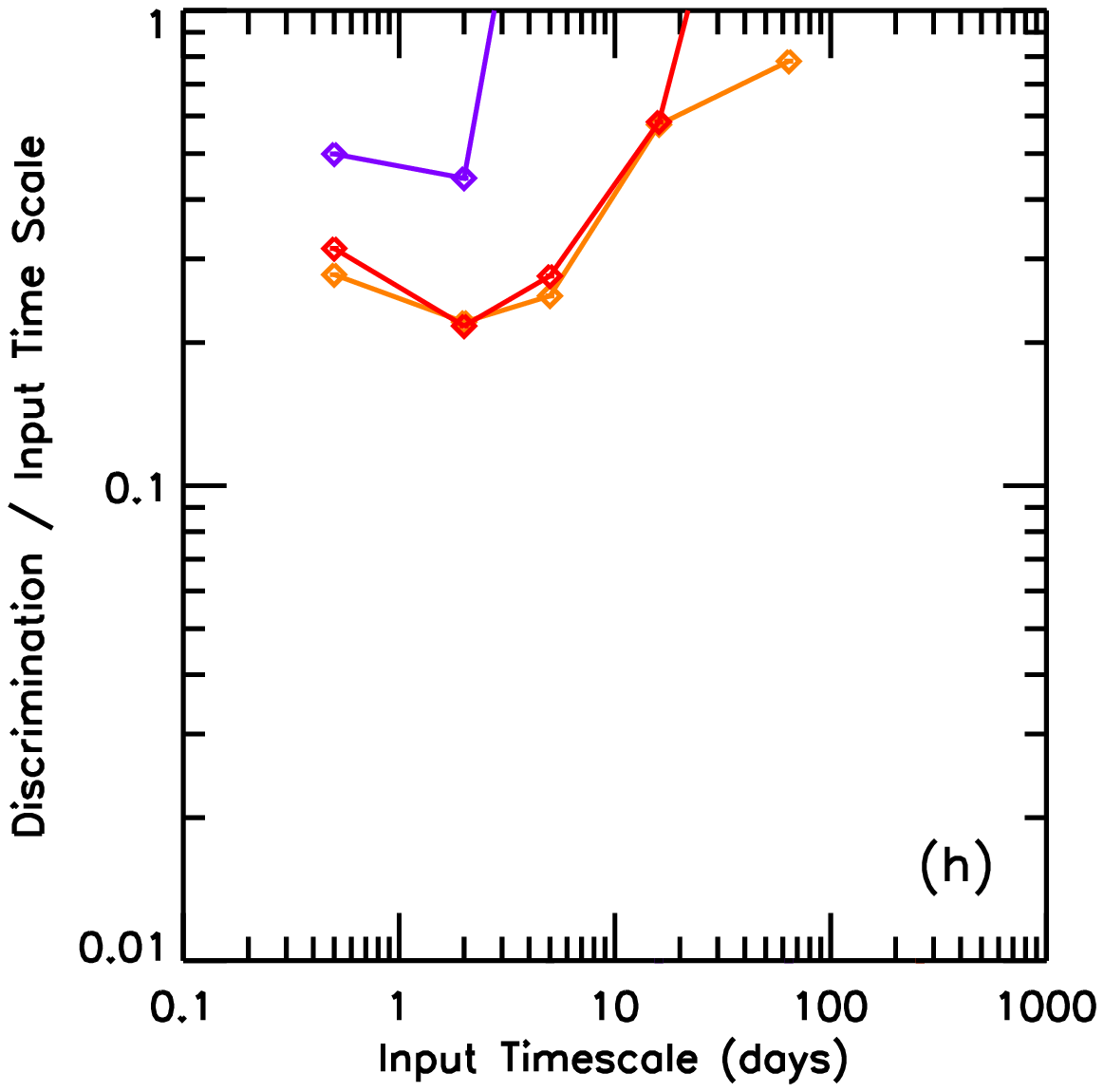}
} {
\includegraphics[width=0.32\textwidth]{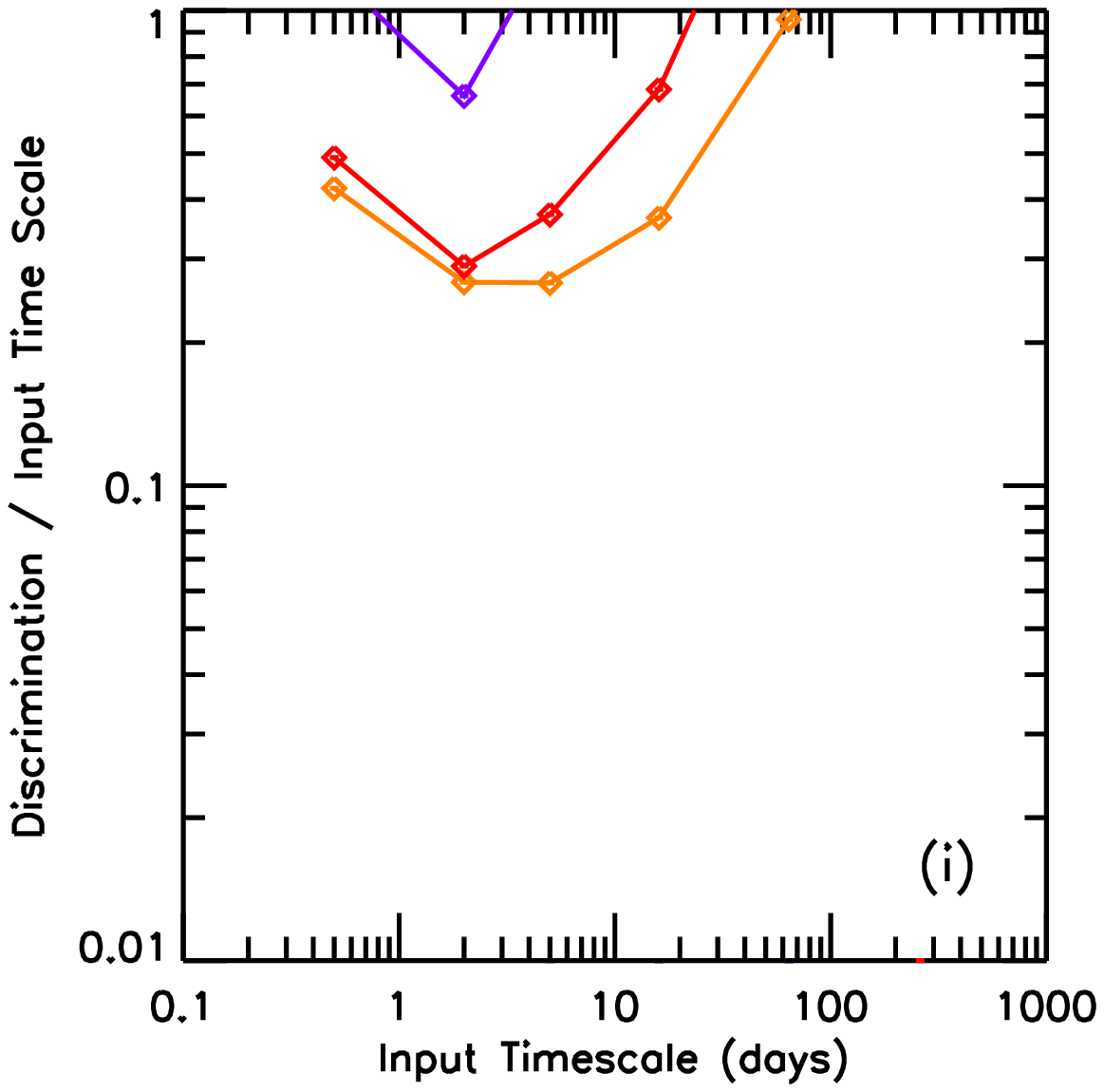}
}
\caption{The timescale calculated from a peak-finding plot, plotted as a function of the true underlying timescale input to the simulation. Columns are, from left to right, for a sinusoidal, squared exponential Gaussian process, and damped random walk model. Black diagonal lines represent the 10:1, 1:1, and 1:10 ratios of output to input. Top panels show the average value of the output timescale. Middle panels show the ratio of the standard deviation to the mean output timescale. Bottom panels show the fractional amount by which the input timescale has to change to significantly affect the output timescale. In all panels, orange represents zero noise, red represents a signal-to-noise ratio of 20, purple a signal-to-noise ratio of 10, and black a signal-to-noise ratio of 4. All light curves have an expected 5-95\% amplitude of 0.5~magnitudes and are sampled using the PTF-NAN Full cadence. Note that the scatter or discrimination statistic exceeds 100\% in some cases, indicating that the metric provides no useful constraint on underlying timescales in those circumstances.} \label{fig_lcmc_peak_snr_ptf}
\end{figure}

Figure~\ref{fig_lcmc_peak_snr_ptf} shows the performance of a peak-finding timescale as a function of the signal-to-noise ratio of the light curve, for light curves sampled at the PTF-NAN Full cadence. The average value of the timescale changes very little between an effectively infinite signal-to-noise and a signal-to-noise ratio of 20 in the case of a sinusoidal signal (panel a), but decreases systematically with signal-to-noise in the case of a damped random walk (panel c). In both cases, at signal-to-noise of 10 or less the peak-finding timescale is barely correlated with the true timescale.

The precision of timescale measurements of a sinusoidal signal (panel d) generally increases with signal-to-noise, as one might expect. The precision of timescale measurements of a damped random walk (panel f), on the other hand, decreases with signal-to-noise. This may indicate that at low signal-to-noise the timescale is strongly affected by a fixed systematic term, which suppresses scatter in the computed timescale.

The discriminating power of the peak-finding timescale roughly follows the precision for a sine wave, but is substantially poorer --- no better than 30\% --- than the precision for a damped random walk (compare panels g and i). As noted above, for signal-to-noise of 10 or less the peak-finding timescale cannot discriminate between short- and long-timescale signals.

\subsubsection{Sensitivity to Cadence}

\begin{figure}[ptb]
\plotnine{
\includegraphics[width=0.32\textwidth]{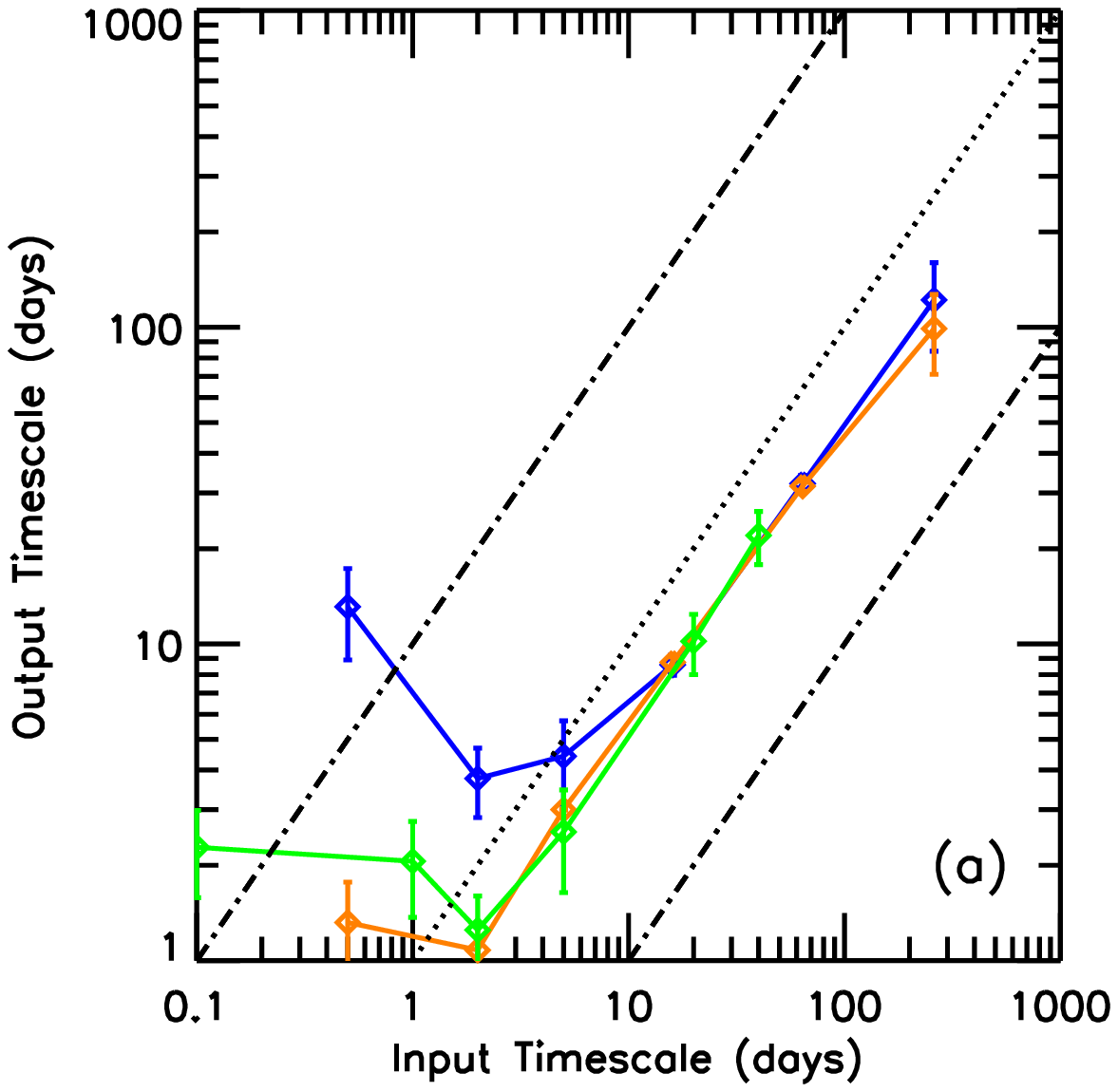}
} {
\includegraphics[width=0.32\textwidth]{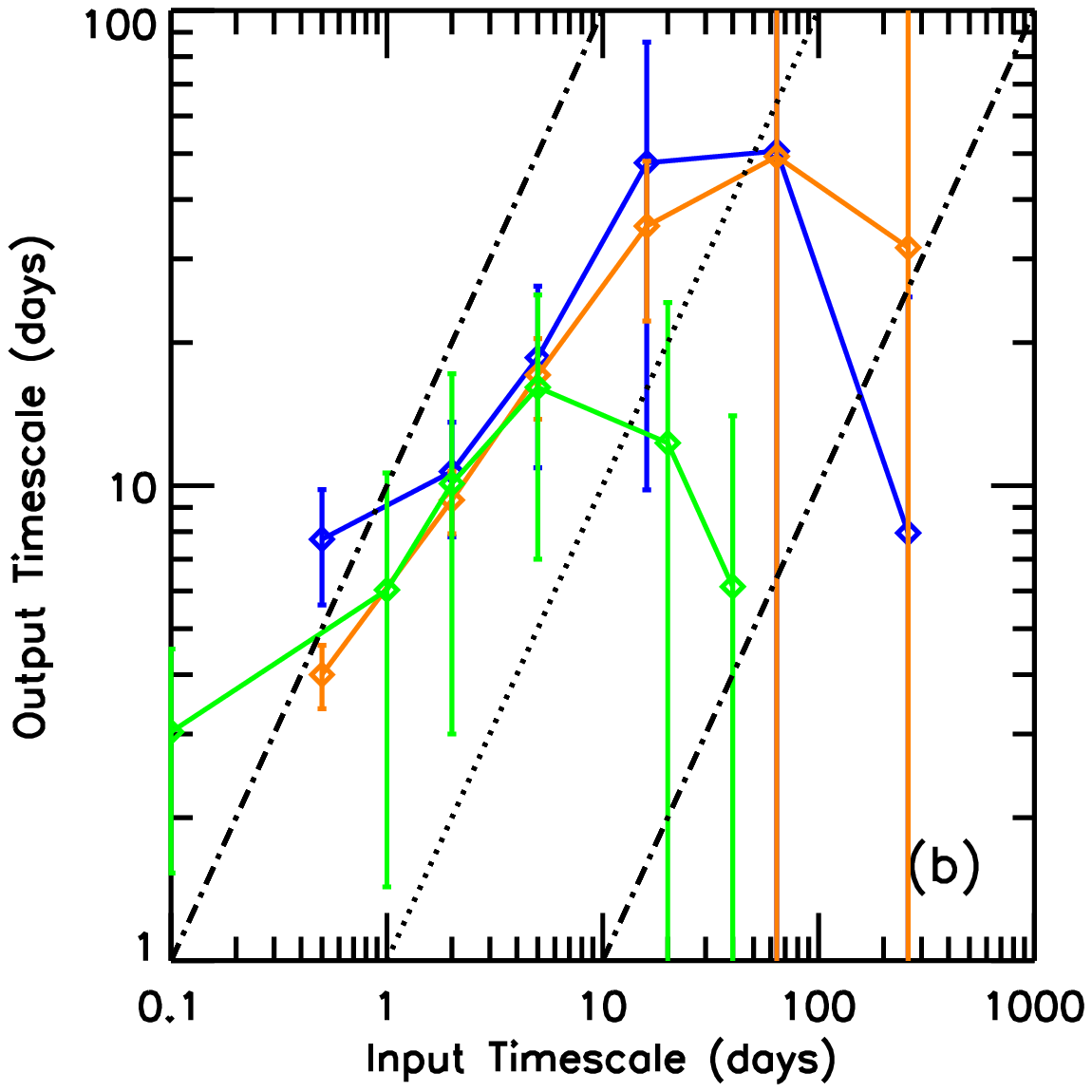}
} {
\includegraphics[width=0.32\textwidth]{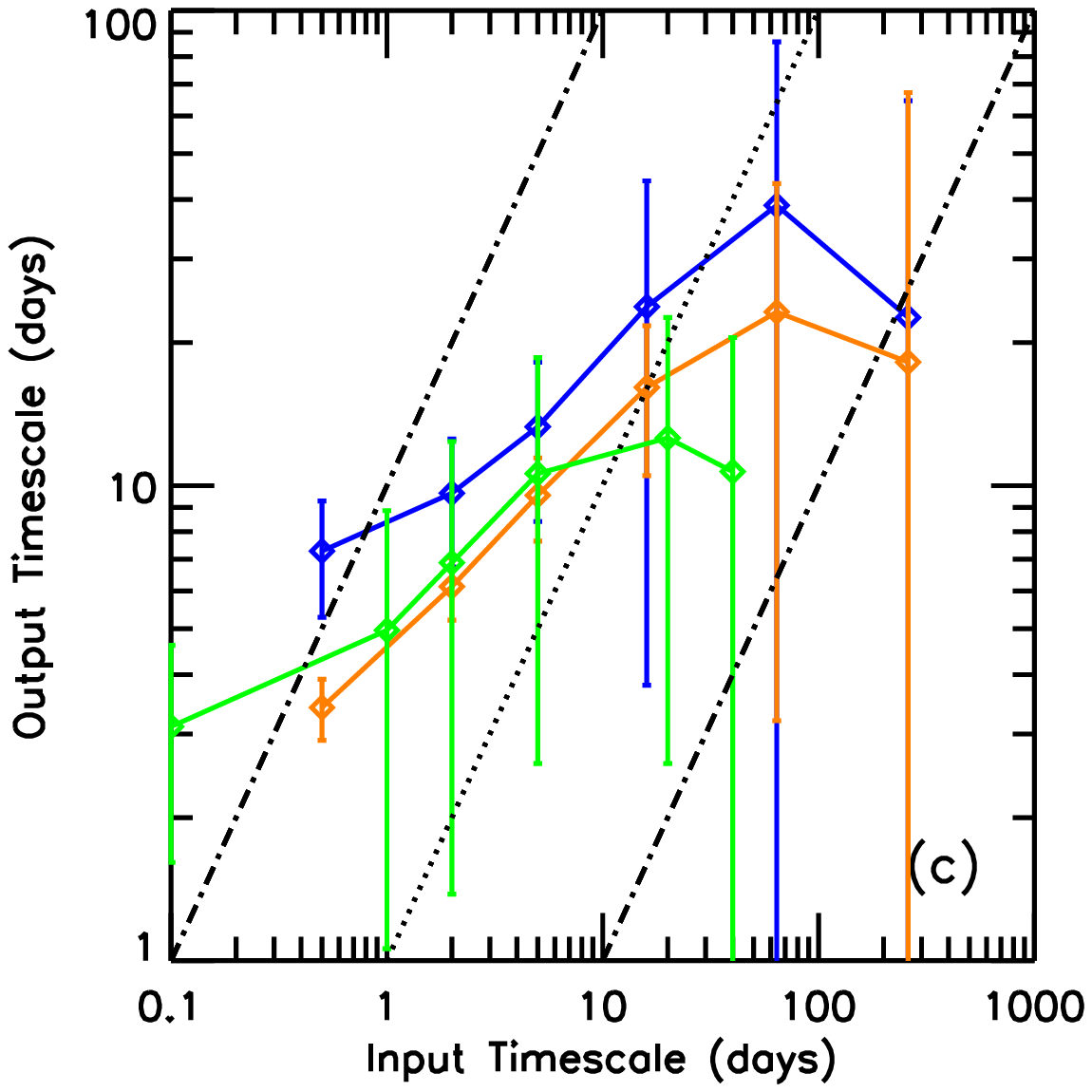}
} {
\includegraphics[width=0.32\textwidth]{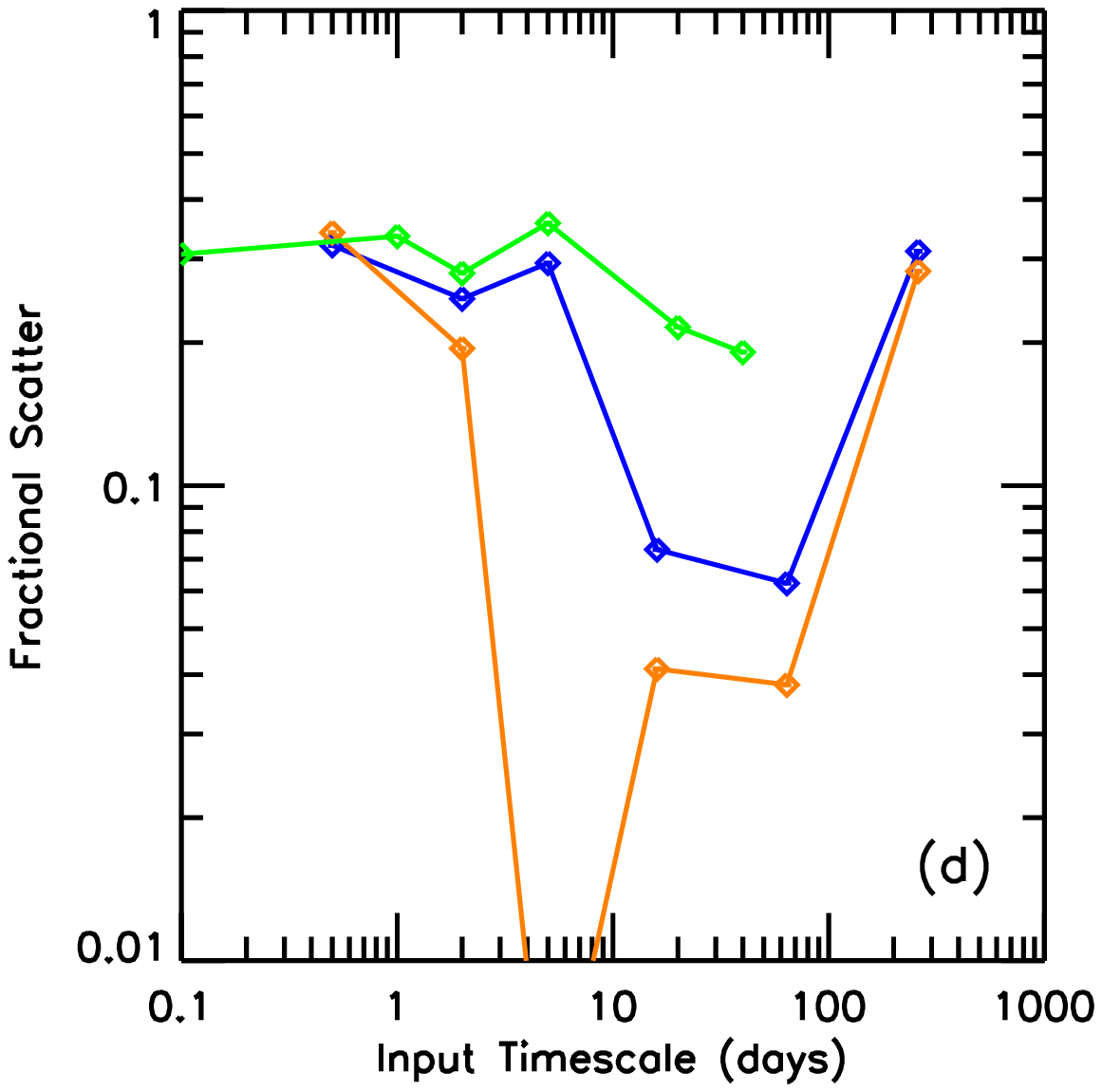}
} {
\includegraphics[width=0.32\textwidth]{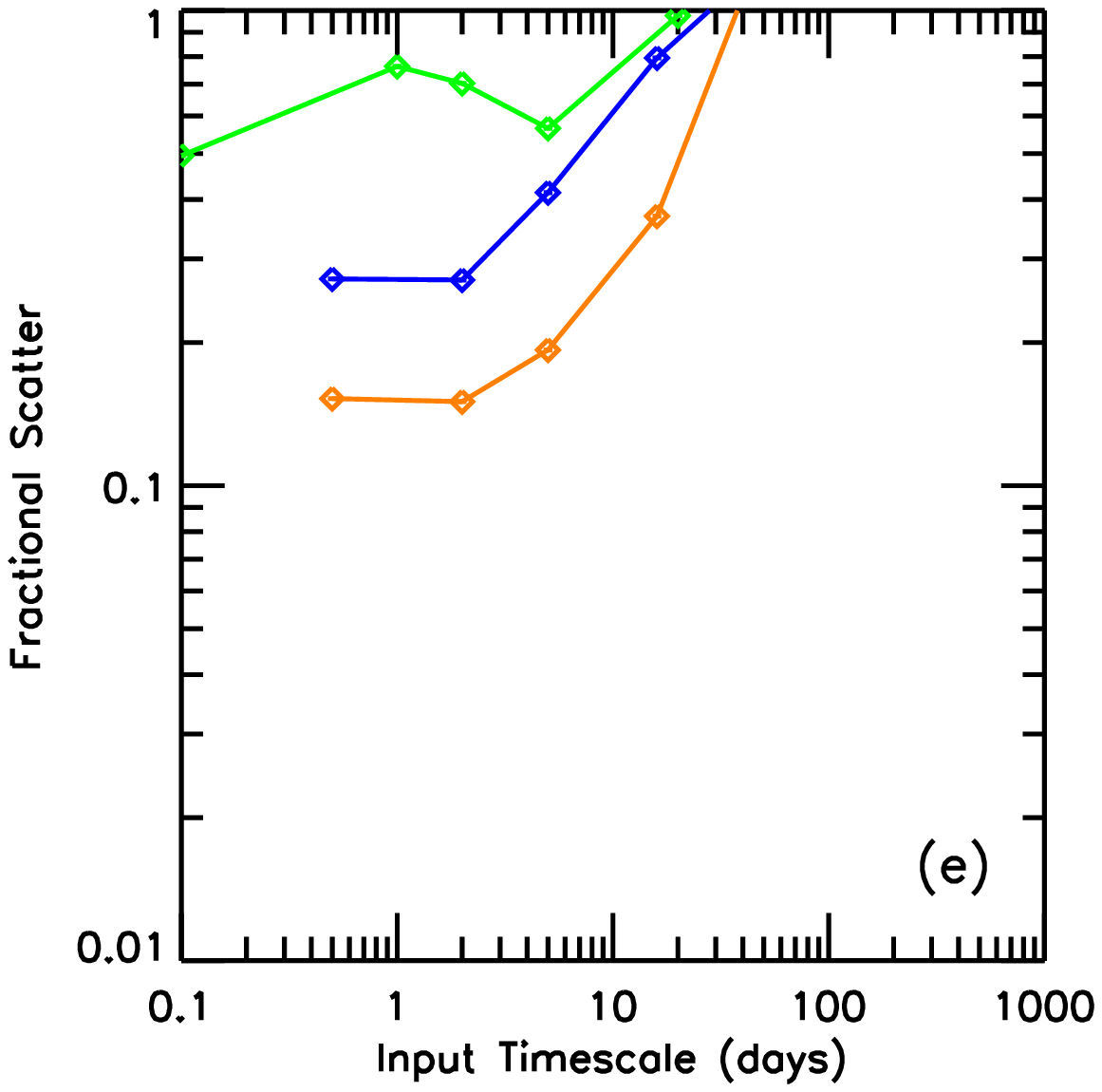}
} {
\includegraphics[width=0.32\textwidth]{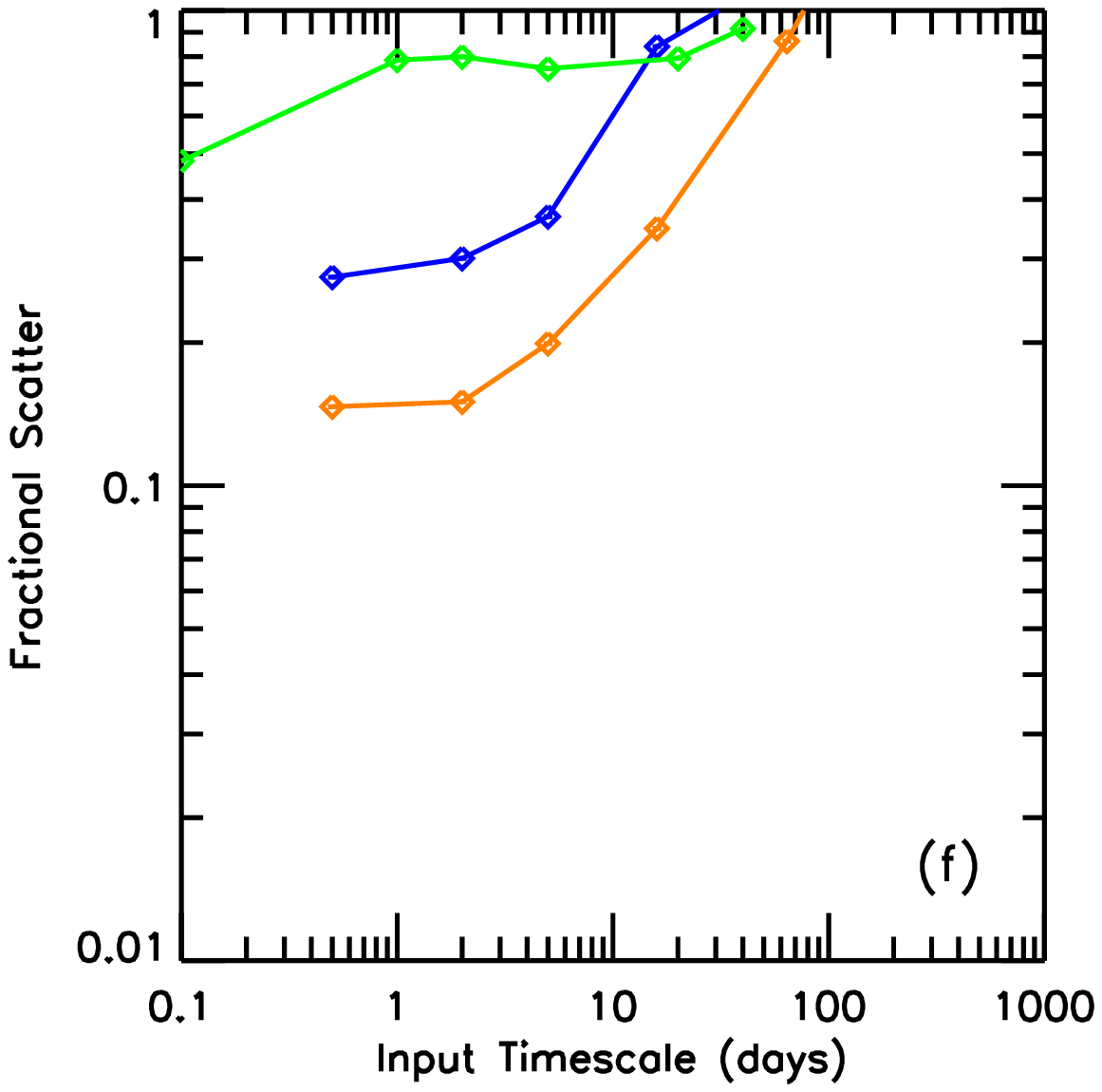}
} {
\includegraphics[width=0.32\textwidth]{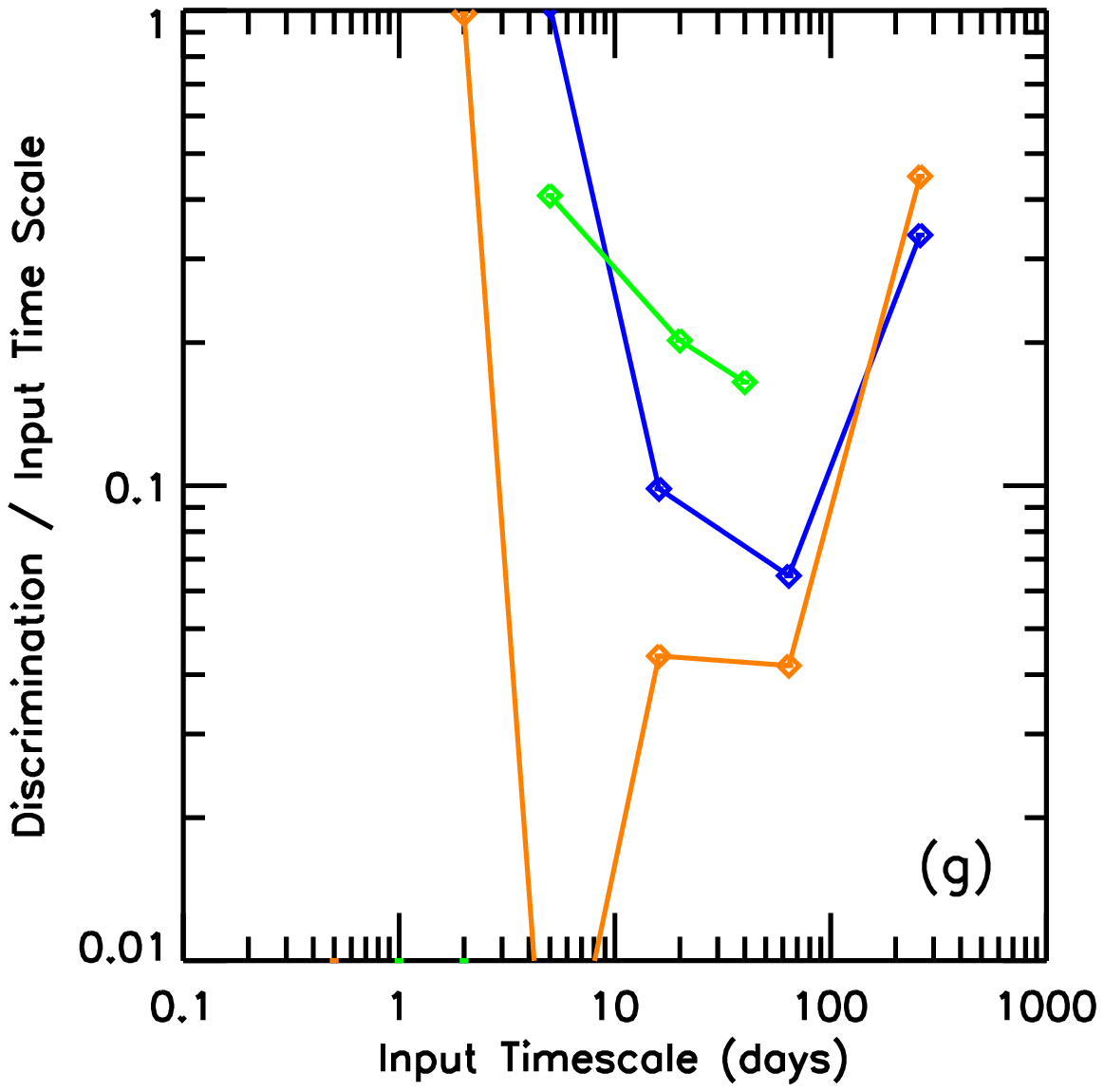}
} {
\includegraphics[width=0.32\textwidth]{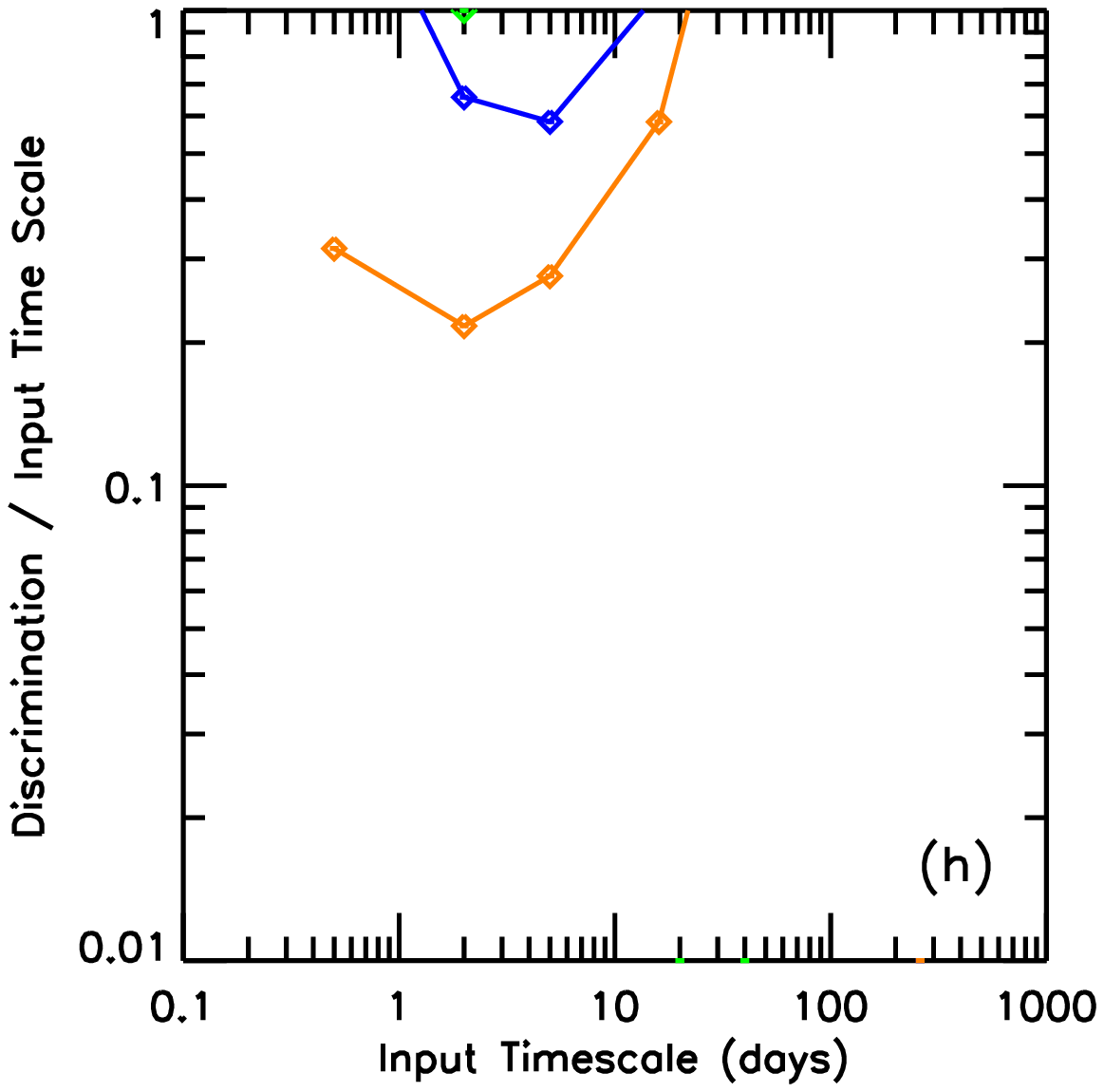}
} {
\includegraphics[width=0.32\textwidth]{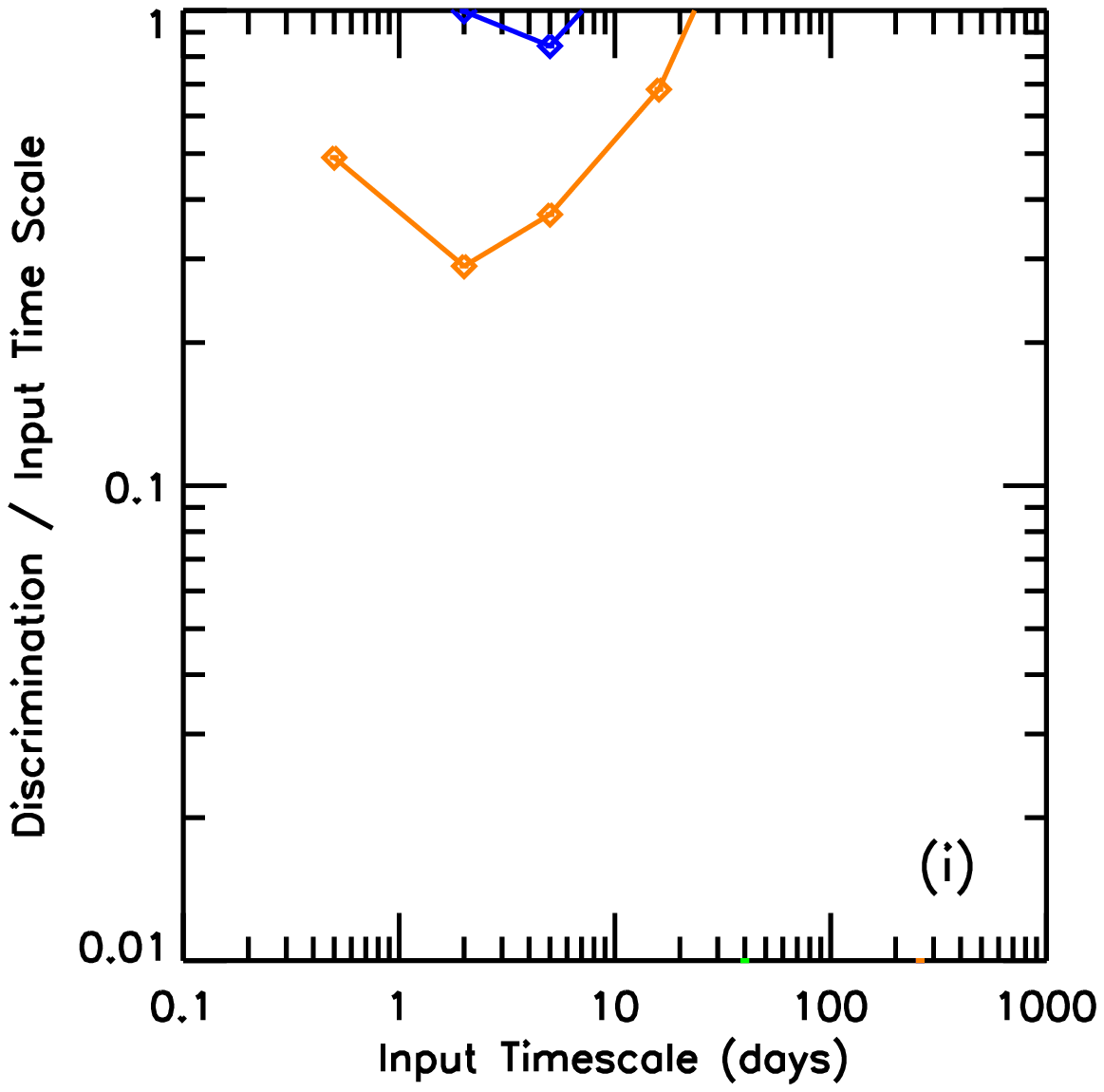}
}
\caption{As Figure~\ref{fig_lcmc_peak_snr_ptf}, but plotting only simulation runs with a signal-to-noise ratio of 20. Blue represents the PTF-NAN Full cadence, green the PTF-NAN 2010 cadence, and orange the YSOVAR 2010 cadence. Note that the scatter or discrimination statistic exceeds 100\% in some cases, indicating that the metric provides no useful constraint on underlying timescales in those circumstances. 
The point at which the inferred timescale first becomes systematically underestimated for long-timescale aperiodic sources, visible as a drop in the upper row of panels, is roughly proportional to the time base line covered by the cadence. Unlike with the \dmdt\ plots shown in Figure~\ref{fig_lcmc_dmdt_bordertimescale}, there is no systematic trend with the frequency of observations.} \label{fig_lcmc_peak_cad_ptf} \label{fig_lcmc_peakslopevscad}
\end{figure}

In Figure~\ref{fig_lcmc_peak_cad_ptf}, we compare the behavior of a peak-finding timescale for both PTF-NAN cadences and the YSOVAR 2010 cadence. For sinusoidal signals sampled at either the PTF-NAN Full cadence or the YSOVAR 2010 cadence (panel a), the timescale is proportional to the period for periods of two days or more. For the more sparsely sampled PTF-NAN 2010 cadence, peak-finding overestimates the period for a 2-day sine by nearly a factor of two. The average behavior of the peak-finding timescale, when applied to a damped random walk (panel c), is qualitatively similar regardless of the light curve cadence.

The peak-finding timescale shows the least scatter for light curves observed with the PTF-NAN Full cadence, more for light curves observed with the PTF-NAN 2010 cadence, and the highest amount of scatter for the YSOVAR 2010 cadence. This is true regardless of whether the timescale is measured for sinusoidal signals or for damped random walks (panels d-f). The discriminating power shows a similar trend with signal-to-noise ratio (panels g-i).

\subsubsection{The CSI~2264 CoRoT Cadence}\label{csi2264}

The CSI~2264 survey \citep{AmcCorot} of the NGC~2264 region combined infrared time series photometry from Spitzer with optical time series photometry from CoRoT to investigate the variability of young stars in both wavelength regimes. \citet{AmcCorot} calculated timescales for CoRoT light curves from the survey using the peak-finding method, which we described in section~\ref{lcmc_pf} and illustrated in Figure~\ref{fig_lcmc_demo_peak}. Since their light curves covered an interval of 39~days at a roughly 10~minute cadence, they represent a qualitatively different observing pattern from the other cadences studied in this section. We therefore simulated peak-finding on the CoRoT cadence using a separate parameter grid, as described in section~\ref{grid}.

\begin{figure}
\plotnine{
\includegraphics[width=0.32\textwidth]{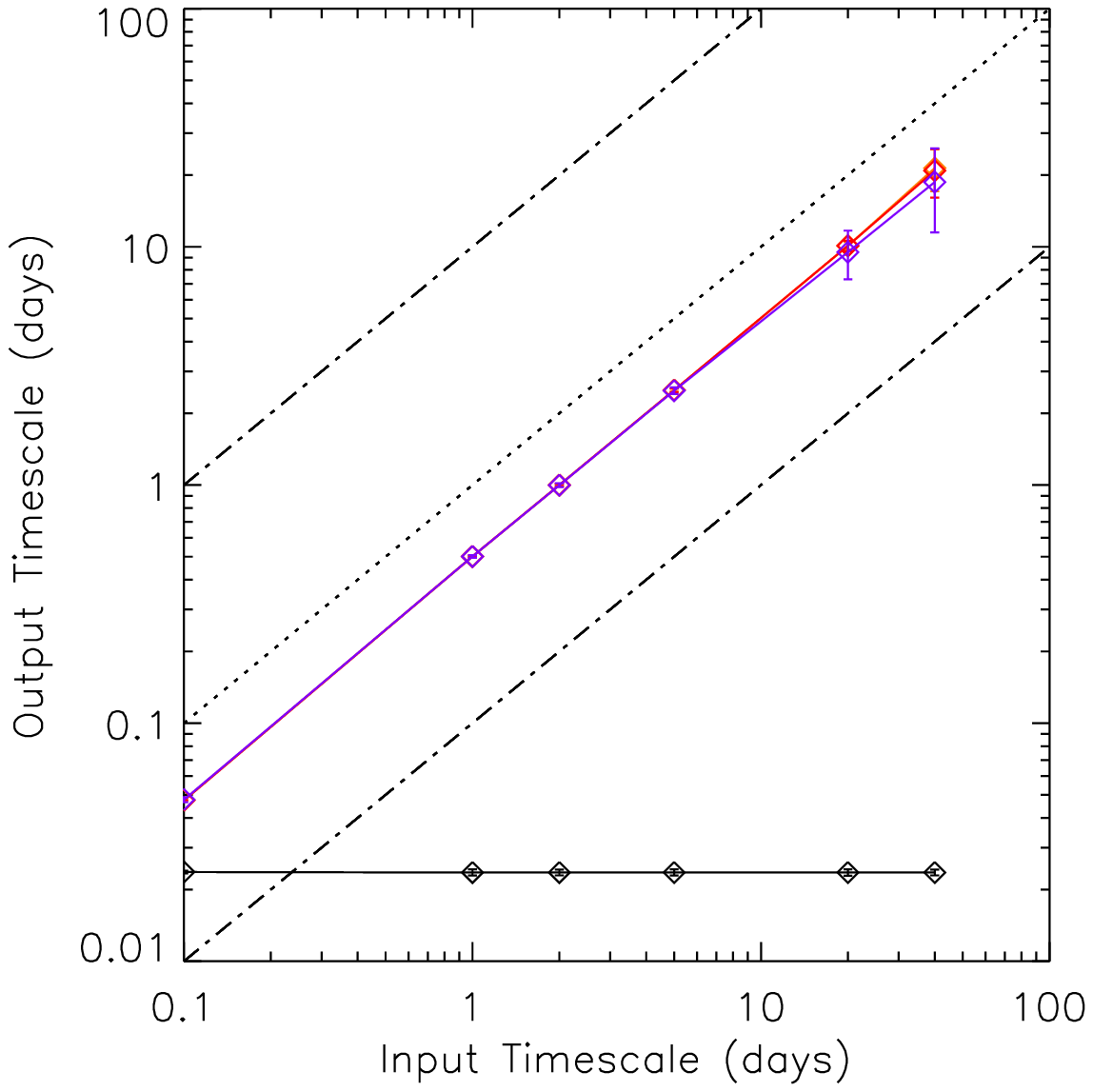}
} {
\includegraphics[width=0.32\textwidth]{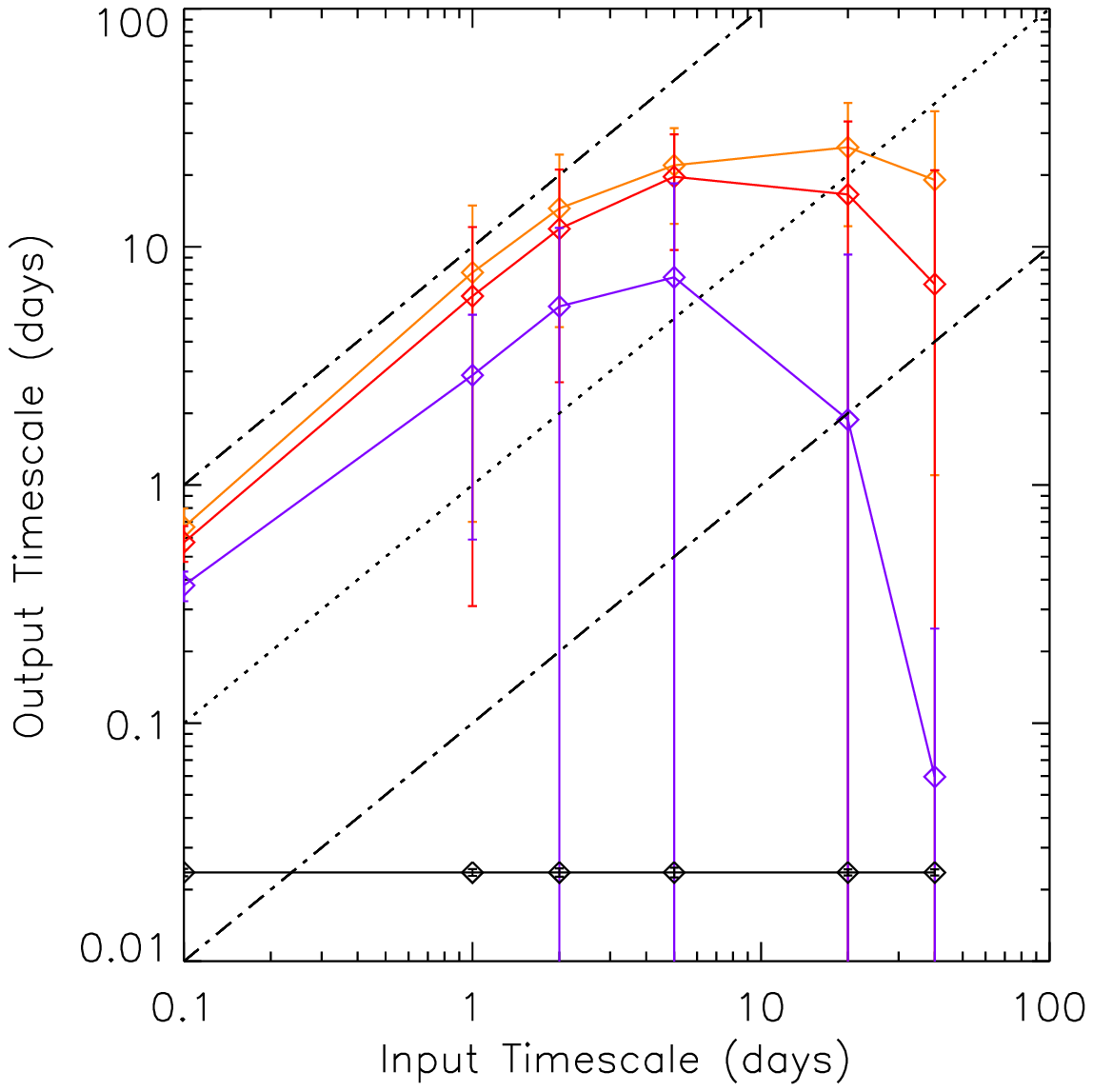}
} {
\includegraphics[width=0.32\textwidth]{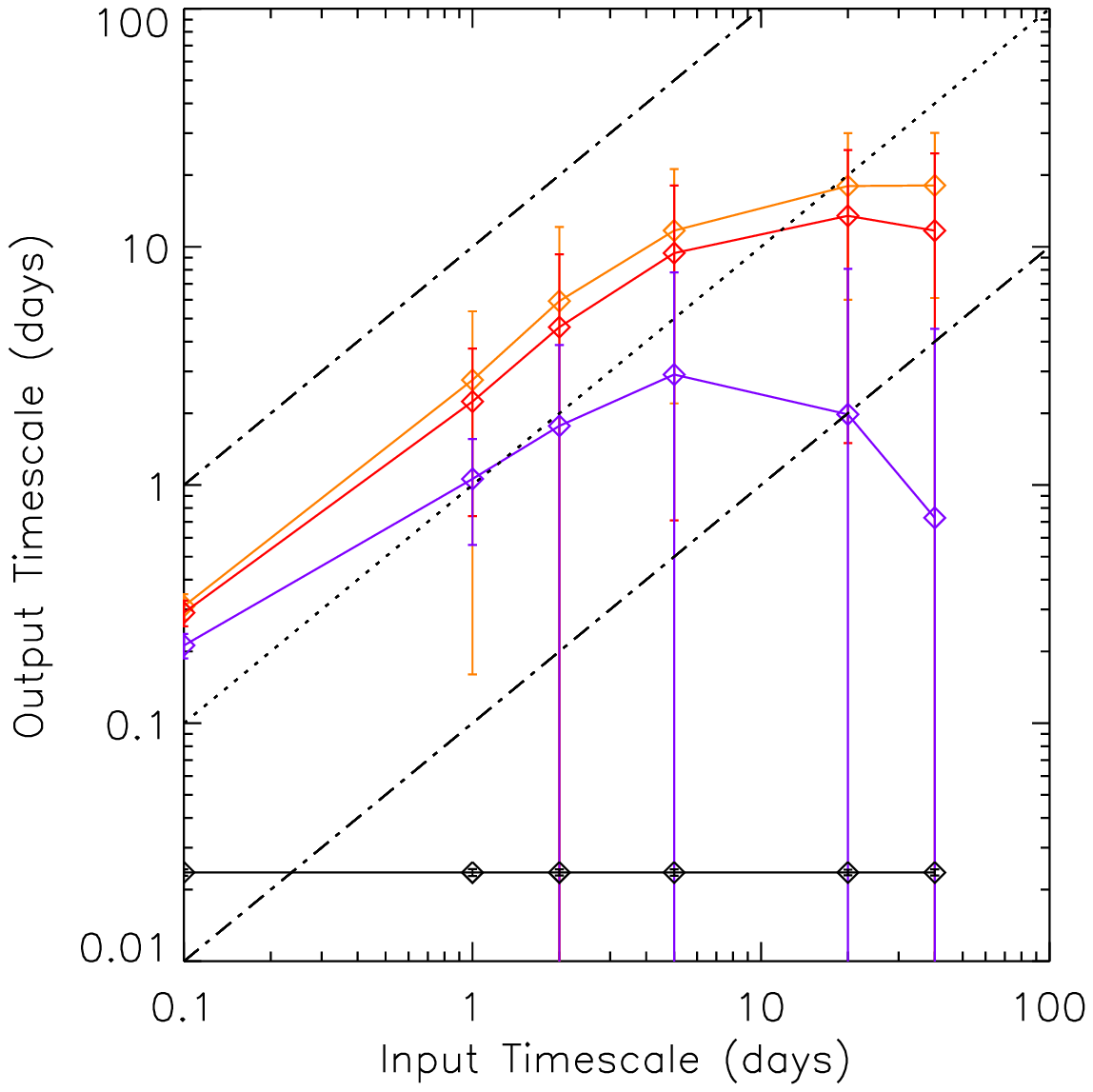}
} {
\includegraphics[width=0.32\textwidth]{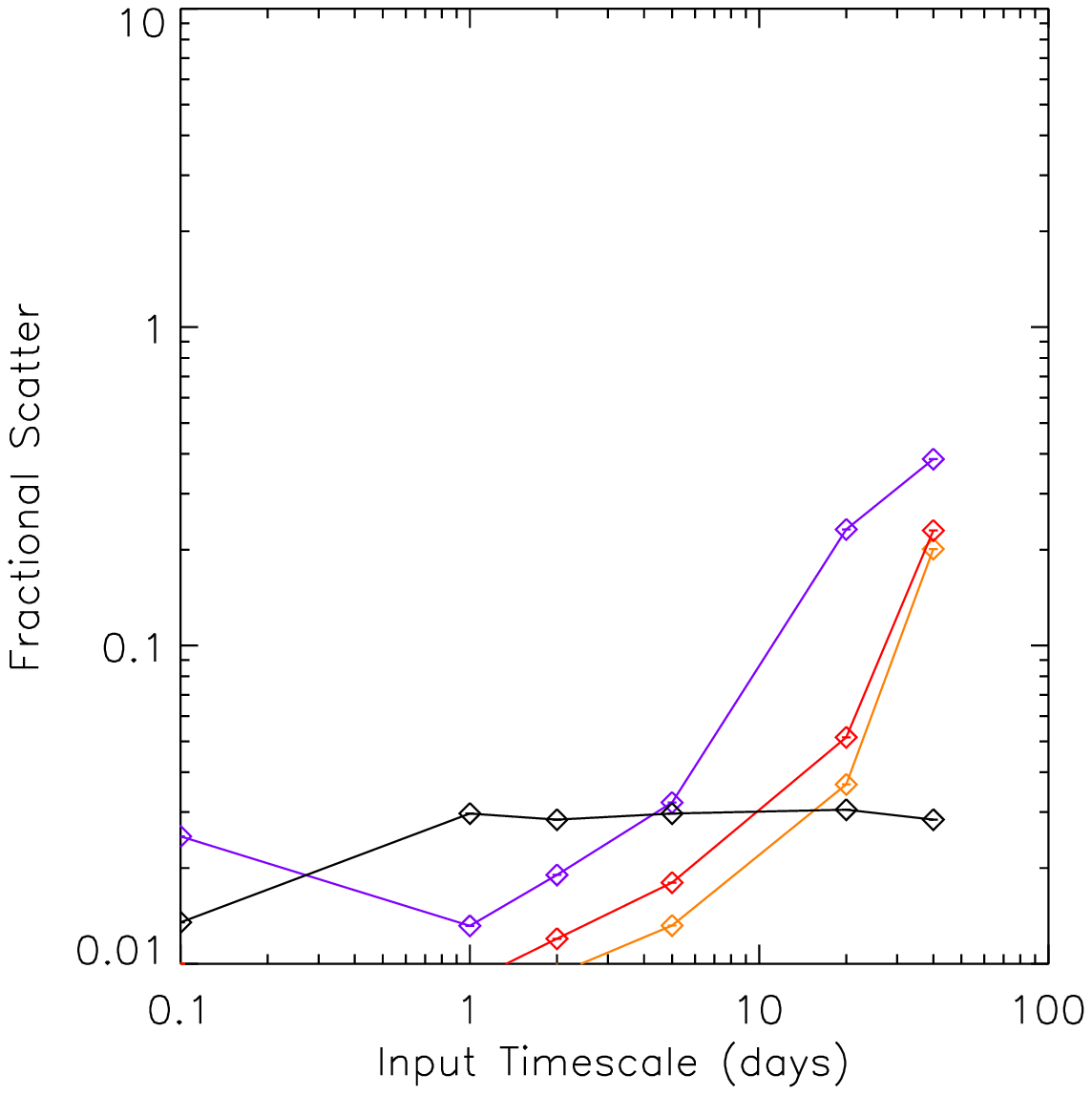}
} {
\includegraphics[width=0.32\textwidth]{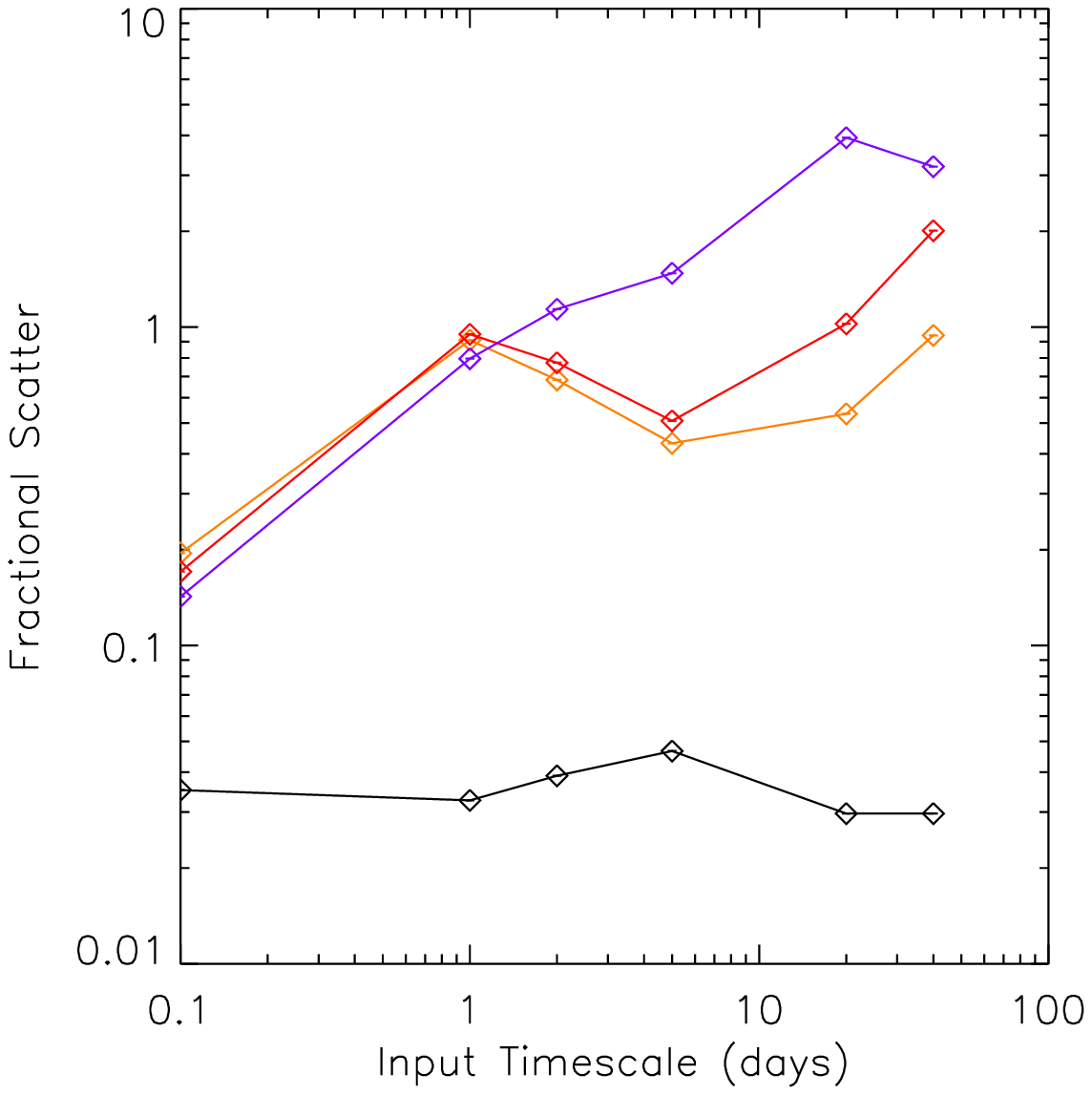}
} {
\includegraphics[width=0.32\textwidth]{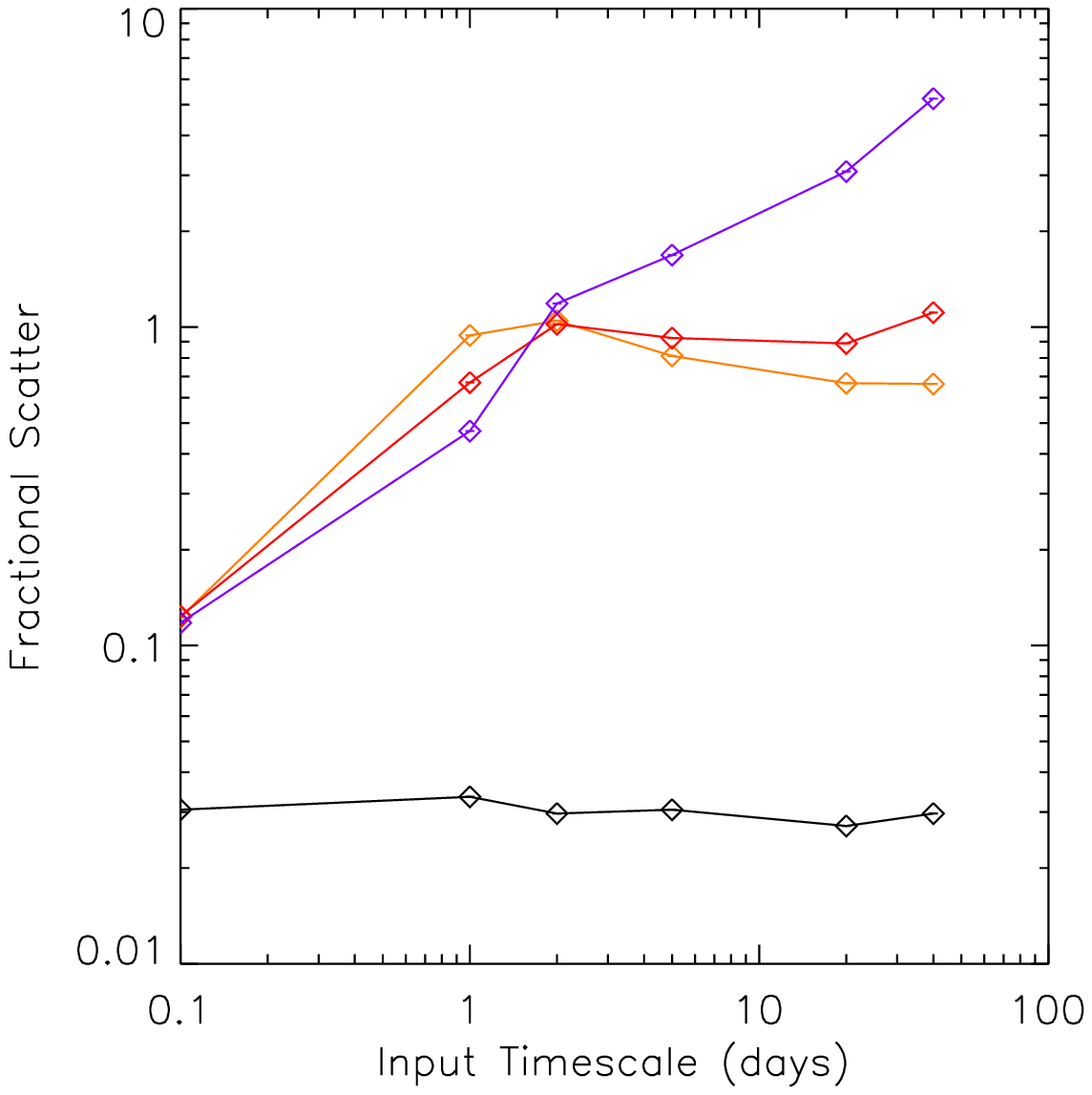}
} {
\includegraphics[width=0.32\textwidth]{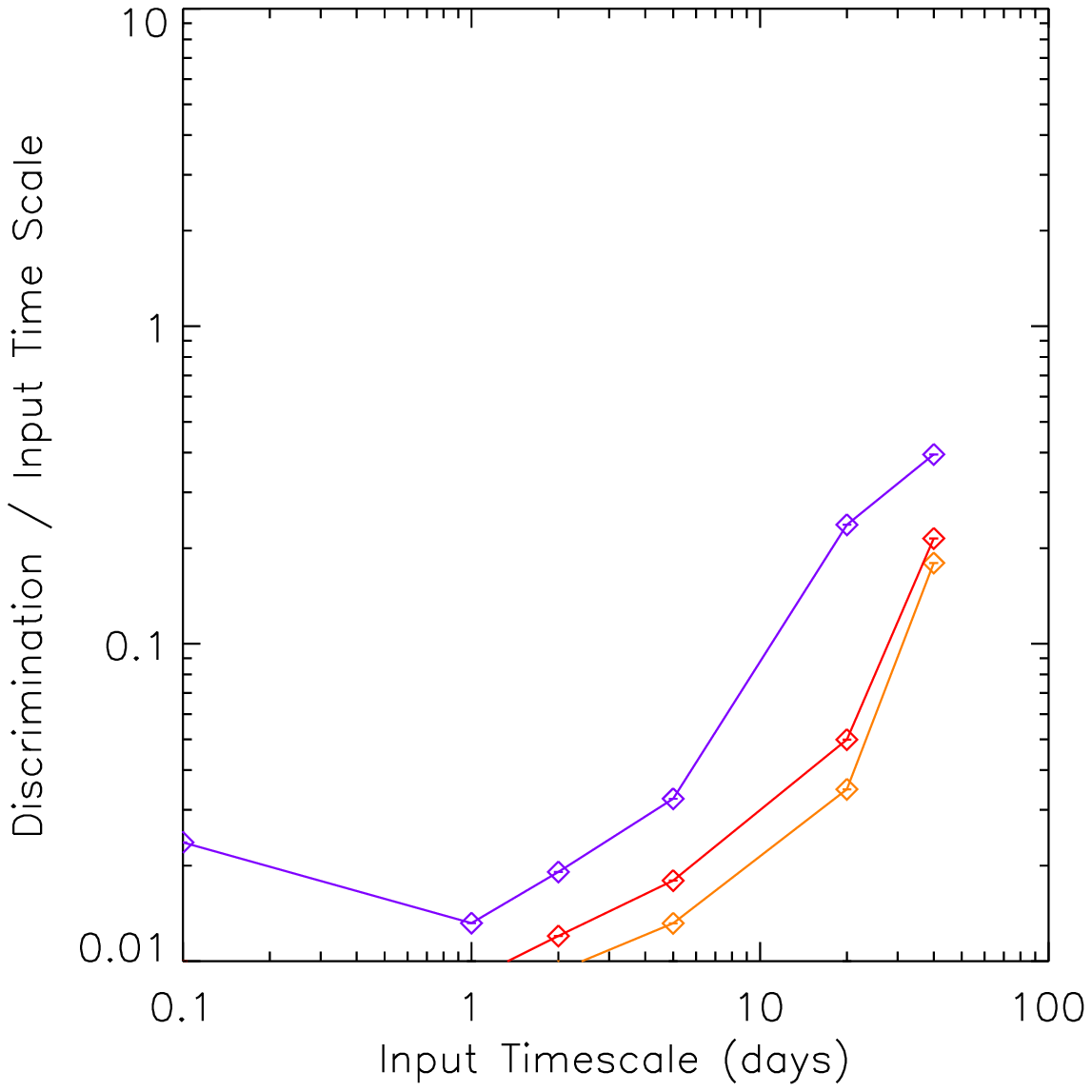}
} {
\includegraphics[width=0.32\textwidth]{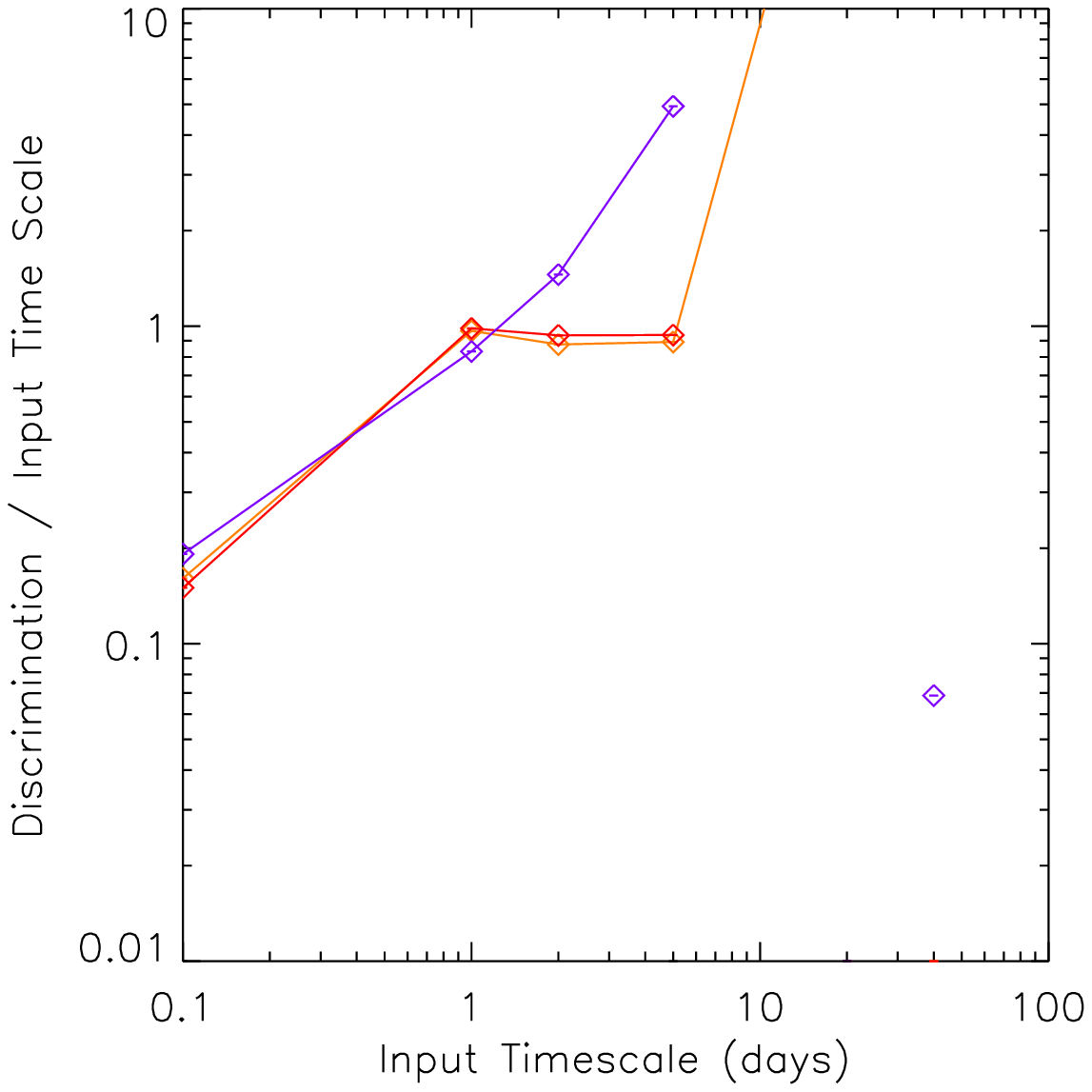}
} {
\includegraphics[width=0.32\textwidth]{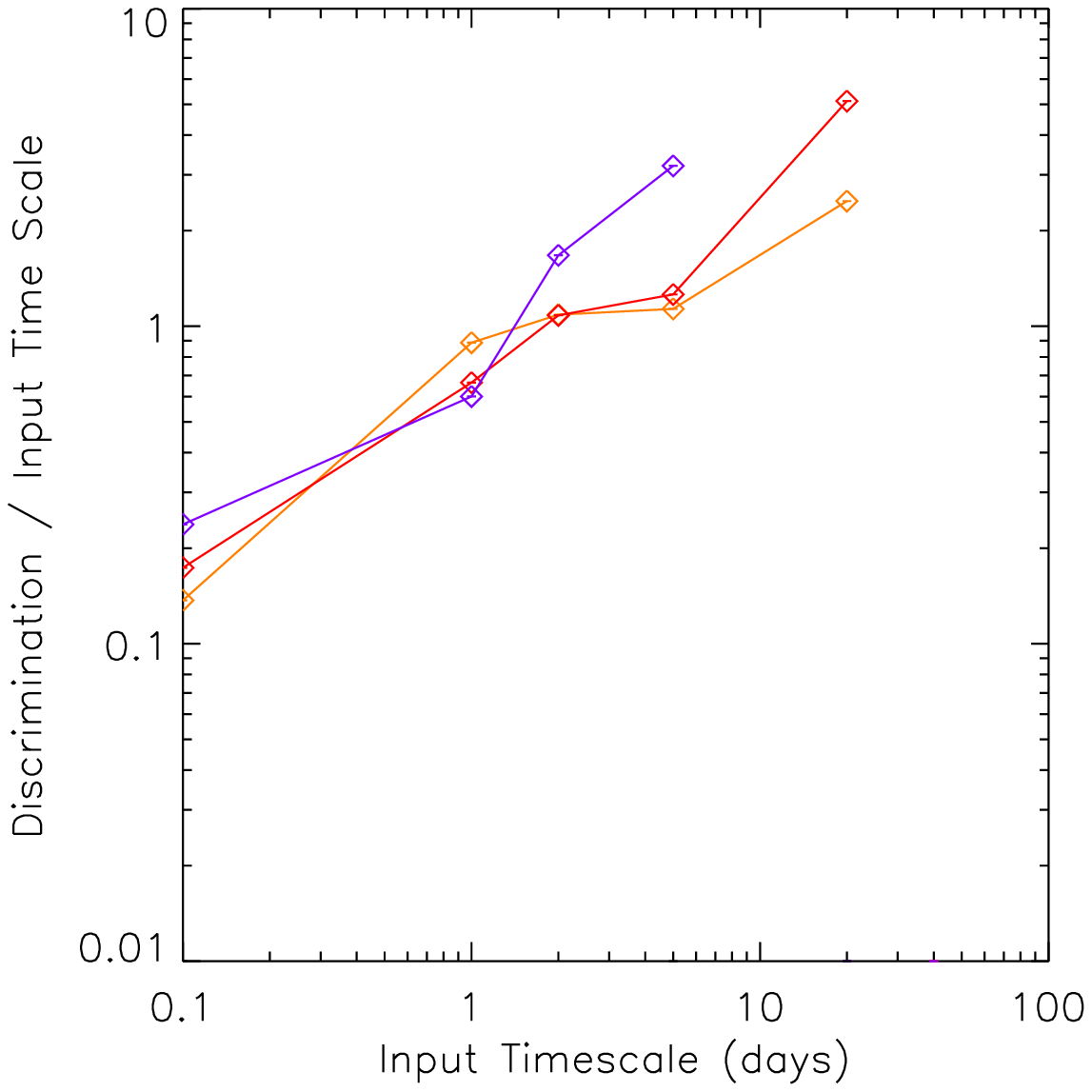}
}
\caption{The timescale calculated from a peak-finding plot in simulated CoRoT observations, plotted as a function of the underlying timescale of the simulated light curve. Columns are, from left to right, for a sinusoidal, squared exponential Gaussian process, and damped random walk model. Black diagonal lines represent the 10:1, 1:1, and 1:10 ratios of output to input. Top panels show the average value of the output timescale, which increases linearly with the true timescale before deviating from linearity at $\sim 10$~days. Middle panels show the ratio of the standard deviation to the mean output timescale; the uncertainty is typically a factor of two for the two aperiodic models. Bottom panels show the fractional amount by which the input timescale has to change to significantly affect the output timescale.
In all panels, orange represents light curves with an expected 5th to 9th percentile amplitude of 0.5~mag, red an amplitude of 0.25~mag, purple an amplitude of 0.1~mag, and black an amplitude of 0.01~mag. All runs have a signal-to-noise ratio of 100.} \label{fig_census_peak_amp_ptf}
\end{figure}

We present the simulation results in Figure~\ref{fig_census_peak_amp_ptf}. The top row of panels, which compares the timescale found by peak-finding to the timescale parameter used in the simulations, show that the peak-finding timescale is proportional to the timescale input to the simulation (the ``true timescale'') for true timescales between 0.1 and 10~days. For longer-term variables, the timescale from the peak-finding analysis appears to level out at roughly 20~days. The bias is much smaller than that found in section~\ref{pf_bias} for peak-finding plots run on the PTF-NAN Full cadence. The middle row of panels shows that, for aperiodic signals, the scatter in timescales is of the same order as the timescale itself, or a factor of two error. This is comparable scatter to that found in section~\ref{pf_scatter}.

\citet{AmcCorot} normalized their peak-finding timescale to twice the value plotted in Figure~\ref{fig_census_peak_amp_ptf} so that it would agree with the period for periodic sources. After we correct for this convention, introduced after the simulations were carried out, our simulations indeed show that the peak-finding timescale found for simulated sinusoidal signals equals the period on average. 


\subsubsection{Summary}

The numerical simulations presented in this section suggest that, while the peak-finding timescale correlates with the input timescale for a variety of light curves, the shape of the scaling is sensitive to the type of light curve. 

For rapidly varying signals the peak-finding timescale shows a scatter of $\sim $10-20\% for long time series such as the PTF-NAN cadences, but somewhat more scatter ($\sim 40\%$) for light curves observed with the YSOVAR 2010 cadence. Regardless of cadence, the scatter grows to a factor of two for aperiodic light curves with input timescales longer than 1/15 to 1/20 the observing base line. Since the peak-finding timescale is $\sim 5$ times larger than the coherence time for a squared exponential Gaussian process or $\sim 3$ times larger than the damping time for a damped random walk, peak-finding timescales larger than 1/5 to 1/3 the observing base line should be treated with caution.

The timescale metric's performance depends on both signal-to-noise and the cadence.
The peak-finding timescales are moderately sensitive to the presence of noise in the light curve. Noise biases the inferred timescale downward; the effect is moderate if the noise RMS is at most one tenth the light curve amplitude, but quickly grows more severe at higher noise levels. Among the PTF-NAN and YSOVAR 2010 cadences, the average behavior of the light curve seems to depend little on the cadence adopted, showing a similar slope (Figure~\ref{fig_lcmc_peakslopevscad}a-c) up to timescales of order the time series base line. While the results for the CoRoT cadence (Figure~\ref{fig_census_peak_amp_ptf}) \emph{do} show a steeper slope than in the other simulation runs, this may be due to the high signal-to-noise ratio of that simulation run. 
As noted above, the timescale needs a long time series to provide precise measurements. 

The peak-finding timescale is best suited for long-term monitoring of short-timescale variability, particularly in densely sampled light curves. The 5-95\% amplitude of the variability should be at least ten times the RMS of the noise to ensure that noise does not confuse the peak-finding algorithm. 

\subsection{Gaussian Process Regression}

\subsubsection{Qualitative Behavior}

\begin{figure}[ptb]
\plotnine{
\includegraphics[width=0.32\textwidth]{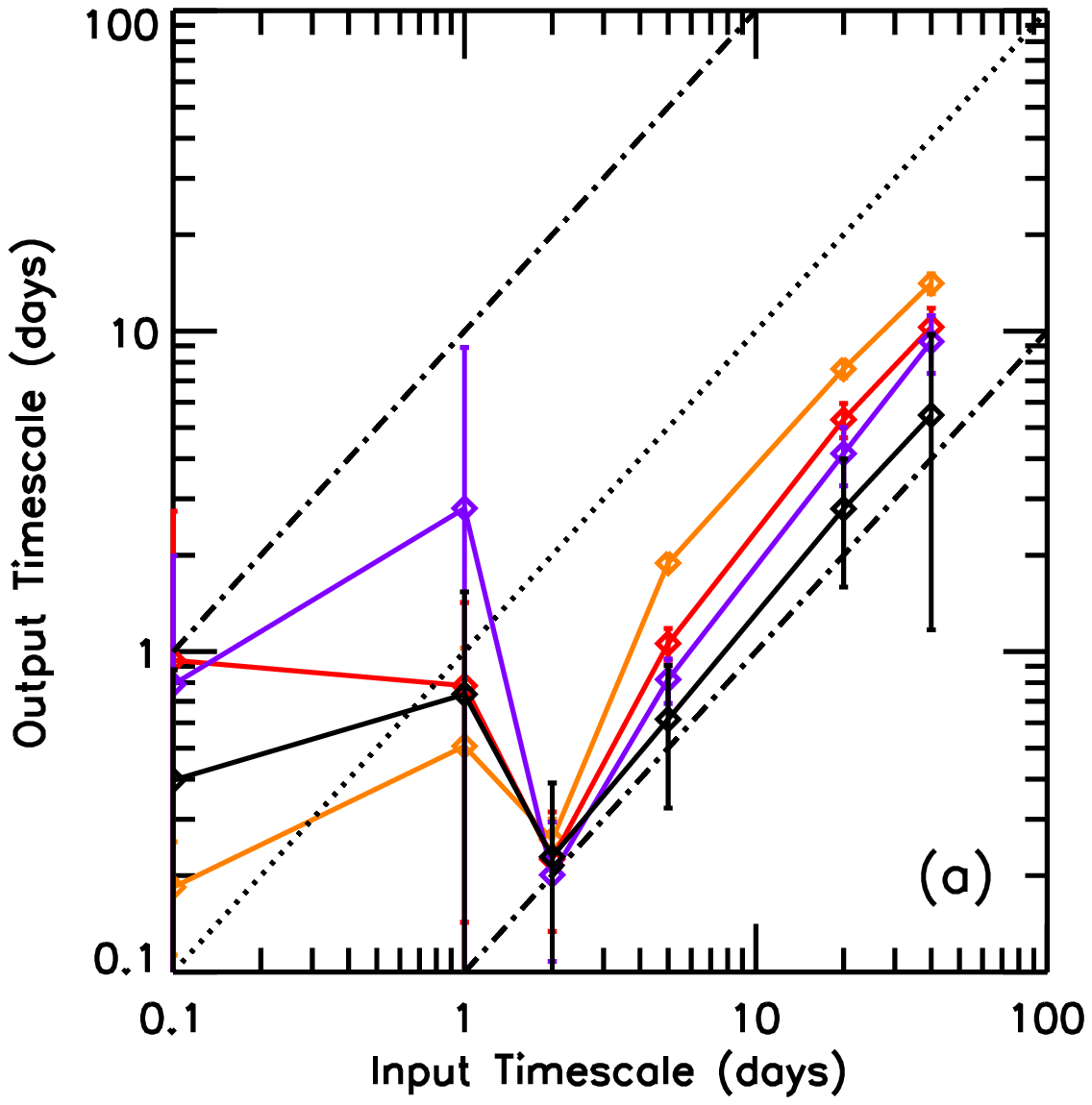}
}{
\includegraphics[width=0.32\textwidth]{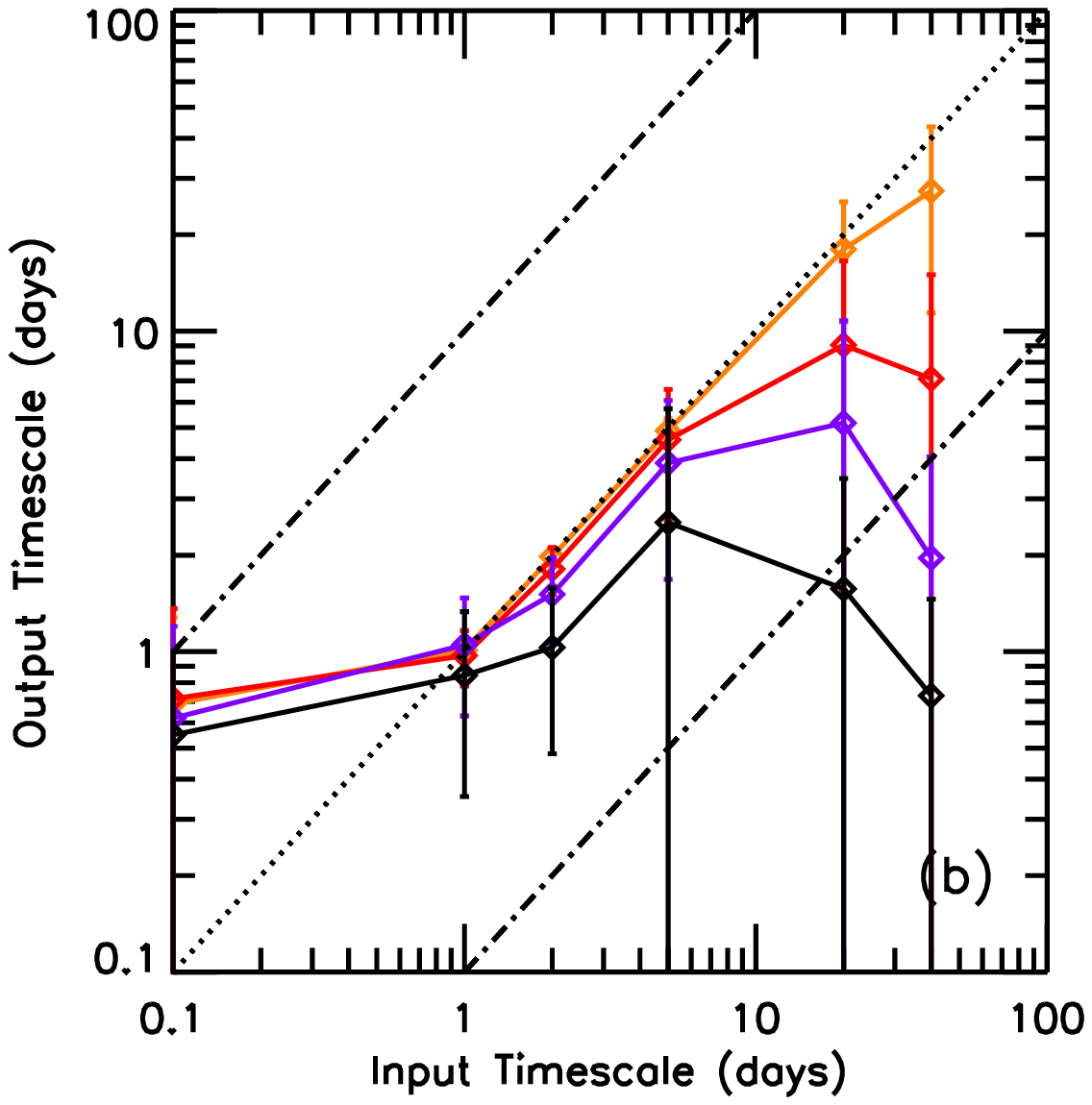}
}{
\includegraphics[width=0.32\textwidth]{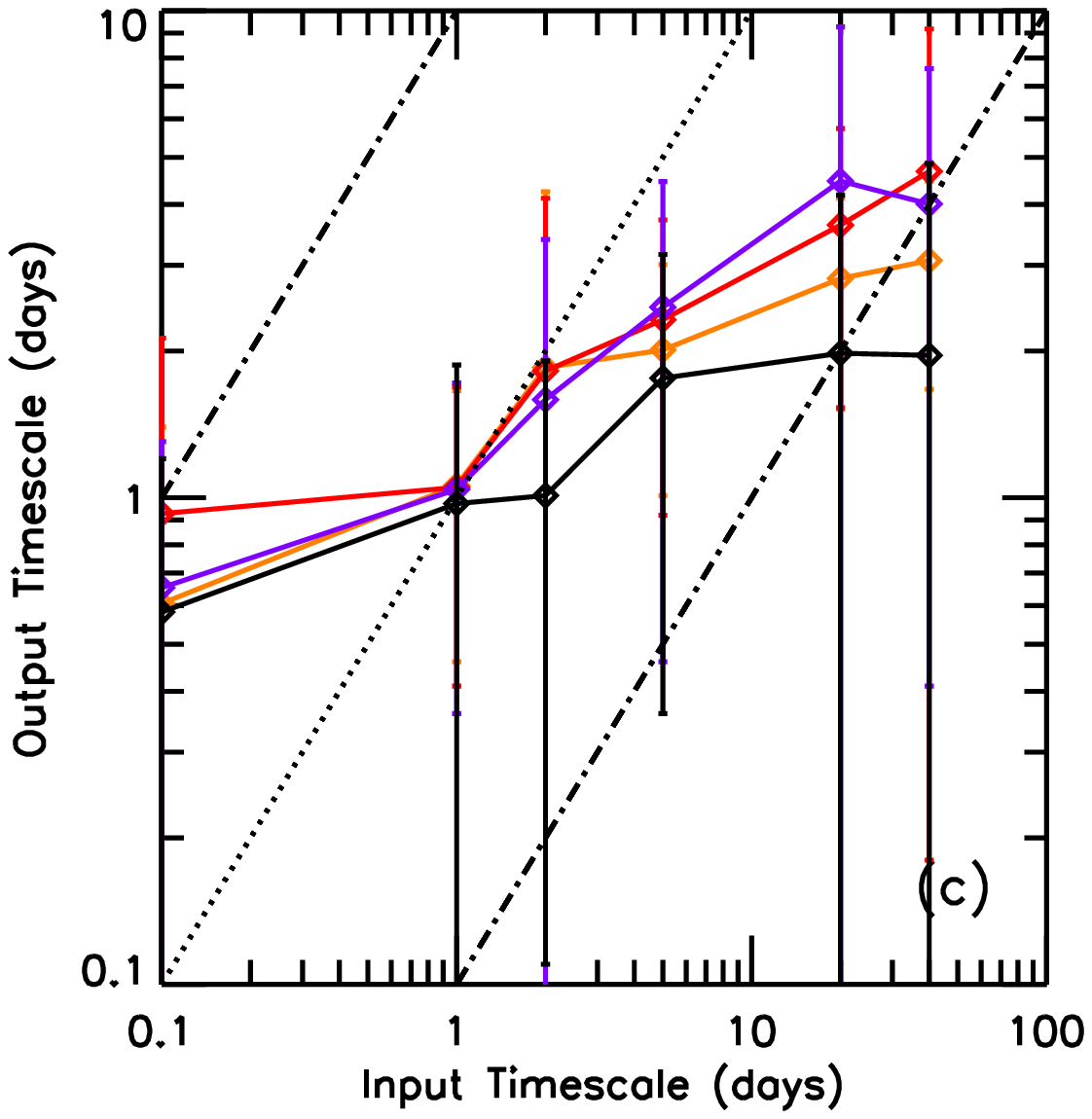}
}{
\includegraphics[width=0.32\textwidth]{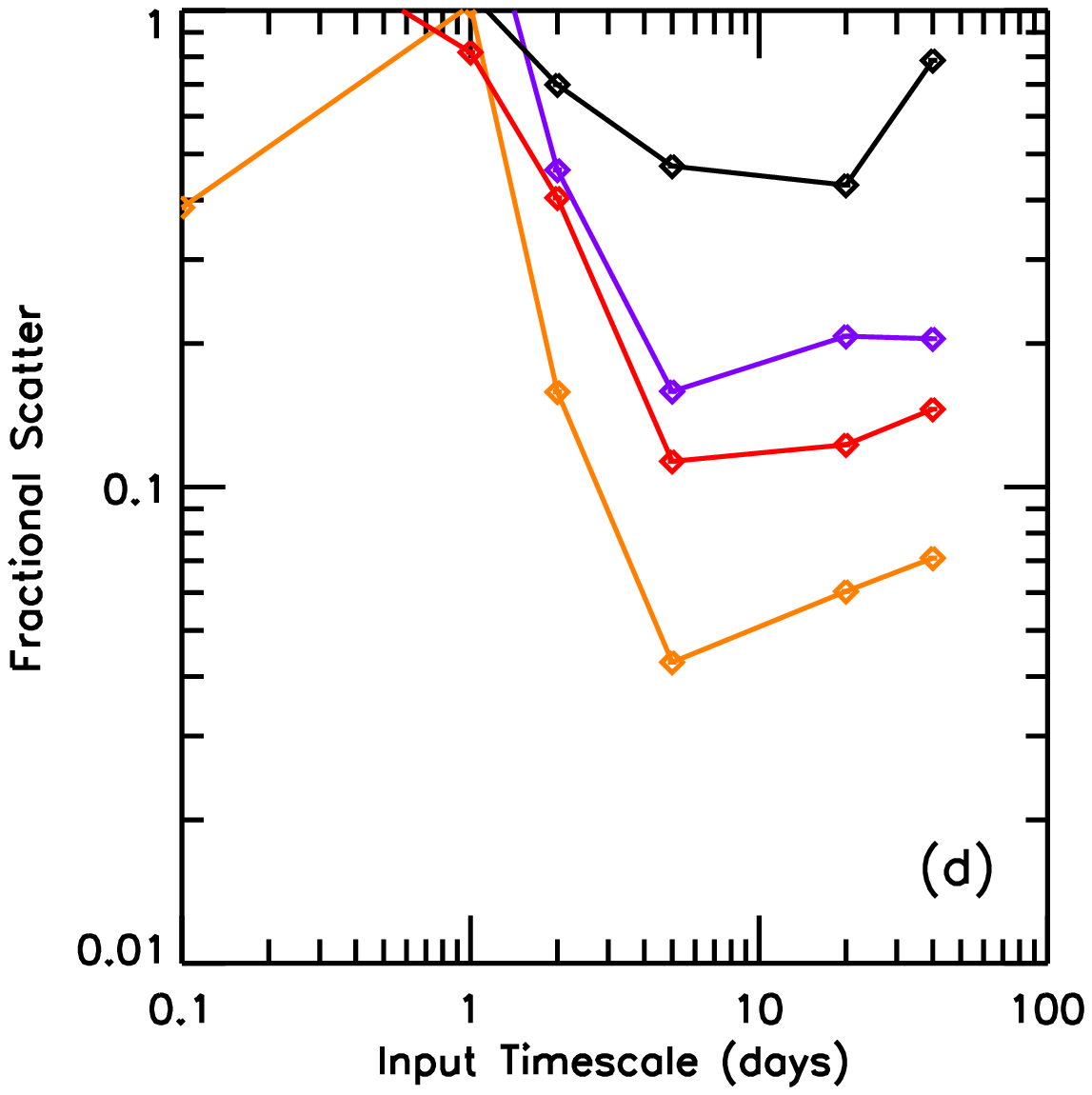}
}{
\includegraphics[width=0.32\textwidth]{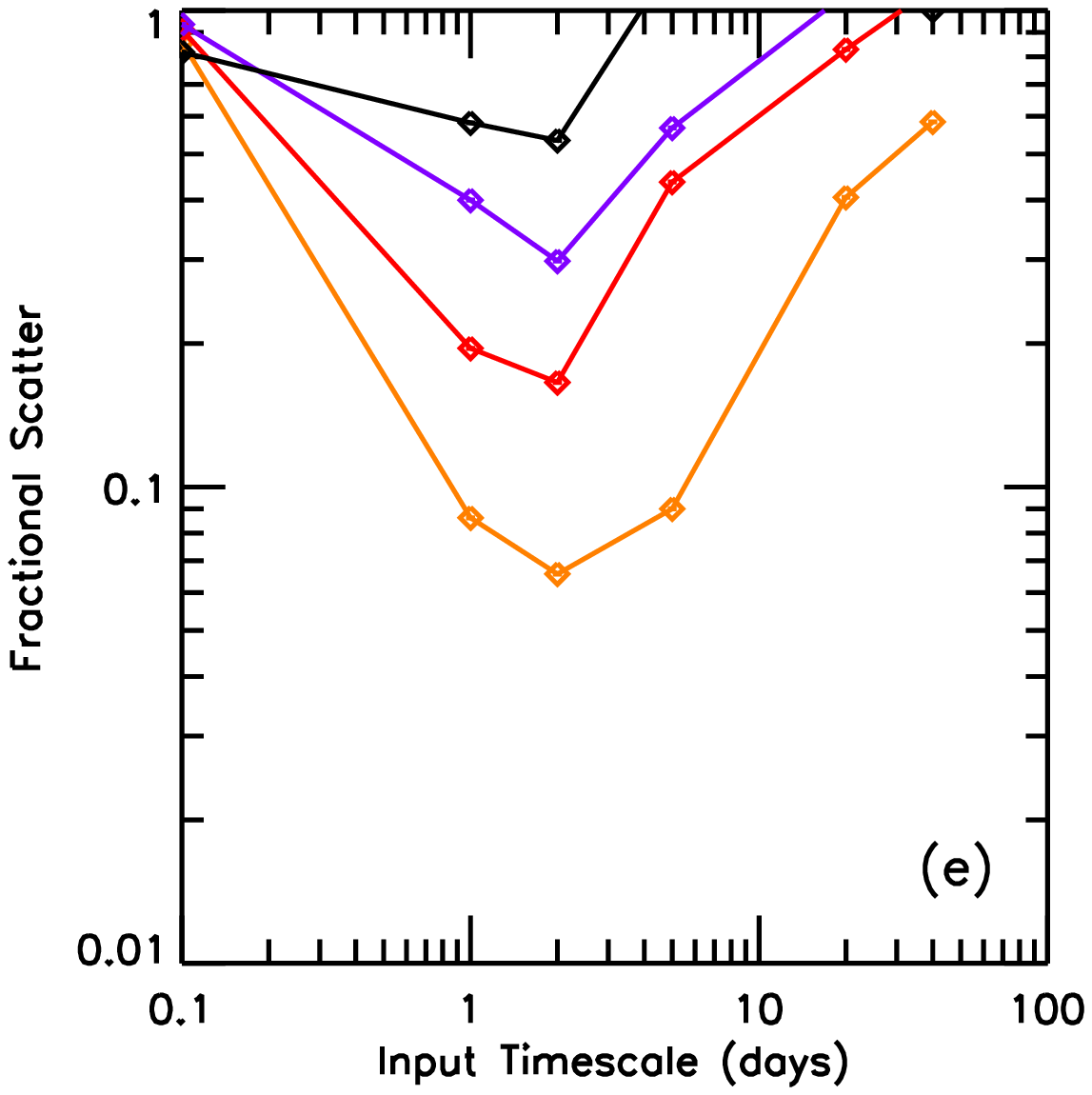}
}{
\includegraphics[width=0.32\textwidth]{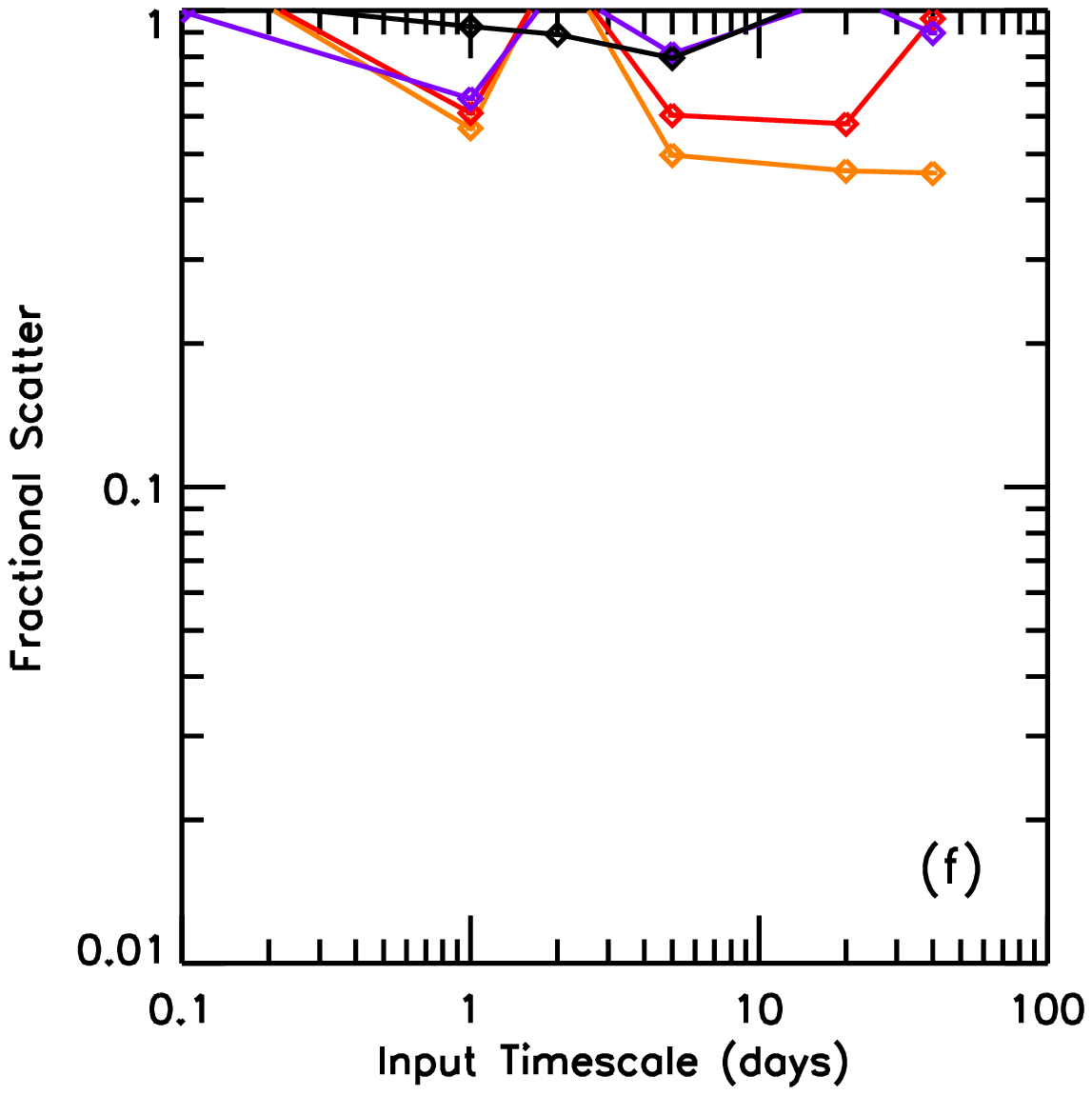}
}{
\includegraphics[width=0.32\textwidth]{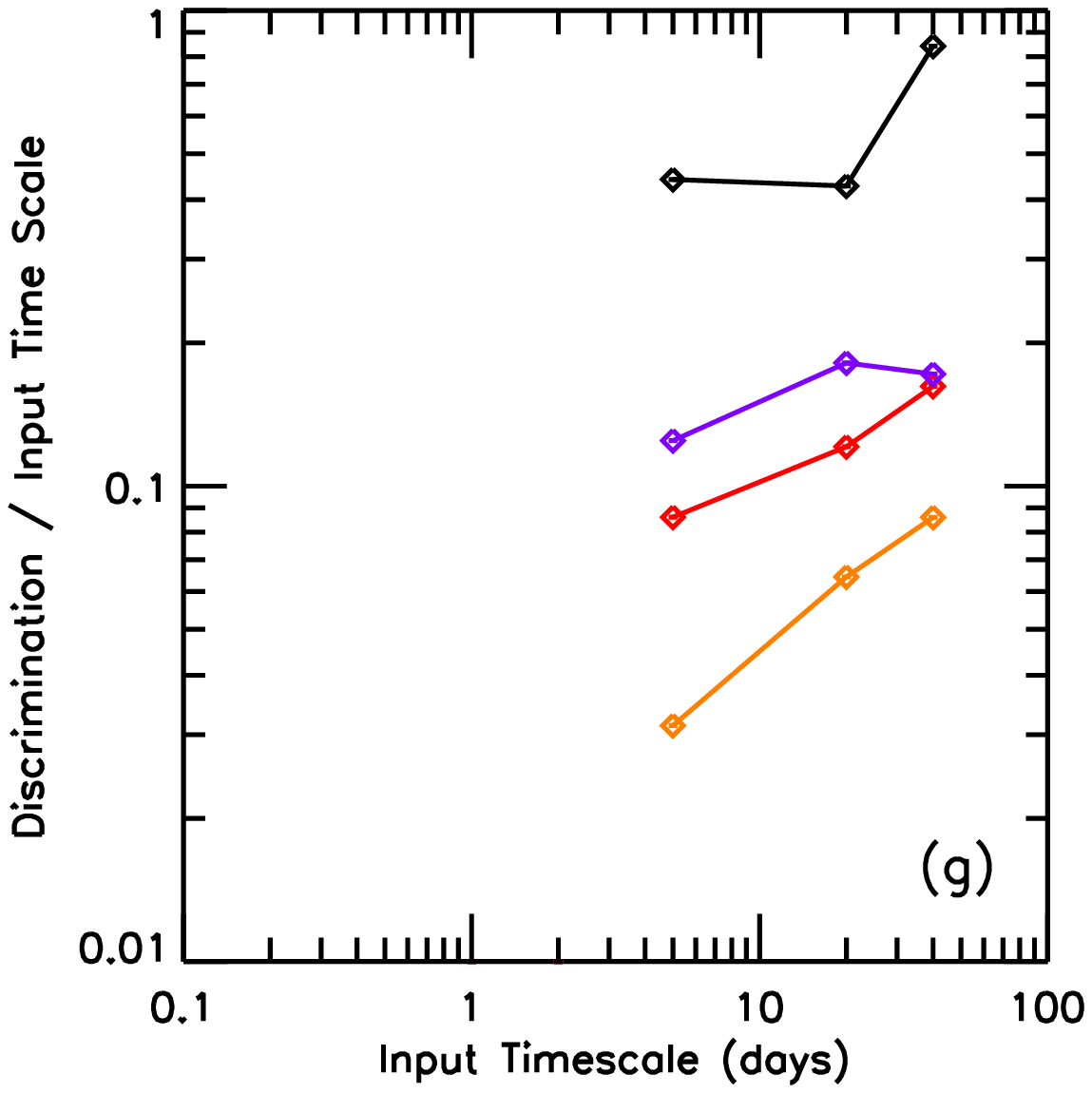}
}{
\includegraphics[width=0.32\textwidth]{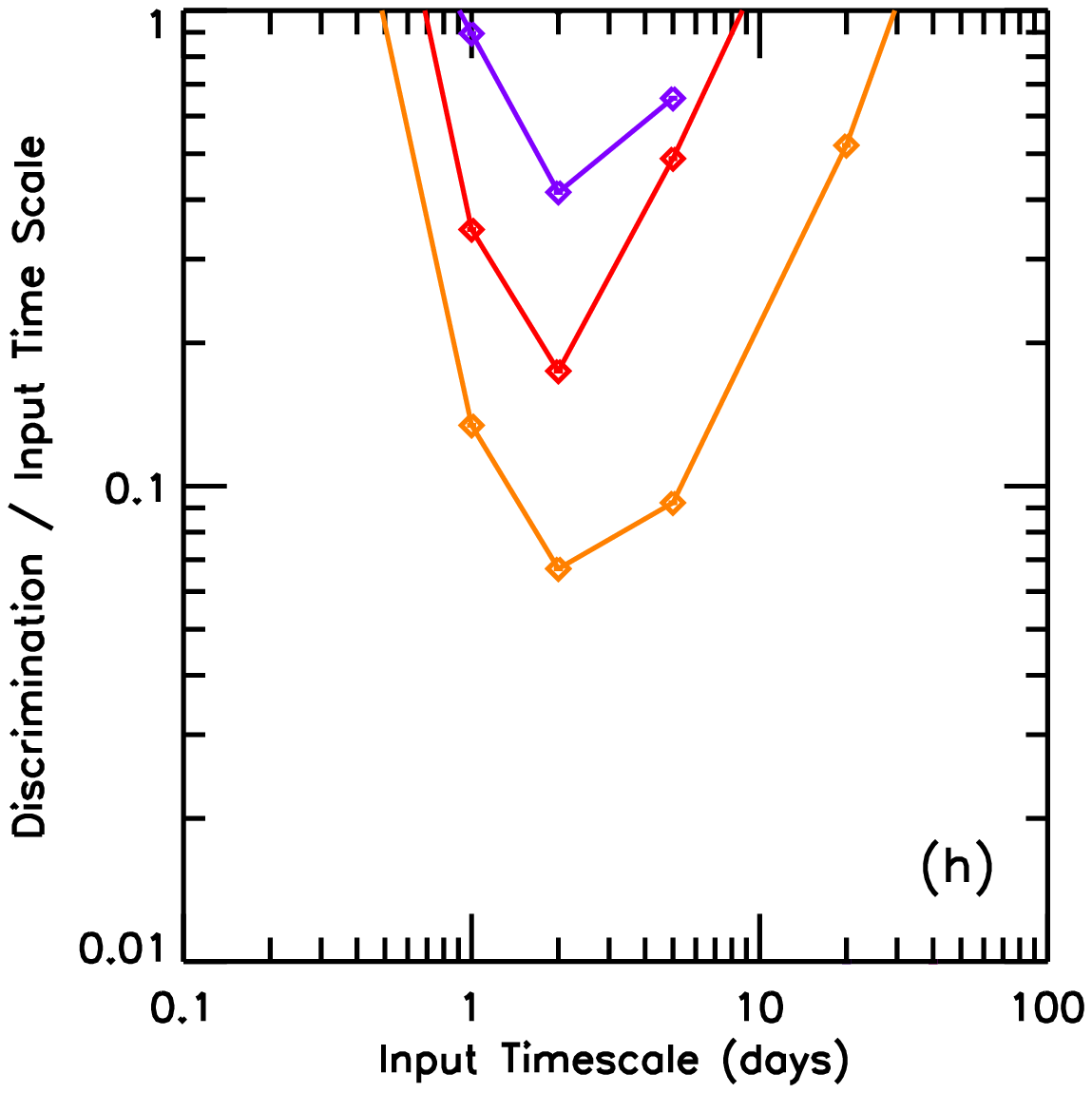}
}{
\includegraphics[width=0.32\textwidth]{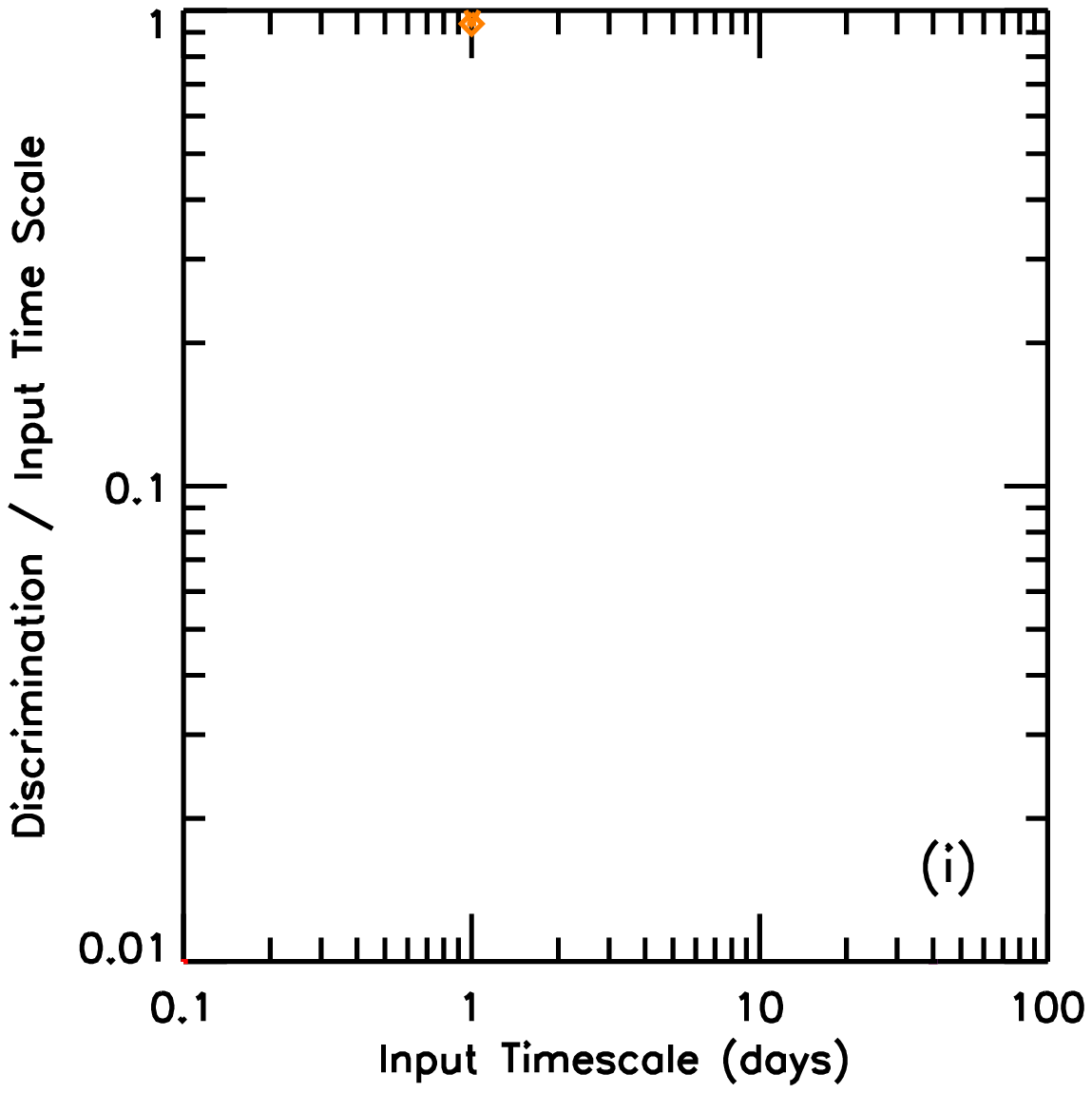}
}
\caption{The timescale calculated from a squared exponential Gaussian process model, plotted as a function of the true underlying timescale input to the simulation. Only runs with an expected 5-95\% amplitude of 0.5~mag, sampled at the YSOVAR cadence, are shown. 
Top panels show the average value of the output timescale. Middle panels show the ratio of the standard deviation to the mean output timescale. Bottom panels show the degree by which the input timescale has to change to significantly affect the output timescale. 
In all panels, orange represents a signal-to-noise ratio of 300, red represents a signal-to-noise ratio of 20, purple a signal-to-noise ratio of 10, and black a signal-to-noise ratio of 4. Note that the scatter or discrimination statistic exceeds 100\% in some cases, indicating that the metric provides no useful constraint on underlying timescales in those circumstances.}\label{fig_lcmc_gpfit_snr}
\end{figure}

Figure~\ref{fig_lcmc_gpfit_snr} shows the performance of timescales derived from Gaussian process fitting, sampled using the YSOVAR 2010 cadence, as a function of the signal-to-noise ratio of the magnitude measurements. 
The fit converged only in 20-60\% of simulations runs with timescales below 1~day, and in 60-100\% of simulations with longer timescales; the curves in Figure~\ref{fig_lcmc_gpfit_snr} are only for those modeling attempts that returned a valid solution.

For sinusoidal signals (panel a), the calculated timescale is proportional to the period for periods of two days or longer, while timescales for shorter sine periods tend to be higher than those for longer periods. For squared exponential Gaussian processes (panel b), the timescale generally increases with the true timescale except in low signal-to-noise simulations. For damped random walks (panel c), on the other hand, the calculated timescale is always 1-4~days, with only a weak dependence on underlying timescale. This might be because the damped random walk has substantial structure on timescales shorter than the characteristic timescale, and Gaussian process fitting is reported to be dominated by the most rapidly varying component (Miller, priv. comm. 2012).

For all three types of light curves, the best fit timescale has a systematic trend in the sense that the timescale is lower for lower signal-to-noise light curves. The same behavior was observed for \dmdt\ plots and for peak-finding, and just as in those cases the likely cause is that the noise (which, by construction, has an infinitesimal timescale) is being mistaken for real variability. Why this confusion should happen when fitting a model that explicitly includes a white noise term, however, is unclear. It may be a bias introduced by the use of maximum-likelihood methods, combined with a partial degeneracy in the model between a short timescale for the main process and a strong white noise component. \citet{DrwSearch} also encountered a bias towards short timescales when fitting damped random walk models to noisy light curves; we may be seeing a related issue here.

Unlike the other timescale metrics described in this work, Gaussian process fitting provides an uncertainty for the best fit timescale. 
To test the accuracy of these uncertainties, for each set of 30 identical simulations we computed 
\begin{displaymath}
\chi^2 = \sum_i \frac{(\hat{\tau}_i - \left<\hat{\tau}_i\right>)^2}{\sigma_{\hat{\tau}_i}^2}
\end{displaymath}
where $i$ denotes one of the 30 light curves in each run, $\hat{\tau}_i$ is the timescale returned by the fit, and $\sigma_{\hat{\tau}_i}$ is the error returned by the fit. 
In general, the formal errors reflect the true uncertainty (i.e., $\chi^2 \approx 30$) in $\hat{\tau}$ for light curves with timescales shorter than 1-2 days, but grossly underestimate it ($\chi^2 \sim 10^3$-$10^5$) for longer timescale light curves, with the discrepancy growing at decreasing signal-to-noise. 
Curiously, even when the light curve is a squared exponential process with noise --- in other words, the model being fitted is perfectly accurate --- the scatter in $\hat{\tau}$ still exceeds that predicted by the formal uncertainties. 
The excess scatter at long input timescales may indicate that the time series no longer covers a long enough interval to sample the full variability; given the steep dependence of running time on light curve length, we did not test any extensions of the YSOVAR 2010 cadence.

\subsubsection{Precision}

The middle three panels of Figure~\ref{fig_lcmc_gpfit_snr} show the scatter in the estimated timescale over multiple simulation runs. For sinusoidal signals and squared exponential Gaussian process signals (panels d-e), the scatter is typically a few tens of percent. The scatter in the timescale is much larger for damped random walks (panel f), on the order of 80\% or more.

\subsubsection{Discrimination}

Because the Gaussian process fitting timescale is linear with the period for a sine, the discrimination (panel g) shows the same behavior as the scatter. Because the timescale grows very slowly with the damping time for a damped random walk, the discriminating power of Gaussian process regression is never better than a factor of two (panel i). The simulations with squared exponential Gaussian process light curves (panel h) are intermediate between these two cases, with good discriminating power at high signal-to-noise but rapid degradation as the data get noisier.

\subsubsection{Completeness}

The Gaussian process fit rarely converges when applied to sines with periods shorter than 2~days, has a roughly 40\% convergence rate at 2~days, and has a high ($> 90\%$) convergence rate at longer periods. This threshold roughly corresponds to the period at which the timescale begins to show a linear dependence on the period. There is no clear trend with signal-to-noise.

When applied to a short timescale squared exponential Gaussian process light curve, the fit converges roughly 50\% of the time, with higher rates for longer true timescales. Curiously, while for timescales of 10 or 20~days the convergence rate is maximized at high signal-to-noise, for timescales of 1 or 2~days the convergence rate is higher for a signal-to-noise ratio of 10 or 20 than at signal-to-noise of 300. As with the sines, the change between these two regimes corresponds to a change in the behavior of the timescale itself: at long periods, the timescale is systematically underestimated at low signal-to-noise, just when the fraction of successful fits falls.

The rate of successful convergence is qualitatively the same for a damped random walk as for a squared Gaussian process, except with a weaker dependence on signal-to-noise. There is no obvious change in the behavior of the timescales at 5~days, when the convergence rate for low signal-to-noise light curves begins to fall.

\subsubsection{Sensitivity to Noise}

Figure~\ref{fig_lcmc_gpfit_snr} shows the performance of Gaussian process fitting as a function of the signal-to-noise ratio of the light curve. When applied to either sinusoidal or squared exponential Gaussian processes (panels d-e), the timescale inferred from Gaussian process fitting decreases systematically with signal-to-noise. When applied to a damped random walk (panel f), the timescale shows no obvious trend with signal-to-noise, but also seems to depend little on the intrinsic properties of the light curve.

The precision of timescale measurements of a sinusoidal signal or a squared exponential Gaussian process signal generally increases with signal-to-noise, as one might expect. The fractional uncertainty is 10-30\% for mid-range timescales (1-10~days), but rises to 100\% at timescales of 1~day or shorter, and 20~days or longer for the Gaussian process. 
The scatter in timescale measurements also increases at low signal-to-noise for a damped random walk, but to a much lesser degree than in the other two cases.

The discriminating power of the Gaussian process timescale roughly follows the precision for a sine wave and for a squared exponential Gaussian process (panels g-h), although in the latter case the degradation with signal-to-noise is much steeper than for the precision. The Gaussian process has almost no discriminating power when applied to a damped random walk (panel i), thanks to the combination of weak average dependence on the damping timescale and high scatter from light curve to light curve.

\subsubsection{Summary}

The numerical simulations presented in this section suggest that the Gaussian process timescale is correlated with the true timescale for some kinds of light curves but not others, nor does it always converge. Therefore, Gaussian process regression is of limited use when characterizing light curves whose statistical properties are not known a priori.

We do not recommend the use of Gaussian process models as a timescale metric, unless the data are known in advance to have only a small number of frequency components. For complex light curves, such as damped random walks, the model results are inaccurate as well as computationally expensive.

\section{Summary of Numerical Results}\label{lcmc_end}

\subsection{The Relative Merits of Candidate Timescale Metrics}\label{competition}

We summarize in Table~\ref{tbl_lcmc_all_sims} the simulated performance of each timescale metric according to the criteria provided in section~\ref{new_crit_list}. In most cases these are necessarily qualitative descriptions, as the timescale metrics' performance can vary as an arbitrary function of cadence, signal-to-noise, and simulated signal timescale. Therefore, we direct the reader to section~\ref{perform} and the accompanying figures for quantitative details.

\begin{table}[bht]
\small
\begin{tabular}{l|c c c}
Criterion			&  \dmdt\ Plots & Peak-Finding & GP Modeling \\
\hline
Precision			& 20-100\%	& 10-100\%	& 10-100\%	\\
Discriminatory power		& 20-100\%	& 25-100\%	& 6-600\%	\\
Sensitivity to Noise		& High iff RMS noise $>$ 1/10 amp.&High iff RMS noise $>$ 1/10 amp.& High	\\
Sensitivity to Cadence		& High		& High		& \nodata \\
Sensitivity to incomplete data	& High 		& Moderate	& \nodata \\
\end{tabular}
\caption{Performance of timescale metrics according to our simulations. See the individual sections of this work for more details. A value of \nodata means the criterion was not tested.}\label{tbl_lcmc_all_sims}
\end{table}

The scatter in timescale measurements between different random realizations of the same signal (``precision'') covers a range from a few tens of percent to a factor of two for all three timescale metrics tested. The precision is usually a strong function of the light curve timescale as well as of signal-to-noise. In general, peak-finding tends to show slightly less scatter than \dmdt~plots do.

The tested timescale metrics do differ in the smallest difference in underlying timescale that can be distinguished by a timescale metric (``discrimination''). The poor correlation between the timescale inferred from Gaussian process regression and the timescale of the simulated light curve manifests as an inability to determine the timescale to better than an order of magnitude. The discrimination of \dmdt~plots and peak-finding are both much lower, and comparable to each other. While peak-finding is more precise than \dmdt~plots, it also suffers from larger systematic effects, and the discrimination score depends on both.

While simulations showed that the scatter of timescale metrics usually rises smoothly with decreasing signal-to-noise, the correlation between inferred and underlying timescale underwent a sharp transition between simulations with a signal-to-noise ratio of 20 and a signal-to-noise ratio of 10, for runs with a 5-95\% amplitude of 0.5~mag. This corresponds to a ratio between the RMS amplitudes of a sinusoid and of the noise of 5-9, and a ratio between the RMS amplitudes of a Gaussian process or damped random walk and of the noise of 1.5-3. For noisier light curves than this threshold, the timescale would be systematically underestimated, while light curves with better signal-to-noise had much more reliable measurements. This was not the case with Gaussian process regression, which tended to produce a too-low timescale when applied to light curves with any noise, and where the size of the bias rose smoothly with decreasing signal-to-noise.

Finally, both the \dmdt\ plots and peak-finding plots are highly dependent on the details of the cadence used, in particular the typical sampling frequency and the total time base line covered. \dmdt\ plots are only linearly correlated with the input timescale over timescales between the survey's characteristic cadence and 1/15 of the survey time base line, and appear to ``saturate'' at 1/5 the survey base line. Peak-finding shows high scatter above 1/20-1/15 the base line and levels out at 1/5 the survey base line, but appears to be more robust when the light curve timescale is comparable to the survey cadence. These numbers are for the timescales input as model parameters, which are typically shorter than more intuitive measures such as periods or peak-finding by a factor of several.

While Gaussian process regression is too unreliable to use in research, the limitations of \dmdt\ plots and peak-finding are modest enough to permit \emph{careful} use in real applications. Both metrics offer a rough characterization of the characteristic timescale of a light curve, provided that the photometry is reasonably clean and noise-free. The most substantial difference between the two metrics is the maximum underlying timescale, relative to the observing cadence, at which the metric gives an unbiased result. 
We recommend the use of peak-finding for signals whose statistical properties are known \textit{a priori} (e.g., they are all well described by a particular model) as well as timescales much shorter than the monitoring base line, and the use of \dmdt\ plots for signals of unknown form but with timescales known to be intermediate between the cadence and the maximum base line.

\subsection{Converting Between Timescale Metrics}

Because they probe different parts of the light curve, timescales based on competing metrics cannot be directly converted to each other; the conversion factor depends on the statistical properties of a signal, which for a real light curve are usually unknown \textit{a priori}. The distinction is similar to the varying conversion factor between RMS amplitude and peak-to-peak amplitude for light curves with different properties.

\begin{table}[hbt]
\small
\setlength{\tabcolsep}{2pt}
\begin{tabular}{l|c c c}
		 & \multicolumn{3}{|c}{Multipliers of the \dmdt\ Timescale} \\
Timescale Metric & Sinusoid & Squared Exponential GP & Damped Random Walk \\
\hline
Period & $\sim 7$ & \nodata & \nodata \\
\dmdt\ 90\% quantile crosses $1/2$ amp & 
	$1$ & 	$1$ & 	$1$ \\
Peak-finding crosses $1/2$ amp & 
	$\sim 3$ & 	$\sim 2 $ & 	$\sim 2 $ \\
Peak-finding crosses 80\% of peak-finding max & 
	$\sim 3 $ & 	$\sim 6 $ & 	$\sim 6$ \\
\end{tabular}
\caption{Output timescale of each metric on simulated data, normalized by the \dmdt\ timescale. This table can be used to convert between different timescale metrics by dividing each column by the timescale to be converted \emph{from}. For example, the period is usually around 7/3 the peak-finding timescale for a periodic signal. Two variants of the peak-finding timescale are listed: finding the time at which the peak-finding curve crosses 80\% of its maximum is the approach adopted by \citet{AmcCorot}, while we prefer to find the time at which it crosses half the light curve amplitude, which gives less scatter in measurements at the cost of systematically lower results at long timescales. All results are for a simulated 0.5~mag peak-to-peak light curve observed at a median signal-to-noise ratio of 20 using the PTF-NAN Full cadence.} \label{tbl_lcmc_tau_convert}
\end{table}

In Table~\ref{tbl_lcmc_tau_convert}, we show the ratio of all simulated timescales to \dmdt\ timescales for each light curve model considered here. We used simulation runs with the PTF-NAN Full cadence, since the cadence has a high dynamic range, and ignored the long- and short-timescale runs where the timescale reported by the metric was no longer proportional to the timescale input to the simulation. Gaussian process timescales are not listed because the metric does not show a clear correlation with light curve properties. While most definitions of timescale are within an order of magnitude of each other, they can still differ by a factor of several. These conversions need to be kept in mind while comparing results from different papers.

The timescale conversion factors between peak-finding and \dmdt\ timescales differ greatly between the sinusoid and the aperiodic models we considered. This difference in scale arises because peak-finding characterizes the most extreme variations in a light curve, while \dmdt\ timescales characterize the most typical variations. 
For a sinusoidal model, we showed in \citet{KpfThesis} that the \dmdt\ timescale is 1/6 the period, consistent with a factor of 7 difference between peak-finding (when adopting \citet{AmcCorot}'s convention of using the distance between consecutive minima or consecutive maxima) and \dmdt\ plots after allowing for errors introduced by noise and cadence. 
For either flavor of Gaussian process model, on the other hand, the probability of an extremum falls faster than exponentially as one moves away from the model mean, so the peaks selected by peak-finding become proportionally rarer. 
The ratio between peak-finding and \dmdt\ timescales for real light curves therefore depends on whether observed fluxes tend to cluster in the middle of a source's variability range (as in a Gaussian process) or towards the edges (as in a sinusoid).

A major result of this section is that, while relative timescales are meaningful when
using the same metric, timescale comparisons across metrics should be avoided without
careful attention to systematics such as those reported in Table 3. Unlike the situation for periodic variability, where timescale is naturally defined by the period, for aperiodic variability the derived timescale is inherently tied to the choice of tuning parameters such as magnitude-difference thresholds. A good choice of threshold may require characterization of other aspects of the light curve, particularly its variability amplitude.

\section{Applications and Conclusions}

Aperiodic variability is a common feature of accreting systems, such as young stars, cataclysmic variables, and active galactic nuclei. In addition, aperiodic variability is a key characteristic of the light curves of massive stars. Researchers in many fields of astronomy can use aperiodic variability to probe physics of interest, complementary to traditional techniques such as spectroscopy or multi-wavelength studies.

However, so far astronomy does not have a standard approach to the treatment of aperiodic variability, analogous to the common use of Fourier analysis and periodograms for periodic signals in all fields. Instead, AGN researchers focus on structure functions \citep[and references therein]{2010MNRAS.404..931E}, massive star researchers prefer periodograms or wavelet analysis \citep[e.g.,][]{1998ApJ...499L.195M,2008ApJ...679L..45M}, and YSO researchers use wavelets or autocorrelation functions \citep[e.g.,][]{CodyMost}. This fragmentation has impaired attempts to solve problems common to time series analysis across astronomical fields.

Periodic time series analysis is at the forefront of extrasolar planet detection, asteroseismology, the Cepheid period-luminosity relation, pulsar timing, and many other fields, despite several key weaknesses of Fourier analysis. In particular, the pattern of observations almost always introduces spurious peaks (``aliases'') in the frequency spectrum, which can confuse measurements of the period. Period analysis has been successful because time domain astronomers are aware of the danger of aliasing, and are prepared to use either simulations or analytical techniques \citep[e.g., following][]{WindowFunction} to identify aliased periods. The well-characterized nature of period analysis contributes as much to its value as a scientific tool as more obvious factors such as its precision.

A well-characterized metric is also essential for accurately analyzing timescales of aperiodic variables. \citet{2010MNRAS.404..931E} showed that invalid conclusions can be drawn from structure functions because of cadence-related artifacts, \citet{deDiego2010} and \citet{deDiego2014} did the same for variance-based tests, and we have shown that the quality and accuracy of \dmdt~plots, peak-finding plots, and Gaussian process regression also depends on the observing cadence. To our knowledge, there is no analytical treatment of the effect of arbitrary sampling patterns on aperiodic timescale analysis, so simulations are the only way at present to constrain the reliability of these timescale metrics.

We have attempted to characterize the dependence in terms of characteristic timescales of an observing pattern, particularly the average cadence and the time base line, but we cannot prove a direct connection between timescale metric performance and these figures. We recommend that future researchers use cadence properties only as a rough estimate of what timescales can be probed by an observing pattern (bearing in mind that \dmdt\ plots are slightly more robust to limited time base lines, while peak-finding is more robust to coarse sampling; see section~\ref{competition}), and run simulations specific to their data sets when analyzing aperiodic timescales.

In addition, we have shown that aperiodic timescale metrics have uncertainties of a few tens of percent or greater, considerably more than is typical for measurements of periods. These large errors are due in part to the absence of any coherence in the light curve; while collecting more data does make \dmdt\ or peak-finding timescales more precise, the uncertainty does not decrease inversely with the observing base line as it does for periods. It is important for researchers to keep these large uncertainties in mind, lest they over-interpret their data to draw erroneous conclusions.

With the growing prominence of time domain astronomy, we will soon have access to a wealth of data for both Galactic and extragalactic aperiodically varying sources. We have outlined two preliminary techniques, \dmdt\ plots and peak-finding plots, for extracting information from irregularly sampled, aperiodic light curves, along with limitations that need to be kept in mind to obtain valid results. We have also presented a conceptual and software framework that can be used to test the suitability of different aperiodic light curve analysis techniques for specific observing cadences. These tools and techniques represent only the first steps towards making the most of the coming wave of time domain astronomy.

\acknowledgments

We thank the referee for valuable corrections and feedback. 
We would also like to thank the Time Domain Forum, a Caltech initiative organized by Ashish Mahabal, for valuable feedback on time series techniques and their characterization. 
K. F. also thanks Adam Miller and Timothy Morton for suggestions they made for the LightcurveMC software.

\bibliographystyle{hapj}
\bibliography{../../../agn_data,../../../history,../../../psych,../../../sfr,../../../stats,../../../ysovar_data,../../../ysovar_methods,../../../ysovar_theory}

\end{document}